\documentclass[11pt,letterpaper]{article}
\usepackage[margin=0.95in,headheight=15pt]{geometry}
\usepackage{amsmath,amssymb,mathtools,amsthm}
\newtheoremstyle{manualplain}{8pt}{6pt}{\itshape}{}{\bfseries}{.}{.5em}{\thmnote{#3}}
\theoremstyle{manualplain}
\newtheorem*{formalstatement}{}
\newenvironment{holantstatement}[1]{\begin{formalstatement}[#1]}{\end{formalstatement}}
\newtheoremstyle{manualdefinition}{8pt}{6pt}{\normalfont}{}{\bfseries}{.}{.5em}{\thmnote{#3}}
\theoremstyle{manualdefinition}
\newtheorem*{formaldefinition}{}
\newenvironment{holantdefinition}[1]{\begin{formaldefinition}[#1]}{\end{formaldefinition}}

\usepackage{fontspec,unicode-math}
\usepackage{booktabs,longtable,array,xcolor,microtype,fancyhdr}
\usepackage[unicode,colorlinks=true,linkcolor=blue!35!black,citecolor=blue!35!black,urlcolor=blue!40!black]{hyperref}
\providecommand{\tightlist}{\setlength{\itemsep}{0pt}\setlength{\parskip}{0pt}}
\allowdisplaybreaks[1]
\title{A Dichotomy for Cubic Bipartite Holant Problems\\with Complex Algebraic Weights}
\author{Yin (Hugh) Liu\\[0.4ex]\normalsize Google\\[0.2ex]\small\href{mailto:hughliu@google.com}{\texttt{hughliu@google.com}}}
\date{}

\usepackage{tikz}
\usetikzlibrary{arrows.meta,positioning,calc}
\usepackage{float}
\usepackage{caption}
\definecolor{leftink}{HTML}{215D80}
\definecolor{rightink}{HTML}{99561C}
\tikzset{
 L/.style={draw=leftink,fill=leftink!8,rectangle,minimum size=8mm,inner sep=2pt,font=\small},
 R/.style={draw=rightink,fill=rightink!8,circle,minimum size=8mm,inner sep=1pt,font=\small},
 boxg/.style={draw,rounded corners=2pt,fill=black!3,minimum width=13mm,minimum height=9mm,inner sep=4pt,font=\small},
 interpolated/.style={boxg,dashed},
 lab/.style={fill=white,inner sep=1.2pt,font=\small},
 holantstep/.style={draw,rounded corners=2pt,align=center,text width=9.4cm,inner sep=6pt,font=\small},
 arr/.style={-{Stealth[length=2mm]},thick},
 every picture/.style={line width=0.65pt}
}

\hypersetup{pdfauthor={Yin (Hugh) Liu},pdftitle={Cubic bipartite Holant with complex algebraic weights},pdfsubject={Holant complexity dichotomy}}
\begin{document}
\maketitle
\vspace{-2em}
\textbf{Abstract.} We classify the exact evaluation of \(\operatorname{Holant}(f\mid=_3)\) for every fixed complex algebraic symmetric Boolean ternary signature \(f\). An input is a cubic bipartite multigraph: every vertex on one side carries \(f\), every vertex on the other side carries ternary equality, and no auxiliary signatures are freely available. The tractable signatures are precisely rank-one tensors, generalized equalities, and equality-preserving cube-root diagonal transformations of six affine signatures, together with nonzero scalings and reversal. Every other signature gives a \#P-hard problem under polynomial-time Turing reductions. We also identify the exact real intersection: it consists of the same tractable families as in the rational classification, with real algebraic parameters. The proof preserves degree exactly three on both sides of every oracle instance. A rank-one matrix extracted by interpolation supplies one unary signature only after its unused factor has been absorbed in triples. Over the complex numbers this absorption has three exceptional projective directions. We combine this constraint with projective matrix-group orbits, explicit ternary replacements, and an exhaustive treatment of finite projective orders. The cases of orders three and five include exact polynomial certificates; the certificate identities and a rational-arithmetic verifier are supplied as supplementary material.

\clearpage
\begingroup\small\setlength{\parskip}{0pt}
\tableofcontents
\endgroup\clearpage
\section{1. Introduction and the dichotomy}\label{introduction-and-the-dichotomy}

Counting complexity asks which weighted combinatorial sums admit exact polynomial-time evaluation. A local description of a sum need not reveal its complexity. For example, the same graph can carry a matching constraint, a parity constraint, or an equality constraint at each vertex, and these choices lead to different counting problems. The Holant framework expresses such sums by assigning a function to every vertex, placing Boolean variables on edges, multiplying the local function values, and summing over all edge assignments. Holographic transformations can expose algebraic structure that is not apparent from the original functions \cite{ref1,ref2}.

Here we impose a particularly restrictive graph structure. Every vertex has degree three; the graph is bipartite; one side carries a single symmetric ternary function \(f\); and the other side carries ternary equality. Equality identifies the three incident edge variables. Consequently this is also a ternary counting constraint satisfaction problem in which every variable occurs \textbf{exactly three times}. A parallel edge represents two occurrences of the same variable in one constraint, and is permitted. The graph is not required to be planar.

The restriction to exactly three occurrences affects reductions as well as the problem statement. Adding a unary constraint, splitting a variable into an equality tree, or composing a gadget with a transpose can change the allowed degree or the prescribed side of a signature. None of those operations is available without a proof. This issue persists even when a dichotomy is known for a more general counting framework: hardness on a larger family of instances need not hold on the cubic bipartite subfamily.

Fan--Cai classified the nonnegative symmetric case \cite{ref4}, and Cai--Fan--Liu classified rational symmetric weights, including mixed signs \cite{ref5}. We extend the latter classification to arbitrary fixed complex algebraic weights. The resulting tractable signatures are familiar rank-one, equality, and affine signatures with equality-preserving diagonal transformations. The principal assertion is that these are the only tractable cases under the strict degree and bipartition requirements. The proof must address arbitrary root-of-unity orders, complex zeros of absorption factors, and singular or finite-order gadget recurrences.

\subsection{1.1 The input problem and the computational model}\label{the-input-problem-and-the-computational-model}

Fix a symmetric Boolean ternary signature

\[f=[f_0,f_1,f_2,f_3]\in\overline{\mathbb Q}^{\,4}.\]

The notation means \(f(i,j,k)=f_{i+j+k}\) for bits \(i,j,k\in\{0,1\}\). The entries may be genuinely complex algebraic numbers. In particular, they are fixed constants of the problem rather than part of the input.

An input is a cubic bipartite multigraph \(G=(L,R,E)\). Parallel edges are allowed, as they are in the underlying cubic bipartite Holant model. Every left vertex carries \(f\); every right vertex carries ternary equality \(e=[1,0,0,1]\). The output is

\[Z_f(G)=\sum_{\sigma:E\to\{0,1\}}
\prod_{v\in L} f_{\sum_{e\ni v}\sigma(e)}
\prod_{w\in R} e\bigl(\sigma|_{E(w)}\bigr).\]

The equality factor is one exactly when all three incident bits agree. Thus it is also possible to regard each right vertex as a Boolean variable of degree exactly three. The left vertices are ternary constraints. Neither arbitrary variable degrees nor free unary signatures are permitted.

The field \(K=\mathbb Q(f_0,f_1,f_2,f_3)\) is fixed. Represent an output by its rational coordinates in a fixed basis of \(K/\mathbb Q\); the rationals are written in binary. A reduction may use a further fixed finite extension containing its eigenvalues and gadget constants. Its degree is independent of the input graph. Thus FP below refers to bit complexity for exact algebraic output, rather than to unit-cost arithmetic on arbitrary complex numbers. Hardness means \#P-hardness under polynomial-time Turing reductions. The signature \(f\) itself is fixed, so Theorem A is a classification of one evaluation problem for each \(f\), not a uniform problem in which an unrestricted algebraic-number encoding is supplied as input.

For real algebraic \(f\), the output field \(K\) is a subfield of \(\mathbb R\). A proof may still use complex eigenvalues in a fixed extension when reconstructing a polynomial coefficient. This does not change an oracle instance: every final vertex still carries the original real \(f\) or equality. Rational-coordinate arithmetic in the larger fixed field is simply part of the reduction. Thus the real restriction below has the same exact-output interpretation, and does not assume access to a complex-weight oracle in addition to its real-weight oracle.

\subsection{1.2 The dichotomy}\label{the-dichotomy}

A rank-one signature has the form

\[\lambda[u_0^3,u_0^2u_1,u_0u_1^2,u_1^3],\]

where the constants are algebraic complex numbers; this includes the zero signature. A generalized equality has the form \([x,0,0,y]\).

For \(\omega^3=1\), define the diagonal transformation

\[\Gamma_\omega[f_0,f_1,f_2,f_3]
=[f_0,\omega f_1,\omega^2f_2,f_3].\]

The matrix \(\operatorname{diag}(1,\omega)\) gives this transformation on the left signatures and preserves ternary equality on the right. We call it an equality-preserving diagonal transformation. Reversal is \(f^{\mathrm{rev}}=[f_3,f_2,f_1,f_0]\).

Let \(\mathcal T\) consist of the rank-one signatures, generalized equalities, and all nonzero scalar multiples of cube-root gauges and reversals of the following six signatures:

\[[1,0,1,0],\quad [1,0,-1,0],\]

\[[1,1,-1,-1],\quad[1,-1,-1,1],\quad
[1,i,1,i],\quad[1,-i,1,-i].\]

The six affine representatives and their three cube-root gauges give eighteen distinct normalized signatures with first entry one. The gauged signatures need not themselves be affine in the original basis; they inherit tractability from their affine representatives. Reversal also includes signatures whose first endpoint is zero. We keep reversal explicit so that no endpoint case is hidden by normalization.

\begin{holantstatement}{Theorem A (complex algebraic cubic bipartite dichotomy)}

For every fixed complex algebraic symmetric ternary signature \(f\), the problem \(\operatorname{Holant}(f\mid e)\) is polynomial-time computable when \(f\in\mathcal T\), and is \#P-hard otherwise.

\end{holantstatement}

The proof is organized by the first actual straddled matrix

\[A_f=\begin{pmatrix}f_0&f_2\\f_1&f_3\end{pmatrix}.\]

When at least one endpoint is nonzero, scaling and reversal let us write \(f=[1,a,b,c]\). Singular \(A_f\), repeated eigenvalues, infinite projective order, and each possible finite projective order are treated separately. The final section treats \(f_0=f_3=0\) without assuming that normalization is available.

\subsection{1.3 The real restriction and the additional complex phases}\label{the-real-restriction-and-the-additional-complex-phases}

The complex classification contains the real algebraic classification as an exact restriction. We spell out the intersection because a complex scalar or diagonal transformation could otherwise obscure which real signatures occur.

\begin{holantstatement}{Corollary A.1 (real algebraic cubic bipartite dichotomy)}

Let \(f\) be a fixed real algebraic symmetric ternary signature. Then \(\operatorname{Holant}(f\mid e)\) is in FP for signatures in the union \(\mathcal T_{\mathbb R}\) of the following families, and is \#P-hard outside that union:

\begin{enumerate}
\def\labelenumi{\arabic{enumi}.}
\tightlist
\item
  real rank-one signatures, including zero;
\item
  real generalized equalities \([u,0,0,v]\);
\item
  nonzero real scalar multiples and reversals of
\end{enumerate}

\[[1,0,1,0],\quad[1,0,-1,0],\quad
[1,1,-1,-1],\quad[1,-1,-1,1].\]

In particular,

\[\mathcal T\cap\mathbb R^4=\mathcal T_{\mathbb R}.\]

\end{holantstatement}

\begin{proof}[Proof]

First, a real signature that is rank one over \(\mathbb C\) also has a real rank-one representation. If its first entry is nonzero, divide by that real entry. Rank one then gives

\[f/f_0=[1,t,t^2,t^3],\qquad t=f_1/f_0\in\mathbb R.\]

If its first entry is zero, a nonzero rank-one ternary tensor has only its last entry nonzero and again has a real representation. The zero tensor is included. A generalized equality that is real has real endpoint parameters, so its intersection causes no further cases.

Consider now a scalar multiple of a diagonally transformed affine representative. Reversal preserves reality and commutes with the diagonal operation according to

\[\bigl(\Gamma_\omega h\bigr)^{\mathrm{rev}}
=\Gamma_{\omega^{-1}}\bigl(h^{\mathrm{rev}}\bigr),\qquad\omega^3=1.\]

We may therefore reverse the entire signature first if necessary. It remains to inspect the six displayed representatives, all of whose first entries are one. Reality forces their nonzero scalar multiplier to be real. For the parity-supported representatives the normalized third entry is \(\pm\omega^2\); its reality forces \(\omega=1\), since 1 is the only real cube root of unity. For the two full-support real representatives the second entry is \(\pm\omega\) and gives the same conclusion. The remaining representatives \([1,i,1,i]\) and \([1,-i,1,-i]\) have nonreal last entry, which the cube-root transformation leaves unchanged. Neither can give a real signature after a real scalar multiplication. This proves that the displayed real families exhaust the intersection; the reverse inclusion follows by taking \(\omega=1\).

Theorem A now gives the dichotomy. The real output convention is the one in Section 1.1: although intermediate calculations may take place in a fixed complex extension, the final oracle graphs use the given real signature throughout.
\end{proof}

This corollary extends the rational theorem \cite{ref5} without adding a new real tractable family. Over the complex numbers there are two distinctions. A primitive cube-root twist such as

\[[1,\omega,-\omega^2,-1]=\Gamma_\omega[1,1,-1,-1]\]

is nonreal but has exactly the same partition function as the real representative on every cubic bipartite graph. The reason is that each right equality vertex contributes either zero or three one-valued incidences, so the total multiplier is one. By contrast,

\[[1,i,1,i]\]

is a genuinely additional complex affine case: it cannot be made real by any combination of nonzero scaling, cube-root diagonal transformations, and reversal. Its endpoint ratio is \(i\); the diagonal transformation and scaling preserve that ratio, and reversal replaces it by \(-i\). Both are nonreal. Its tractability instead follows from its quadratic Boolean phase, as proved below. Thus the complex theorem is more than the real theorem together with equality-preserving twists of real signatures.

Appendix B records several consequences specific to real weights: the smaller exceptional set in unary absorption, an elementary real-matrix construction of an interpolating recurrence, and examples showing why particular shortcuts fail. These arguments clarify the real restriction without supplying a second proof of the entire dichotomy.

\subsection{1.4 Relation to earlier dichotomies}\label{relation-to-earlier-dichotomies}

There are three distinct comparisons with the existing literature.

First, the present problem has the same input model as the rational-weight theorem of Cai--Fan--Liu \cite{ref5}. Several of its graph constructions remain valid over the complex numbers; we retain these constructions and derive their matrices explicitly in Section 2. Fan--Cai's nonnegative theorem \cite{ref4} provides the Exact-One hardness base. The planar cubic bipartite work of Cai--Fan \cite{ref6} supplies a further ternary construction. Li--Huang--Zheng also classify nonnegative ternary functions without the symmetry restriction \cite{ref20}. Their theorem removes the symmetry restriction while retaining nonnegative weights. The present theorem retains symmetry and extends the rational mixed-sign classification to complex algebraic weights.

The splitting of straddled binary signatures originates with Fan and Cai \cite{ref4} and is also used by Li, Huang, and Zheng \cite[Lemmas 4 and 5]{ref20}. Their subsequent reduction to unrestricted counting CSP uses positive unary weights and weighted binary equalities \cite[Lemmas 10 and 11]{ref20}. For example, positivity ensures that a parameter \(\beta^{3/2}\) is not a root of unity when \(\beta>0\) and \(\beta\ne1\). With complex weights, interpolation ratios can instead have finite order, and the contraction used to dispose of an unused factor can vanish. Our interpolation and absorption lemmas state the required nonvanishing conditions explicitly; the later case analysis supplies legal gadgets when those conditions initially fail.

Second, Huang--Lu classify symmetric complex-weighted \(\#\operatorname{CSP}^{d}\), in which each variable occurs a positive multiple of \(d\) times \cite[Theorem 4.1]{ref8}. Their tractable condition predicts the diagonal affine transformations appearing here. When \(d=3\), a twelfth-root diagonal multiplier splits into a cube-root multiplier and a fourth-root multiplier. The fourth-root part is an affine linear phase, whereas the cube-root part preserves ternary equality. Lin's subsequent classification treats general degree-divisible signatures \cite{ref9}. Neither statement imposes degree exactly three. Their degree-divisibility reductions may add occurrences of a variable, and pinning or binary signatures used in an auxiliary problem need to be justified separately in the strict model. Our absorption argument supplies that justification for one fixed unary type at a time.

Third, ordinary Boolean Holant allows signatures to occupy unrestricted vertices and permits ordinary edge connections. The ordinary ternary classification of Yang--Huang--Fu \cite{ref19}, the odd-arity dichotomy of Meng--Wang--Xia--Zheng \cite{ref12}, and the full complex-valued manuscript of Liu--Meng--Wang \cite{ref13} concern this larger setting. Section 15 gives an explicit signature for which the strict problem is in FP but the corresponding ordinary signature-set problem is \#P-hard. Thus an ordinary Holant hardness criterion cannot simply be restricted to the present input graphs. The binary theorem of Kowalczyk--Cai \cite{ref7}, by contrast, has exactly the auxiliary interface that we use: a symmetric binary function on one side and ternary equality on the other.

Yang, Huang, and Fu also classify \(k\)-regular asymmetric spin systems with complex binary edge functions \cite{ref21}. This is a separate result from their ordinary ternary classification \cite{ref19}. In a spin system each edge carries an ordered binary function, and the order may vary from edge to edge. Their cubic case recalls the classification of Cai, Kowalczyk, and Williams \cite{ref11}; their higher-degree proof uses binary equality and gadgets built around equality vertices of degree four or more. A decomposition of a ternary function into two tensor cubes instead produces a cubic bipartite spin instance with every edge ordered from the left part to the right part. Example 23 gives an explicit edge matrix that is tractable under this restriction but hard in the unrestricted spin model. The latter hardness theorem therefore does not by itself establish Theorem A.

Some intermediate reductions do extend to complex weights. Suppose an invertible binary signature on the left and a nonsingular diagonal binary signature on the right have already been made jointly available by a legal reduction. Diagonal normalization and binary interpolation then give binary equality on both sides. Two copies of the resulting weighted ternary equality give a weighted arity-four equality, whose nonzero weight can be removed by the double-edge chain of \cite[Lemma 2.2]{ref21}, including when the weight is a root of unity. The complex \(\#\operatorname{CSP}^{2}\) classification \cite[Theorem 2.6]{ref21} then applies. The task in the strict model is to obtain the required auxiliary signatures without adding a free unary or changing a degree. Our contribution is this step and the resulting classification; the auxiliary binary and counting-CSP classifications are published inputs.

Matrix groups already have an established role in Holant reductions. Cai--Kowalczyk--Williams use gadgets, anti-gadgets, and projective matrix groups \cite{ref11}. More recently, Huang--Fu use Schur's theorem, finite subgroups of \(\operatorname{SL}_2(\mathbb C)\), and stable subgroup sequences for ordinary complex-valued quaternary Holant \cite{ref14}. Xia also studies binary-group organization in Boolean tensor networks \cite{ref15}. We use these classical group principles in the strict cubic bipartite setting. The additional work is the graph interface: extracting a rank-one straddled tensor, absorbing its column without freely available unaries, and resolving the exceptional finite-order parameters using actual cubic bipartite gadgets. We do not claim the matrix-group strategy or the affine algorithms as new.

\subsection{1.5 Proof organization and notation}\label{proof-organization-and-notation}

The proof separates the construction of reductions from the exhaustion of parameter cases. We first prove the algorithms for the tractable set \(\mathcal T\). For hardness, fix \(f\notin\mathcal T\). If an endpoint is nonzero, scaling and reversal give \(f=[1,a,b,c]\), and the first actual matrix is

\[A_f=\begin{pmatrix}1&b\\a&c\end{pmatrix},\qquad
\det A_f=c-ab,\qquad \Delta_f=(1-c)^2+4ab.\]

These quantities give the complete case division below. When \(A_f\) is invertible and has distinct eigenvalues, put \(\rho=\lambda_1/\lambda_2\). Its projective order is the least positive \(n\) for which \(A_f^n\) is a nonzero scalar matrix, if such an \(n\) exists; in the distinct-eigenvalue case this is precisely the order of \(\rho\). The singular and repeated-root loci intersect, so the diagram tests singularity first. The proofs of both boundary theorems retain the intersection because it is reused later.

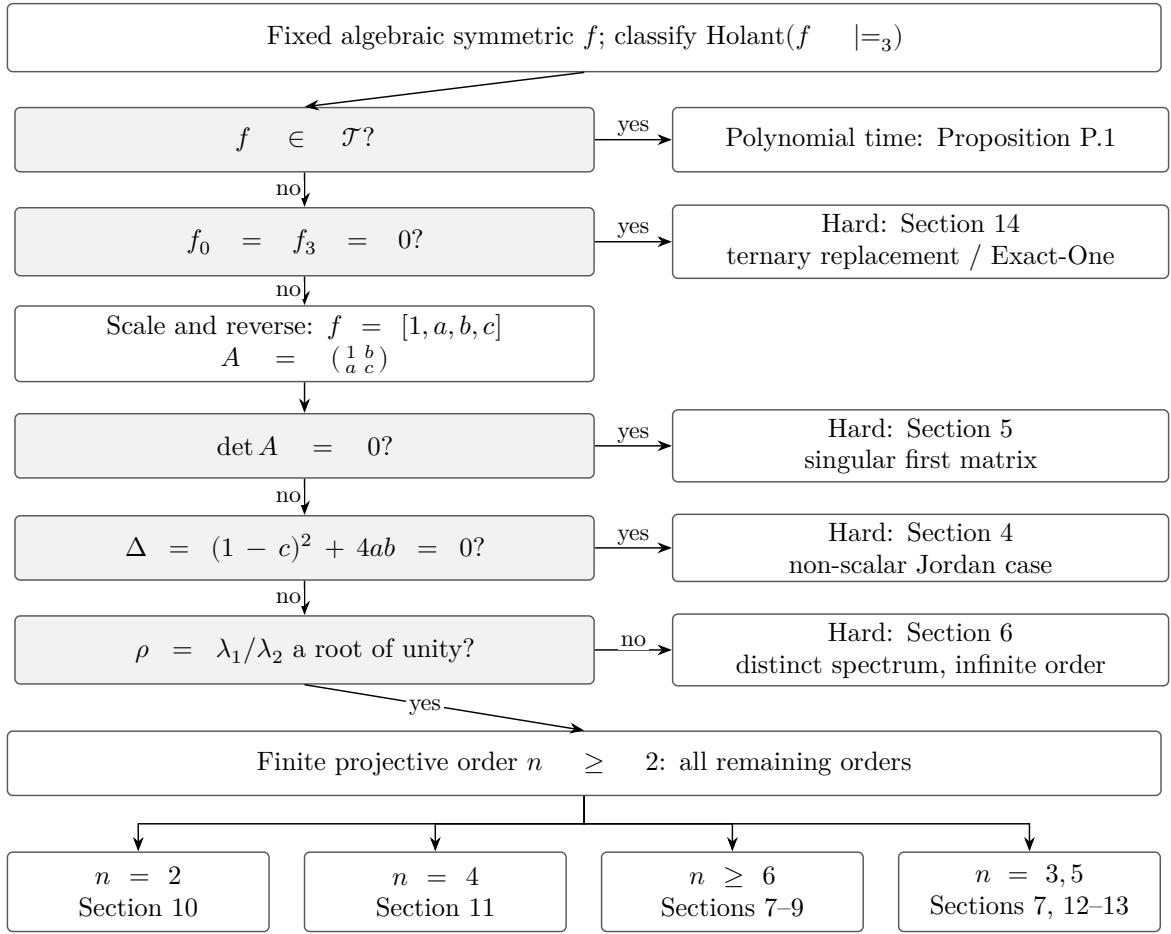
\begin{figure}[htbp]
\centering
\begin{tikzpicture}[x=1cm,y=-1cm,>=Stealth,font=\small]
\tikzset{mapnode/.style={draw=black!65,rounded corners=2pt,align=center,inner sep=3pt,line width=.55pt},maparrow/.style={->,line width=.55pt}}
\node[mapnode,fill=white,text width=15.05cm,minimum height=0.9cm] (start) at (7.9,0.65) {Fixed algebraic symmetric $f$; classify $\operatorname{Holant}(f\mid =_3)$};
\node[mapnode,fill=black!5,text width=7.45cm,minimum height=0.85cm] (easy) at (4.2,2) {$f\in\mathcal T$?};
\node[mapnode,fill=white,text width=6.35cm,minimum height=0.85cm] (fp) at (12.35,2) {Polynomial time: Proposition P.1};
\node[mapnode,fill=black!5,text width=7.45cm,minimum height=0.9cm] (endpoints) at (4.2,3.35) {$f_0=f_3=0$?};
\node[mapnode,fill=white,text width=6.35cm,minimum height=0.9cm] (zero) at (12.35,3.35) {Hard: Section 14\\ternary replacement / Exact-One};
\node[mapnode,fill=white,text width=7.45cm,minimum height=0.9cm] (norm) at (4.2,4.7) {Scale and reverse: $f=[1,a,b,c]$\\$A=\left(\begin{smallmatrix}1&b\\a&c\end{smallmatrix}\right)$};
\node[mapnode,fill=black!5,text width=7.45cm,minimum height=0.85cm] (det) at (4.2,6.05) {$\det A=0$?};
\node[mapnode,fill=white,text width=6.35cm,minimum height=0.85cm] (singular) at (12.35,6.05) {Hard: Section 5\\singular first matrix};
\node[mapnode,fill=black!5,text width=7.45cm,minimum height=0.85cm] (disc) at (4.2,7.4) {$\Delta=(1-c)^2+4ab=0$?};
\node[mapnode,fill=white,text width=6.35cm,minimum height=0.85cm] (repeated) at (12.35,7.4) {Hard: Section 4\\non-scalar Jordan case};
\node[mapnode,fill=black!5,text width=7.45cm,minimum height=0.9cm] (ratio) at (4.2,8.75) {$\rho=\lambda_1/\lambda_2$ a root of unity?};
\node[mapnode,fill=white,text width=6.35cm,minimum height=0.9cm] (infinite) at (12.35,8.75) {Hard: Section 6\\distinct spectrum, infinite order};
\node[mapnode,fill=white,text width=15.05cm,minimum height=0.85cm] (finite) at (7.9,10.25) {Finite projective order $n\ge2$: all remaining orders};
\node[mapnode,fill=white,text width=3.25cm,minimum height=1.05cm] (two) at (2,11.95) {$n=2$\\Section 10};
\node[mapnode,fill=white,text width=3.25cm,minimum height=1.05cm] (four) at (5.93,11.95) {$n=4$\\Section 11};
\node[mapnode,fill=white,text width=3.25cm,minimum height=1.05cm] (large) at (9.86,11.95) {$n\ge6$\\Sections 7--9};
\node[mapnode,fill=white,text width=3.25cm,minimum height=1.05cm] (small) at (13.79,11.95) {$n=3,5$\\Sections 7, 12--13};
\draw[maparrow] (start.south) -- (easy.north);
\draw[maparrow] (easy.east) -- (fp.west) node[midway,above,fill=white,inner sep=1pt,font=\footnotesize]{yes};
\draw[maparrow] (easy.south) -- (endpoints.north) node[midway,left,fill=white,inner sep=1pt,font=\footnotesize]{no};
\draw[maparrow] (endpoints.east) -- (zero.west) node[midway,above,fill=white,inner sep=1pt,font=\footnotesize]{yes};
\draw[maparrow] (endpoints.south) -- (norm.north) node[midway,left,fill=white,inner sep=1pt,font=\footnotesize]{no};
\draw[maparrow] (norm.south) -- (det.north);
\draw[maparrow] (det.east) -- (singular.west) node[midway,above,fill=white,inner sep=1pt,font=\footnotesize]{yes};
\draw[maparrow] (det.south) -- (disc.north) node[midway,left,fill=white,inner sep=1pt,font=\footnotesize]{no};
\draw[maparrow] (disc.east) -- (repeated.west) node[midway,above,fill=white,inner sep=1pt,font=\footnotesize]{yes};
\draw[maparrow] (disc.south) -- (ratio.north) node[midway,left,fill=white,inner sep=1pt,font=\footnotesize]{no};
\draw[maparrow] (ratio.east) -- (infinite.west) node[midway,above,fill=white,inner sep=1pt,font=\footnotesize]{no};
\draw[maparrow] (ratio.south) -- (finite.north) node[midway,left,fill=white,inner sep=1pt,font=\footnotesize]{yes};
\draw[maparrow] (finite.south) -- (7.9,11.05) -| (two.north);
\draw[maparrow] (finite.south) -- (7.9,11.05) -| (four.north);
\draw[maparrow] (finite.south) -- (7.9,11.05) -| (large.north);
\draw[maparrow] (finite.south) -- (7.9,11.05) -| (small.north);
\end{tikzpicture}
\caption{Coverage of Theorem A. After the first negative test, $f$ is outside the stated tractable set, so every subsequent terminal branch proves hardness. The singular branch is tested before the repeated-root branch; the two loci overlap in the parameter space. Orders three and five are divided further according to singularity of $B$ and invariant points or pairs, as described in the text.}
\label{fig:classification-map}
\end{figure}

\subsubsection{1.5.1 How the reductions reach a known hard problem}\label{how-the-reductions-reach-a-known-hard-problem}

The principal target is a hard symmetric binary signature. A right unary row \(v=(r,s)\) attached to one port of \(f\) gives

\[F_f(v)=[r+as,\ ar+bs,\ br+cs].\]

The unary is not free. Sections 2--3 first construct actual straddled matrices, then justify interpolation of an auxiliary matrix throughout an entire network. A rank-one factorization \(M=uv\) supplies its right row only after the unused columns have been disposed of. Grouping three columns at a right equality vertex contributes

\[Q(u)=u_0^3+u_1^3.\]

When \(Q(u)\ne0\), division by its known power makes the row available by a Turing reduction. Otherwise, an actual matrix word must first move the column away from the three projective zeros of \(Q\). A direction means a nonzero row or column up to nonzero scalar multiplication.

For a non-rank-one, non-equality \(f\), at most six row directions give tractable binary contractions; the bound is five when \(A_f\) is invertible (Lemma 4). It therefore suffices to obtain one suitable row, or enough distinct candidates to ensure one succeeds. Several branches instead use a holographic transformation and degree counting to express the value as a known nonzero factor times the number of perfect matchings. These two routes are displayed below.

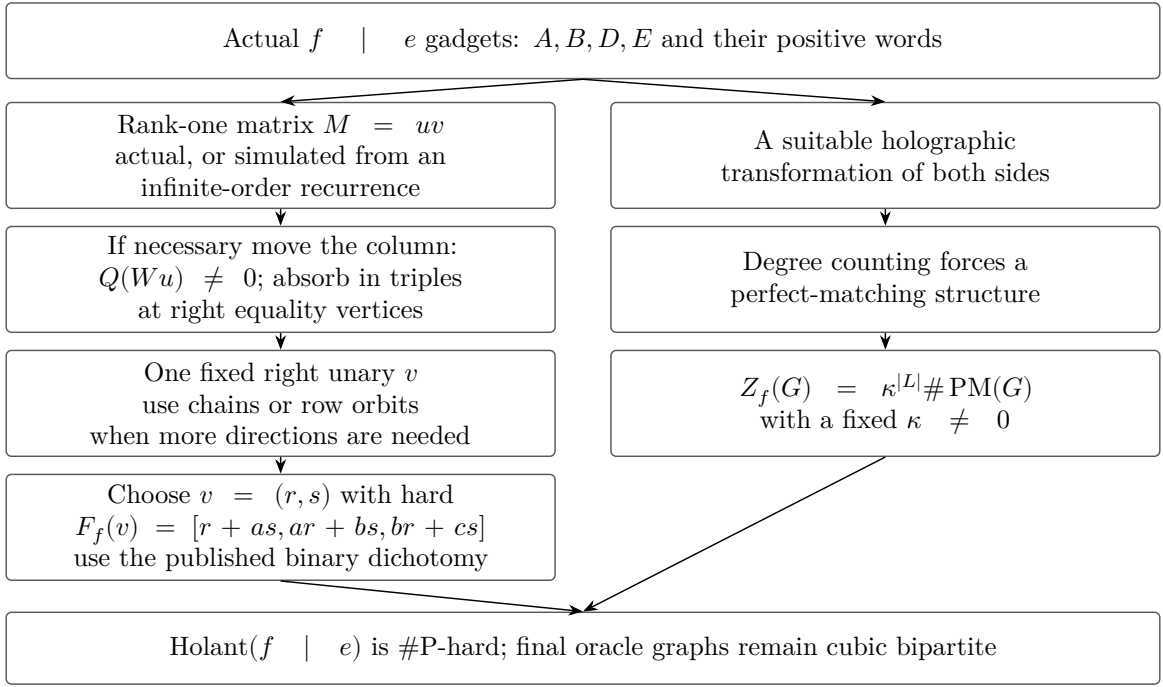
\begin{figure}[H]
\centering
\begin{tikzpicture}[x=1cm,y=-1cm,>=Stealth,font=\small]
\tikzset{mapnode/.style={draw=black!65,rounded corners=2pt,align=center,inner sep=3pt,line width=.55pt},maparrow/.style={->,line width=.55pt}}
\node[mapnode,fill=white,text width=15.05cm,minimum height=1cm] (gadgets) at (7.9,0.533) {Actual $f\mid e$ gadgets: $A,B,D,E$ and their positive words};
\node[mapnode,fill=white,text width=7.05cm,minimum height=1.4cm] (rankone) at (3.9,2.05) {Rank-one matrix $M=uv$\\actual, or simulated from an\\infinite-order recurrence};
\node[mapnode,fill=white,text width=7.05cm,minimum height=1.4cm] (basis) at (11.9,2.05) {A suitable holographic\\transformation of both sides};
\node[mapnode,fill=white,text width=7.05cm,minimum height=1.4cm] (absorb) at (3.9,3.69) {If necessary move the column:\\$Q(Wu)\ne0$; absorb in triples\\at right equality vertices};
\node[mapnode,fill=white,text width=7.05cm,minimum height=1.4cm] (count) at (11.9,3.69) {Degree counting forces a\\perfect-matching structure};
\node[mapnode,fill=white,text width=7.05cm,minimum height=1.4cm] (row) at (3.9,5.33) {One fixed right unary $v$\\use chains or row orbits\\when more directions are needed};
\node[mapnode,fill=white,text width=7.05cm,minimum height=1.4cm] (pm) at (11.9,5.33) {$Z_f(G)=\kappa^{|L|}\#\operatorname{PM}(G)$\\with a fixed $\kappa\ne0$};
\node[mapnode,fill=white,text width=7.05cm,minimum height=1.4cm] (binary) at (3.9,6.97) {Choose $v=(r,s)$ with hard\\$F_f(v)=[r+as,ar+bs,br+cs]$\\use the published binary dichotomy};
\node[mapnode,fill=white,text width=15.05cm,minimum height=0.95cm] (hard) at (7.9,8.568999999999999) {$\operatorname{Holant}(f\mid e)$ is \#P-hard; final oracle graphs remain cubic bipartite};
\draw[maparrow] (gadgets.south) -- (rankone.north);
\draw[maparrow] (gadgets.south) -- (basis.north);
\draw[maparrow] (rankone.south) -- (absorb.north);
\draw[maparrow] (absorb.south) -- (row.north);
\draw[maparrow] (row.south) -- (binary.north);
\draw[maparrow] (basis.south) -- (count.north);
\draw[maparrow] (count.south) -- (pm.north);
\draw[maparrow] (binary.south) -- (hard.north);
\draw[maparrow] (pm.south) -- (hard.north);
\end{tikzpicture}
\caption{The two principal hardness routes. Each arrow is justified under the hypotheses of the corresponding branch; an algebraic factorization alone does not supply a unary. Ternary replacements allow a branch to reuse a previously proved hard case. All auxiliary signatures are removed before the final oracle query.}
\label{fig:reduction-map}
\end{figure}

\subsubsection{1.5.2 Why finite orders require a separate argument}\label{why-finite-orders-require-a-separate-argument}

If \(\rho\) is not a root of unity, the powers of \(A_f\) provide distinct interpolation nodes. If \(\rho\) has finite order, those powers repeat projectively. We then use several actual matrices and their positive words. Lemma 10 proves that an infinite generated projective group with no invariant point or pair provides both an infinite-order recurrence and enough row and column directions. This combines interpolation with the finite exceptional sets just described.

The hypotheses of that group lemma must still be established. Sections 8--9 handle common invariant directions and complete every order at least six. Sections 10 and 11 settle orders two and four. At orders three and five, Section 13 handles singular \(B\) and groups with an invariant point or pair; Lemma 10 handles infinite groups with neither; Section 12 handles finite groups with neither. Section 8 supplies special-family theorems used by these arguments, rather than an additional branch of the initial spectral partition. Section 14 then handles zero endpoints and assembles Theorem A.

\subsubsection{1.5.3 Contribution and use of earlier results}\label{contribution-and-use-of-earlier-results}

The main result extends the rational classification \cite{ref5} to every fixed complex algebraic symmetric ternary signature in the same strict cubic bipartite model. The tractable algorithms, binary dichotomy, rank-one splitting technique, and classical projective-group results are established inputs. The proof here supplies their graph-theoretic hypotheses for complex weights: it avoids vanishing absorption factors, obtains enough unary directions without free auxiliary signatures, and closes all finite-order and singular boundaries with actual gadgets. The order-three and order-five calculations are accompanied by exact identities and reproducible coefficient certificates in Appendix A.

The exact real intersection is a corollary of the complex theorem, not a separate unproved extension. Section 15 gives explicit separations from ordinary Holant and from unrestricted edge orderings in asymmetric spin systems. These examples explain why the broader published dichotomies do not directly imply the present strict-model result; they do not assert that every possible alternative reduction has been excluded.

The term \emph{degenerate} means rank one throughout. For \(f=[1,a,b,c]\), this is equivalent to \(b=a^2,c=a^3\). A generalized equality is \([x,0,0,y]\). Matrix rank and tensor rank are different: on the singular surface \(c=ab\), the matrix \(A_f\) has rank one even when the ternary tensor \(f\) does not.

\subsection{1.6 Polynomial-time algorithms}\label{polynomial-time-algorithms}

\begin{holantstatement}{Proposition P.1}

For each fixed signature \(f\in\mathcal T\), the exact evaluation problem \(\operatorname{Holant}(f\mid e)\) belongs to FP.

\end{holantstatement}

\begin{proof}[Proof]

A cubic bipartite graph has the same number \(n\) of vertices on each side. For a rank-one left signature, equality groups its incident factors, giving

\[Z_f(G)=\lambda^n(u_0^3+u_1^3)^n.\]

For a generalized equality, all bits in a connected component agree. If the component has \(m\) left vertices, its contribution is \(x^m+y^m\). Multiply these contributions over components.

The two parity-supported representatives have support \(x_1+x_2+x_3=0\) over \(\mathbb F_2\), and phase either one or \((-1)^{x_1x_2+x_1x_3+x_2x_3}\). The other four representatives can be written as

\[i^{r(x_1+x_2+x_3)}(-1)^{x_1x_2+x_1x_3+x_2x_3},\qquad r=0,1,2,3.\]

Thus equality and these signatures give linear support equations over \(\mathbb F_2\) and a quadratic phase over \(\mathbb Z/4\mathbb Z\), whose cross-term coefficients are even. Gaussian elimination parametrizes the support. Substituting Boolean affine forms preserves this phase class, because an exclusive-or expands modulo four into a linear term plus twice pairwise products. The remaining sum is a quadratic Gauss sum and can be evaluated in polynomial time.

For completeness, an odd linear coefficient gives a direct elimination step. If a variable \(y\) occurs as \(a y+2y\ell(z)\) with \(a\) odd and \(\ell\) Boolean affine, then

\[\sum_{y=0}^1 i^{a y+2y\ell(z)}
=(1+i^a)(-i^a)^{\ell(z)}.\]

The resulting exponent is again quadratic modulo four. If all linear coefficients are even, the sum reduces to a quadratic \(\{1,-1\}\) phase over \(\mathbb F_2\), which is evaluated by eliminating nonzero bilinear pairs and then summing linear variables. Each step removes at least one variable.

A nonzero scalar multiplying \(f\) contributes its known \(n\)th power. Reversal bijects assignments by complementing every edge bit. Finally, \(\Gamma_\omega\) multiplies a local left assignment by \(\omega\) to the number of incident one-edges. At each right equality vertex that number is either zero or three, so the product of all gauge factors is one. Hence the cube-root gauge leaves every closed network value unchanged. These observations give algorithms for all of \(\mathcal T\).
\end{proof}

\section{2. Actual gadgets and their matrices}\label{actual-gadgets-and-their-matrices}

A gadget is a finite graph over the allowed signatures, with dangling edges. Its tensor is obtained by fixing the dangling bits and summing the internal bits. A straddled gadget has one dangling edge incident to a left vertex and one incident to a right vertex. Its row index is the left-port bit and its column index is the right-port bit, both ordered zero, one. Its matrix need not be symmetric.

In the following diagrams a square denotes a left signature and a circle denotes right equality. Port labels name Boolean bits. The first constructions occur in the cubic bipartite reductions of Cai--Fan--Liu \cite{ref5}; the second ternary construction is from Cai--Fan \cite{ref6}. We give their contraction formulas here so that their validity for complex weights can be checked directly.

\subsection{2.1 The first and second matrices}\label{the-first-and-second-matrices}

\begin{figure}[H]
\centering
\begin{tikzpicture}[x=1.05cm,y=1.05cm]
\coordinate (l) at (0,0);
\coordinate (r) at (3,0);
\coordinate (i) at (-1.5,0);
\coordinate (j) at (4.5,0);
\draw (i) to[] node[midway,above,lab] {$i$} (l);
\draw (r) to[] node[midway,above,lab] {$j$} (j);
\draw (l) to[bend left=28] node[midway,above,lab] {$p$} (r);
\draw (l) to[bend right=28] node[midway,below,lab] {$q$} (r);
\node[L] at (l) {$h$};
\node[R] at (r) {$=_3$};
\end{tikzpicture}
\caption{The gadget $A_h$. Squares are left vertices carrying $h$; circles are right equality vertices. The two curved lines are distinct parallel edges. Only $i,j$ are external inputs; $p,q$ are summed over.}
\label{fig:gadget_a-1}
\end{figure}
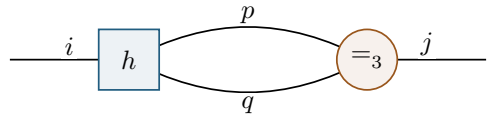

For a generic ternary \(h=[w,x,y,z]\), two parallel edges between \(h\) and equality give

\[A_h(i,j)=\sum_{p,q\in\{0,1\}}h(i,p,q)e(j,p,q)=h(i,j,j),\]

because equality forces \(p=q=j\). Therefore

\[A_h=\begin{pmatrix}w&y\\x&z\end{pmatrix},\qquad
A=A_f=\begin{pmatrix}1&b\\a&c\end{pmatrix}.\]

For example, \(A_h(0,1)=h(0,1,1)=y\), whereas \(A_h(1,0)=h(1,0,0)=x\). The repeated \(j\) in \(h(i,j,j)\) is forced by the two internal edges at equality. It does not change the domain of the Boolean function \(h\) and does not assert that the matrix is symmetric.

\begin{figure}[H]
\centering
\begin{tikzpicture}[x=1.05cm,y=1.05cm]
\coordinate (l0) at (0,0);
\coordinate (r0) at (4,0);
\coordinate (r1) at (1,-1.8);
\coordinate (l1) at (3,-1.8);
\coordinate (i) at (-1.3,0);
\coordinate (j) at (5.3,0);
\draw (i) to[] node[midway,above,lab] {$i$} (l0);
\draw (r0) to[] node[midway,above,lab] {$j$} (j);
\draw (l0) to[] node[midway,above,lab] {$j$} (r0);
\draw (l0) to[] node[midway,above,lab] {$t$} (r1);
\draw (l1) to[] node[midway,above,lab] {$j$} (r0);
\draw (r1) to[bend left=28] node[midway,above,lab] {$t$} (l1);
\draw (r1) to[bend right=28] node[midway,below,lab] {$t$} (l1);
\node[L] at (l0) {$h$};
\node[R] at (r0) {$=_3$};
\node[R] at (r1) {$=_3$};
\node[L] at (l1) {$h$};
\end{tikzpicture}
\caption{The gadget $B_h$. Vertex shape specifies the bipartition, even when a square is drawn to the right of a circle. Equality has already forced the top and right edges to carry $j$, and the three lower-left incident edges to carry the same bit $t$. Sum over $t$.}
\label{fig:gadget_b-2}
\end{figure}
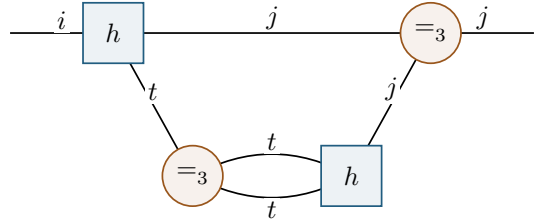

The second pictured construction has two left and two right vertices. Equality leaves one bit \(t\) to sum over:

\[B_h(i,j)=\sum_{t=0}^1h(i,j,t)h(j,t,t).\]

Consequently

\[B_h=\begin{pmatrix}w^2+xy&x^2+yz\\wx+y^2&xy+z^2\end{pmatrix},\qquad
B=B_f=\begin{pmatrix}1+ab&a^2+bc\\a+b^2&ab+c^2\end{pmatrix}.\]

\subsection{2.2 The ternary construction and its closure}\label{the-ternary-construction-and-its-closure}

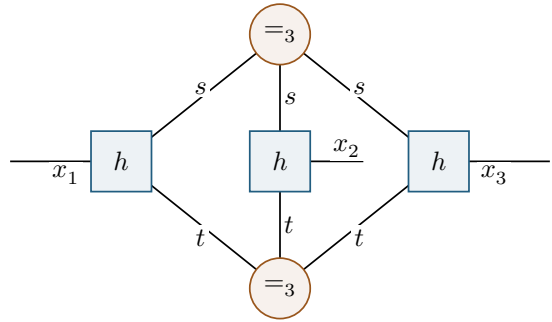
\begin{figure}[H]
\centering
\begin{tikzpicture}[x=1.05cm,y=1.05cm]
\coordinate (l0) at (-2,0);
\coordinate (l1) at (0,0);
\coordinate (l2) at (2,0);
\coordinate (r0) at (0,1.6);
\coordinate (r1) at (0,-1.6);
\coordinate (x0) at (-3.4,0);
\coordinate (x1) at (1.05,0);
\coordinate (x2) at (3.4,0);
\draw (x0) to[] node[midway,below,lab] {$x_1$} (l0);
\draw (l1) to[] node[pos=0.8,above,lab] {$x_2$} (x1);
\draw (l2) to[] node[midway,below,lab] {$x_3$} (x2);
\draw (l0) to[] node[midway,above,lab] {$s$} (r0);
\draw (l0) to[] node[midway,below,lab] {$t$} (r1);
\draw (l1) to[] node[midway,right,lab] {$s$} (r0);
\draw (l1) to[] node[midway,right,lab] {$t$} (r1);
\draw (l2) to[] node[midway,above,lab] {$s$} (r0);
\draw (l2) to[] node[midway,below,lab] {$t$} (r1);
\node[L] at (l0) {$h$};
\node[L] at (l1) {$h$};
\node[L] at (l2) {$h$};
\node[R] at (r0) {$=_3$};
\node[R] at (r1) {$=_3$};
\end{tikzpicture}
\caption{The ternary gadget $\Phi(h)$: three left $h$ vertices and two right equality vertices. All three edges at the upper circle carry $s$; all three at the lower circle carry $t$. The three dangling inputs are $x_1,x_2,x_3$. The central dangling stub ends inside the drawing and does not touch the right square.}
\label{fig:gadget_phi-3}
\end{figure}

Join three left copies of \(h\) once to each of two right equality vertices, leaving one dangling edge at each left vertex. If their bits are \(x_1,x_2,x_3\), the two equality bits \(s,t\) give

\[\Phi(h)(x_1,x_2,x_3)=\sum_{s,t\in\{0,1\}}\prod_{j=1}^3h(x_j,s,t).\]

This is symmetric in the external bits. For their Hamming weight \(k\),

\[\Phi(h)_k=w^{3-k}x^k+2x^{3-k}y^k+y^{3-k}z^k.\]

Thus

\[\Phi(h)=[w^3+2x^3+y^3,\ w^2x+2x^2y+y^2z,\ wx^2+2xy^2+yz^2,\ x^3+2y^3+z^3].\]

\begin{figure}[H]
\centering

\begin{tikzpicture}[x=1.2cm,y=1cm]
\node[boxg,minimum height=16mm] (p) at (0,0) {$\Phi(h)$};
\node[R] (e) at (3,0) {$=_3$};
\draw (-1.8,0) -- node[above,lab] {$i$} (p.west);
\draw (p.north east) to[bend left=25] node[above,lab] {$j$} (e.north west);
\draw (p.south east) to[bend right=25] node[below,lab] {$j$} (e.south west);
\draw (e.east) -- node[above,lab] {$j$} (4.4,0);
\end{tikzpicture}

\caption{$D_h=A_{\Phi(h)}$. The box abbreviates the ternary gadget with three left $h$ vertices and two shared right equality vertices; it is not a new freely available signature. Attach its second and third ports to one new equality vertex. Expanding the box gives three original left vertices and three right vertices.}
\label{fig:gadget_d-4}
\end{figure}
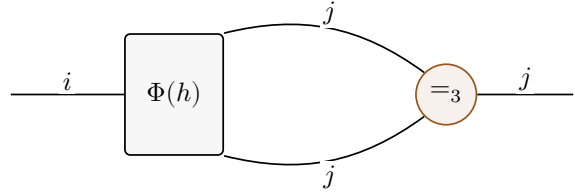

Applying the first construction to \(\Phi(f)\) gives

\[D=A_{\Phi(f)}=\begin{pmatrix}
1+2a^3+b^3&a^2+2ab^2+bc^2\\
a+2a^2b+b^2c&a^3+2b^3+c^3
\end{pmatrix}.\]

A whole left ternary gadget can replace any left vertex. In particular \(\operatorname{Holant}(\Phi(f)\mid e)\le_T\operatorname{Holant}(f\mid e)\). All internal and external vertices retain the required bipartition and degree.

\subsection{2.3 Serial composition}\label{serial-composition}

\begin{figure}[H]
\centering

\begin{tikzpicture}[x=1cm,y=1cm]
\node[boxg] (a) at (0,0) {$A$};\node[boxg] (b) at (3,0) {$B$};
\draw (-1.8,0)--node[above,lab] {$i$}(a.west);
\draw (a.east)--node[above,lab] {$k$}(b.west);
\draw (b.east)--node[above,lab] {$j$}(4.8,0);
\node[below=2mm of a] {\small L port\quad R port};
\node[below=2mm of b] {\small L port\quad R port};
\node at (1.5,-1.7) {$(AB)_{ij}=\displaystyle\sum_{k=0}^1 A_{ik}B_{kj}$};
\node[R,dashed] (v) at (0,-3.2) {$v$};\node[boxg] (h) at (3,-3.2) {$H$};
\draw (v.east)--node[above,lab] {$i$}(h.west);
\draw (h.east)--node[above,lab] {$j$}(4.8,-3.2);
\node at (1.5,-4.3) {$(vH)_j=\displaystyle\sum_{i=0}^1 v_iH_{ij}$};
\end{tikzpicture}

\caption{Port order determines matrix order. Each box has its L port on the left and R port on the right. Joining an R port to an L port is legal. The lower picture assumes that the RHS unary $v$ has already been simulated; its circular shape indicates its side, not that it is an equality vertex.}
\label{fig:composition-5}
\end{figure}
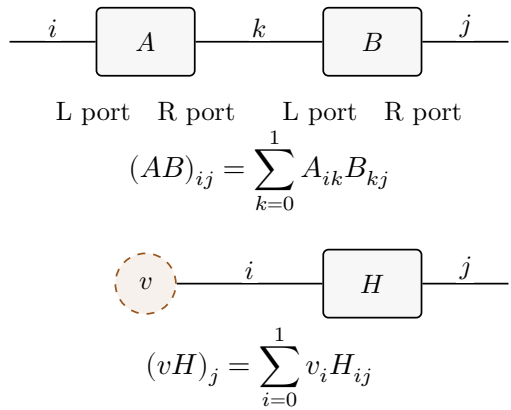

Joining the right port of a matrix \(X\) to the left port of \(Y\) gives

\[(XY)_{ij}=\sum_{k=0}^1X_{ik}Y_{kj}.\]

A \emph{word} in the matrices \(A,B,\ldots\) is an ordered product, such as \(AB^2A=ABBA\), whose length is four. It represents the serial composition of the corresponding four gadgets. A \emph{positive word} uses only these matrices, without inverse symbols; its entries need not be positive or real. Positive words in actual matrices are therefore actual gadgets.

A word chosen once for the fixed signature \(f\) has constant length with respect to the input graph size. Its length may depend on \(f\); no common bound for all signatures is implied. By contrast, an interpolation chain \(H^k\), with \(k\) growing polynomially with the number of marked locations, has polynomial length. Both constructions preserve the left/right port convention.

If a right unary row \(v\) is available through a reduction, attaching an actual matrix \(X\) gives the row \(vX\), since its entry at the remaining bit \(j\) is \(\sum_i v_iX_{ij}\). Matrix subtraction is not a graph operation: expressions such as \(AB-BA\) are used only to analyze actual matrices.

\subsection{2.4 A second ternary construction}\label{a-second-ternary-construction}

\begin{figure}[htbp]
\centering
\definecolor{svgcolor0}{HTML}{344054}
\definecolor{svgcolor1}{HTML}{e7f1fb}
\definecolor{svgcolor2}{HTML}{20578a}
\definecolor{svgcolor3}{HTML}{fff1dc}
\definecolor{svgcolor4}{HTML}{996022}
\begin{tikzpicture}[x=0.463636pt,y=-0.463636pt]
\path[use as bounding box] (0,0) rectangle (600,550);
\path[fill=white,draw=none,line width=0.4636pt] (0,0) rectangle (600,550);
\path[fill=black,draw=svgcolor0,line width=1.1591pt] (300,245) -- (300,125);
\path[fill=black,draw=svgcolor0,line width=1.1591pt] (300,245) -- (405,320);
\path[fill=black,draw=svgcolor0,line width=1.1591pt] (300,245) -- (195,320);
\path[fill=black,draw=svgcolor0,line width=1.1591pt] (465,160) -- (300,125);
\path[fill=black,draw=svgcolor0,line width=1.1591pt] (465,160) -- (405,320);
\path[fill=black,draw=svgcolor0,line width=1.1591pt] (300,440) -- (405,320);
\path[fill=black,draw=svgcolor0,line width=1.1591pt] (300,440) -- (195,320);
\path[fill=black,draw=svgcolor0,line width=1.1591pt] (135,160) -- (195,320);
\path[fill=black,draw=svgcolor0,line width=1.1591pt] (135,160) -- (300,125);
\path[fill=black,draw=svgcolor0,line width=1.1591pt] (465,160) -- (530,115);
\node[anchor=base,inner sep=0pt,text=black,font=\rmfamily\fontsize{10.664}{12.796}\selectfont] at (530,106) {x};
\path[fill=black,draw=svgcolor0,line width=1.1591pt] (300,440) -- (300,510);
\node[anchor=base,inner sep=0pt,text=black,font=\rmfamily\fontsize{10.664}{12.796}\selectfont] at (300,537) {y};
\path[fill=black,draw=svgcolor0,line width=1.1591pt] (135,160) -- (70,115);
\node[anchor=base,inner sep=0pt,text=black,font=\rmfamily\fontsize{10.664}{12.796}\selectfont] at (70,106) {z};
\path[fill=svgcolor1,draw=svgcolor2,line width=0.9273pt] (280,225) rectangle (320,265);
\node[anchor=base,inner sep=0pt,text=black,font=\rmfamily\fontsize{10.664}{12.796}\selectfont] at (300,252) {f};
\path[fill=svgcolor3,draw=svgcolor4,line width=0.9273pt] (300,125) circle[radius=9.2727pt];
\node[anchor=base,inner sep=0pt,text=black,font=\rmfamily\fontsize{10.664}{12.796}\selectfont] at (300,132) {=};
\path[fill=svgcolor3,draw=svgcolor4,line width=0.9273pt] (405,320) circle[radius=9.2727pt];
\node[anchor=base,inner sep=0pt,text=black,font=\rmfamily\fontsize{10.664}{12.796}\selectfont] at (405,327) {=};
\path[fill=svgcolor3,draw=svgcolor4,line width=0.9273pt] (195,320) circle[radius=9.2727pt];
\node[anchor=base,inner sep=0pt,text=black,font=\rmfamily\fontsize{10.664}{12.796}\selectfont] at (195,327) {=};
\path[fill=svgcolor1,draw=svgcolor2,line width=0.9273pt] (445,140) rectangle (485,180);
\node[anchor=base,inner sep=0pt,text=black,font=\rmfamily\fontsize{10.664}{12.796}\selectfont] at (465,167) {f};
\path[fill=svgcolor1,draw=svgcolor2,line width=0.9273pt] (280,420) rectangle (320,460);
\node[anchor=base,inner sep=0pt,text=black,font=\rmfamily\fontsize{10.664}{12.796}\selectfont] at (300,447) {f};
\path[fill=svgcolor1,draw=svgcolor2,line width=0.9273pt] (115,140) rectangle (155,180);
\node[anchor=base,inner sep=0pt,text=black,font=\rmfamily\fontsize{10.664}{12.796}\selectfont] at (135,167) {f};
\node[anchor=base,inner sep=0pt,text=black,font=\rmfamily\fontsize{9.736}{11.684}\selectfont] at (300,38) {A planar ternary gadget with three exterior ports};
\end{tikzpicture}
\caption{The actual four-left-vertex ternary gadget Psi.}
\end{figure}
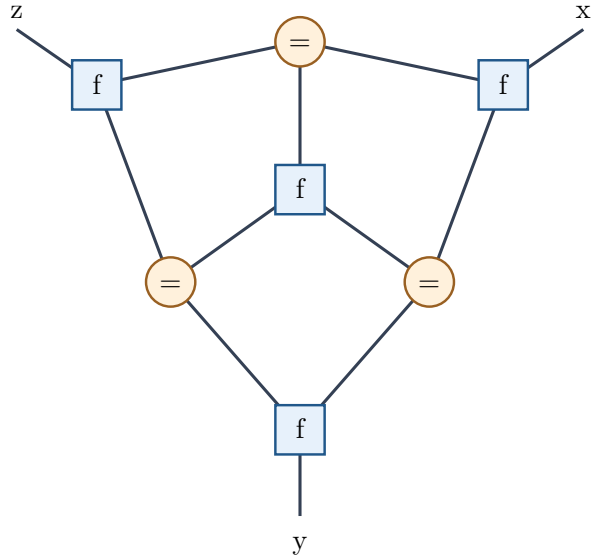

We also need the existing ternary gadget \(\Psi\). It has one central left vertex, three right equality vertices around it, and three outer left vertices. The central vertex joins all three equalities; each outer left vertex joins two consecutive equalities and carries one external port. Summing their three common bits gives

\[\Psi(f)(x,y,z)=\sum_{i,j,k\in\{0,1\}}f(i,j,k)f(x,i,j)f(y,j,k)f(z,k,i).\]

The triangle symmetries make this output symmetric. If \(\Psi(f)=[w,x',y,z']\), direct evaluation gives

\[\begin{aligned}
w&=1+3a^3+3a^2b^2+b^3c,\\
x'&=a+a^4+2a^2b+a^2bc+2ab^3+b^2c^2,\\
y&=a^2+ab^2+2a^3b+b^4+2ab^2c+bc^3,\\
z'&=a^3+3a^2b^2+3b^3c+c^4.
\end{aligned}\]

Closing two ports at equality gives \(E=A_{\Psi(f)}=\left(\begin{smallmatrix}w&y\\x'&z'\end{smallmatrix}\right)\). This is the Cai--Fan construction \cite{ref6}. In Section 12 it supplies a nonzero rank-one matrix at two finite-group parameter configurations where the preceding matrices do not suffice.

\subsection{2.5 Holographic changes of basis and the orientation convention}\label{holographic-changes-of-basis-and-the-orientation-convention}

We use bilinear tensor contraction throughout: joining two ports sums products of their coordinates without complex conjugation. For an invertible matrix \(T\), replace every left ternary tensor by \(T^{\otimes3}f\) and every right tensor by \((T^{-T})^{\otimes3}e\), treating both as column tensors. On each internal edge,

\[\sum_{i=0}^1 T_{i\alpha}(T^{-T})_{i\beta}=\delta_{\alpha\beta}.\]

All edge factors therefore cancel, and the closed partition function is unchanged. This is the bipartite form of the holographic transformation \cite{ref1,ref3}. It does not assert that either transformed signature is individually realizable in the original basis. The transformation changes both sides simultaneously.

The uncontracted indices of a straddled gadget give the matrix transformation

\[M\longmapsto TMT^{-1}.\]

It follows that rank, trace, determinant, eigenvalue ratios, and projective order can be analyzed after a simultaneous change of matrix basis. Column factors transform by \(u\mapsto Tu\), and row factors by \(v\mapsto vT^{-1}\). When checking the absorption factor \(Q(u)=u_0^3+u_1^3\), however, we always use the column in the original equality basis. A temporary eigenbasis is used only to understand the matrix action.

For \(T=\operatorname{diag}(1,\omega)\) with \(\omega^3=1\), the right equality is unchanged and the left signature becomes \(\Gamma_\omega f\). This explains the value-preserving diagonal transformations in Theorem A. The bit-swap matrix gives reversal. General basis changes used in perfect-matching reductions will display both transformed signatures explicitly.

\section{3. Interpolation, absorption, and binary hardness}\label{interpolation-absorption-and-binary-hardness}

The reductions use two different kinds of additional signatures. An \textbf{actual gadget} is a finite bipartite graph over \(f\) and \(e\) with the prescribed dangling ports. Replacing a vertex by such a graph is an exact graph substitution. An additional signature obtained by \textbf{interpolation} need not be the signature of any one graph. Its use is justified by a polynomial-time Turing reduction.

\begin{holantdefinition}{Definition P.0 (simulation with a prescribed port type)}

Fix \(f\) and \(e\). For a fixed matrix \(M\), let \(\mathcal H_f[M]\) be the bipartite tensor-network evaluation problem allowing the original ternary vertices together with any number of boxes labelled \(M\), each having one left port and one right port. Each left port is joined to a right port. We say that \(M\) is simulable if

\[\mathcal H_f[M]\le_T\operatorname{Holant}(f\mid e).\]

For a fixed row \(v\), let \(\mathcal H_f[v]\) be the analogous problem with arbitrarily many right unary vertices labelled \(v\), and no other additional signatures. We say that \(v\) is simulable if

\[\mathcal H_f[v]\le_T\operatorname{Holant}(f\mid e).\]

The same convention defines simulation of a left unary column. All reductions are exact and polynomial-time. The fixed tensor and its field of coefficients are not part of the input.

\end{holantdefinition}

For example, if a fixed matrix \(M\) is simulable and \(W\) is actual, then \(WM\) is simulable: attach the graph of \(W\) at every occurrence of the \(M\) box and apply the reduction for \(M\). The word ``simulable'' will always refer to this explicit reduction, rather than to an allowed linear combination of gadgets. In particular, an equation such as \(M=H-\lambda I\) is only an algebraic identity until an interpolation proof establishes the reduction.

We will usually simulate a single right unary type. If a later argument supplies a different row by attaching an actual chain to that same unary, its copies are removed by the already proved reduction. Separate simulations for two unrelated rows do not, by themselves, assert that both rows are simultaneously available in one oracle instance.

\subsection{3.1 Two forms of matrix interpolation}\label{two-forms-of-matrix-interpolation}

\begin{holantstatement}{Lemma P.2 (spectral and Jordan interpolation)}

Let \(H\) be an invertible actual straddled matrix.

\begin{enumerate}
\def\labelenumi{\arabic{enumi}.}
\tightlist
\item
  If its distinct eigenvalues \(\lambda_1,\lambda_2\) have ratio not a root of unity, each rank-one spectral projector of \(H\) is simulable.
\item
  If \(H=\lambda I+N\) with \(\lambda\ne0\), \(N\ne0\), and \(N^2=0\), the rank-one matrix \(N\) is simulable.
\end{enumerate}

\end{holantstatement}

\begin{proof}[Proof]

In the first case the two eigenvalues are distinct, so \(H\) is diagonalizable over a fixed algebraic extension. The projectors are

\[P_1=\frac{H-\lambda_2I}{\lambda_1-\lambda_2},\qquad
P_2=\frac{H-\lambda_1I}{\lambda_2-\lambda_1}.\]

If \(Hw=\lambda_jw\), the displayed formula gives \(P_iw=\delta_{ij}w\). The two eigenvectors form a basis, so these actions prove that each \(P_i\) has rank one and that

\[P_i^2=P_i,\qquad P_1P_2=P_2P_1=0,\qquad P_1+P_2=I,\qquad P_iH=HP_i=\lambda_iP_i.\]

Consequently

\[H^k=\lambda_1^kP_1+\lambda_2^kP_2.\]

Consider a network containing \(m\) marked copies of \(P_1\). Replace all of them by the same actual chain \(H^k\). For each \(j=0,\ldots,m\), let \(C_j\) be the sum of values obtained by placing \(P_1\) at exactly \(j\) marked locations and \(P_2\) at the remaining locations. These are coefficients in the multilinear expansion, not extra oracle queries. The queried value is

\[Z_k=\sum_{j=0}^m C_j\lambda_1^{kj}\lambda_2^{k(m-j)},\qquad
\lambda_2^{-km}Z_k=\sum_{j=0}^m C_j\bigl((\lambda_1/\lambda_2)^k\bigr)^j.\]

There are only \(m+1\) unknown coefficients, even though the expansion has \(2^m\) projector placements. The \(m+1\) choices \(k=1,\ldots,m+1\) give distinct evaluation nodes because the ratio is not a root of unity. Vandermonde interpolation recovers all coefficients. The coefficient \(C_m\) is the requested all-\(P_1\) value, and \(C_0\) gives the all-\(P_2\) value. This conclusion remains valid if either requested value is zero.

In the second case, \(N^2=0\) gives the binomial identity

\[H^k=\lambda^k\left(I+\frac{k}{\lambda}N\right).\]

Fix a target network with \(m\) marked copies of \(N\), and denote its value by \(Z_N\). Let \(F(t)\) be the value of the same outer network with \(I+(t/\lambda)N\) at every marked location. Multilinearity gives

\[F(t)=\sum_{j=0}^m c_jt^j,\qquad c_m=\lambda^{-m}Z_N.\]

The coefficient of \(t^m\) selects \(N/\lambda\) at all \(m\) locations; the other coefficients collect the placements selecting fewer copies of \(N\). The legal query replacing every location by the actual chain \(H^k\) has value

\[Z_k=\lambda^{km}F(k).\]

Divide by the known nonzero scalar \(\lambda^{km}\) and interpolate \(F\) at \(k=1,\ldots,m+1\). The requested answer is \(Z_N=\lambda^m c_m\), including when \(c_m=0\). All chains and all query networks have polynomial size. The matrices \(I+(t/\lambda)N\) organize the algebraic expansion; only the chains \(H^k\) are submitted to the oracle.
\end{proof}

A projective matrix class has infinite order precisely when an invertible representative falls into one of these two cases. A diagonalizable representative has finite projective order exactly when its eigenvalue ratio is a root of unity.

\subsection{3.2 Rank-one absorption with one unary type}\label{rank-one-absorption-with-one-unary-type}

The following argument uses the straddled-signature splitting technique of Fan and Cai \cite{ref4}; see also its nonnegative formulation in Li, Huang, and Zheng \cite[Lemma 4]{ref20}. We include the proof because the discarded factor need not be nonzero over \(\mathbb C\), and because only one fixed unary type is introduced at a time.

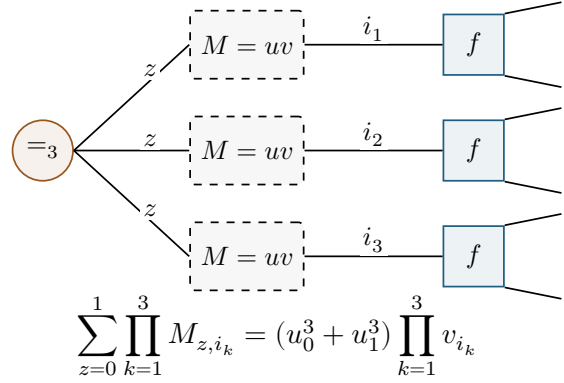
\begin{figure}[H]
\centering

\begin{tikzpicture}[x=1cm,y=1cm]
\node[R] (e) at (-2.7,0) {$=_3$};
\foreach \k/\yy in {1/1.4,2/0,3/-1.4}{
 \node[interpolated] (m\k) at (0,\yy) {$M=uv$};
 \node[L] (f\k) at (3,\yy) {$f$};
 \draw (e.east)--node[pos=0.66,above,lab] {$z$}(m\k.west);
 \draw (m\k.east)--node[above,lab] {$i_{\k}$}(f\k.west);
 \draw (f\k.north east)--++(0.75,0.15);
 \draw (f\k.south east)--++(0.75,-0.15);
}
\node[align=center] at (0.4,-2.45) {$\displaystyle\sum_{z=0}^1\prod_{k=1}^3 M_{z,i_k}
 =(u_0^3+u_1^3)\prod_{k=1}^3 v_{i_k}$};
\end{tikzpicture}

\caption{Absorbing three leftover columns. The three right-hand squares represent attachment sites in the rest of the instance. One new equality vertex contracts the L ports of three copies of $M$. This reproduces three RHS unaries, multiplied by $Q(u)$. All copies of the dashed matrix are eliminated together by the assumed reduction for networks containing $M$. For a spectral projector, this reduction is the chain interpolation already proved.}
\label{fig:absorption-7}
\end{figure}

\begin{holantstatement}{Lemma P.3 (one-unary absorption)}

Suppose the fixed rank-one matrix \(M=uv\) is simulable, where \(u\) is a nonzero column and \(v\) is a nonzero row. If

\[Q(u)=u_0^3+u_1^3\ne0,\]

then the problem with the single additional right unary type \(v\) reduces to \(\operatorname{Holant}(f\mid e)\).

\end{holantstatement}

\begin{proof}[Proof]

In an input with \(L_3\) ternary left vertices, \(R_3\) ternary right vertices and \(m\) right unary vertices, degree counting gives

\[3L_3=3R_3+m,\]

so \(m\) is divisible by three. Replace each right unary by a copy of the full matrix \(M=uv\), attaching its right port to the old unary's left neighbor. Its left port is initially unused. Group these unused left ports in triples and attach each triple to a new right equality vertex. If the three old external bits are \(i_1,i_2,i_3\), their contraction is

\[\sum_{z=0}^1 M_{z,i_1}M_{z,i_2}M_{z,i_3}
=\sum_{z=0}^1u_z^3v_{i_1}v_{i_2}v_{i_3}
=Q(u)v_{i_1}v_{i_2}v_{i_3}.\]

The resulting value is therefore \(Q(u)^{m/3}\) times the target value. Evaluate its copies of \(M\) using their simulation and divide by that nonzero known scalar. Each new edge joins a left port to a right port. Arbitrary triples may be grouped because no planarity restriction is imposed. Thus every final oracle query is a legal cubic bipartite graph; the row and column factors were used to analyze the contraction, not inserted as separately available signatures.
\end{proof}

The same argument works for \(WM=(Wu)v\) when \(W\) is actual. The particular homogeneous cubic \(Q(X,Y)=X^3+Y^3\) has exactly three projective zero directions over \(\mathbb C\), namely \([1:-1]\), \([1:-\omega]\), and \([1:-\omega^2]\) for a primitive cube root \(\omega\). Four distinct column directions therefore guarantee one choice with nonzero absorption factor.

Each application introduces one fixed unary type. A later chain changes that type by an actual graph substitution. We never infer simultaneous availability of unrelated unary types from divisibility of their combined occurrence count.

The left-unary version will also be used on the singular surface. We state it explicitly because the scalar to divide out is then a contraction with \(f\), rather than with equality.

\begin{holantstatement}{Lemma P.3a (left-unary absorption)}

Suppose \(M=uv\) is a fixed simulable rank-one straddled matrix, with nonzero column \(u\) and nonzero row \(v=(v_0,v_1)\). Define

\[C_f(v)=\sum_{i,j,k\in\{0,1\}}f(i,j,k)v_iv_jv_k
=f_0v_0^3+3f_1v_0^2v_1+3f_2v_0v_1^2+f_3v_1^3.\]

If \(C_f(v)\ne0\), then the one fixed left unary \(u\) is simulable.

\end{holantstatement}

\begin{proof}[Proof]

For an input with \(m\) left unary vertices, degree counting gives \(3L_3+m=3R_3\), so \(3\mid m\). Replace each unary by the full matrix \(M\), retaining its left port at the old attachment. Attach its unused right ports in groups of three to fresh left \(f\) vertices. The contraction at each new vertex is

\[\sum_{i,j,k\in\{0,1\}}f(i,j,k)v_iv_jv_k=C_f(v).\]

Consequently the resulting value is \(C_f(v)^{m/3}\) times the target. Simulate all occurrences of \(M\) and divide by this fixed nonzero factor. The final oracle graphs have ternary vertices on both sides. In particular, the proof never assumes that the two factors of \(M\) are separately available merely because \(M\) has rank one.
\end{proof}

\subsection{3.3 Unary interpolation}\label{unary-interpolation}

\begin{holantstatement}{Lemma P.4}

Let \(H\) be an actual straddled gadget over the fixed \(f\) and equality, with two distinct nonzero eigenvalues whose ratio is not a root of unity. Suppose a nonzero right unary row \(v\) is simulable and \(vH\) is not a scalar multiple of \(v\). Then every fixed algebraic right unary row \(w\) is individually simulable.

\end{holantstatement}

\begin{proof}[Proof]

Let \(P_1,P_2\) be the spectral projectors of this \(H\) and put \(v_i=vP_i\). Each \(v_i\) is a row vector, not an entry of \(v\). Since \(P_1+P_2=I\), we have \(v=v_1+v_2\), and \(v_iH=\lambda_i v_i\). If one \(v_i\) were zero, the nonzero row \(v\) would be an eigenrow, contrary to the hypothesis. Thus both are nonzero. They belong to distinct one-dimensional row eigenspaces and form a basis. Write the target row uniquely as \(w=\alpha v_1+\beta v_2\).

For an input with \(m\) occurrences of \(w\), regard its value as the homogeneous polynomial

\[P(X,Y)=\sum_{j=0}^m c_jX^jY^{m-j}\]

obtained by putting \(Xv_1+Yv_2\) at all marked locations. Replace every marked row by \(vH^k=\lambda_1^kv_1+\lambda_2^kv_2\). After division by \(\lambda_2^{km}\), the answer is

\[P((\lambda_1/\lambda_2)^k,1).\]

For \(k=1,\ldots,m+1\), these are evaluations at distinct points. Polynomial interpolation recovers all \(c_j\), hence also \(P(\alpha,\beta)\), including cases where \(\alpha\) or \(\beta\) is zero. Each query has only copies of the original row \(v\) after removing the actual attached chains; apply its given simulation. Thus only a single unary type is used at each stage.
\end{proof}

\subsection{3.4 The binary problem used in the reductions}\label{the-binary-problem-used-in-the-reductions}

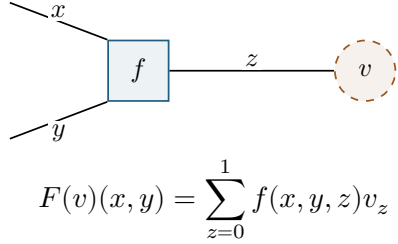
\begin{figure}[H]
\centering

\begin{tikzpicture}[x=1cm,y=1cm]
\node[L] (f) at (0,0) {$f$};\node[R,dashed] (v) at (3,0) {$v$};
\draw (-1.7,0.9)--node[above,lab] {$x$}(f.north west);
\draw (-1.7,-0.9)--node[below,lab] {$y$}(f.south west);
\draw (f.east)--node[above,lab] {$z$}(v.west);
\node[align=center] at (1,-1.7) {$F(v)(x,y)=\displaystyle\sum_{z=0}^1 f(x,y,z)v_z$};
\end{tikzpicture}

\caption{Attach one simulated RHS unary to a left ternary vertex. Its two other ports remain free, giving a binary signature. The dashed unary must later be eliminated by its established Turing reduction; it is not assumed available in the original input.}
\label{fig:contraction-8}
\end{figure}

Attaching \(v=(r,s)\) to one input of \(f=[1,a,b,c]\) gives

\[F_f(v)=[r+as,\ ar+bs,\ br+cs].\]

Its first and last entries are exactly the row \((r,s)A_f\). The notation means a tensor contraction, not an evaluation of \(f\) on non-Boolean inputs.

We use the complex binary dichotomy of Kowalczyk--Cai \cite{ref7} for \(\operatorname{Holant}(g\mid e)\), with \(g=[g_0,g_1,g_2]\). It is tractable precisely in these cases:

{\def\LTcaptype{table} 
\begin{longtable}[]{@{}
  >{\raggedright\arraybackslash}p{(\linewidth - 2\tabcolsep) * \real{0.5000}}
  >{\raggedright\arraybackslash}p{(\linewidth - 2\tabcolsep) * \real{0.5000}}@{}}
\toprule\noalign{}
\begin{minipage}[b]{\linewidth}\raggedright
Binary type
\end{minipage} & \begin{minipage}[b]{\linewidth}\raggedright
Algebraic condition
\end{minipage} \\
\midrule\noalign{}
\endhead
\bottomrule\noalign{}
\endlastfoot
Rank one & \(g_0g_2=g_1^2\) \\
Diagonal & \(g_1=0\) \\
Disequality & \(g_0=g_2=0\) \\
Additional affine types & \(g_1\ne0\), \(g_0g_2=-g_1^2\), \((g_0/g_1)^{12}=1\) \\
\end{longtable}
}

The last row consists of twelve projective points, all on the same conic \(g_0g_2+g_1^2=0\). This is the exact complex classification; restricting to only the real affine rays would be incorrect here. A binary left vertex and ternary equality right vertices are the standard edge-function formulation on cubic graphs, so the published binary theorem applies to precisely this auxiliary model.

\begin{holantstatement}{Lemma P.5 (binary contraction reduction)}

If the one right unary \(v\) is simulable and \(F_f(v)\) is outside the table, then \(\operatorname{Holant}(f\mid e)\) is \#P-hard.

\end{holantstatement}

\begin{proof}[Proof]

Replace each binary left vertex of an instance of \(\operatorname{Holant}(F_f(v)\mid e)\) by one \(f\) vertex carrying one attached right unary \(v\). This is a legal instance with that one unary type. Its value is unchanged by the definition of contraction. Simulate the unary using its stated reduction.
\end{proof}

Lemma 4 proves the following useful geometric count, which will be invoked whenever several rows are available: if \(f\) is neither rank one nor generalized equality, at most six row directions give an easy binary; if \(A_f\) is invertible, at most five do. The reason is that the image of \(F_f\) is a projective line, with at most two intersections with each of the rank-one and affine conics, and at most one diagonal direction. Invertible \(A_f\) excludes the disequality point entirely. The detailed proof appears as Lemma 4.

\subsection{3.5 Exact arithmetic and fixed choices}\label{exact-arithmetic-and-fixed-choices}

The signature \(f\) is fixed; the input is the graph. Each gadget and each auxiliary matrix word selected in a proof is chosen once for that \(f\). When a lemma simulates a specified fixed target unary, its coordinates are fixed as well. A constant-length word has length independent of the graph size, whereas interpolation uses chains of polynomially bounded length, as defined in Section 2.3.

For any one reduction, all signature entries, eigenvalues, selected vectors, and target unary coordinates belong to one fixed finite extension \(K/\mathbb Q\). Choose a rational basis of this field. Field addition and multiplication then use a fixed table of rational structure constants; a nonzero field element can be inverted by solving a fixed-dimensional rational linear system. This is exact arithmetic, not floating-point approximation.

Powers of fixed field elements with polynomially bounded exponents have polynomial bit length in this basis. The value of a polynomial-size network does as well: every summand is a product of polynomially many fixed weights, and the exponential number of assignments increases the logarithm of the coefficient bound only polynomially. This is a bound on the representation length of an oracle answer, not an algorithm for computing the counting sum directly.

The Vandermonde systems used above have polynomial dimension and entries of polynomial bit length. Their nodes are proved distinct before inversion, so exact rational linear algebra in the fixed field takes polynomial time. The same bounds apply to division by nonzero powers of absorption constants. A fixed number of composed reductions remains polynomial-time.

For the remaining cases, all parameters are fixed complex algebraic numbers. Unless a homogeneous signature is displayed explicitly, \(f=[1,a,b,c]\), \(e=[1,0,0,1]\), and \(A,B,D,E\) denote the actual matrices of Section 2. Every hardness assertion concerns the same cubic bipartite problem with no free unaries.

\subsection{3.6 Common criteria used in the case analysis}\label{common-criteria-used-in-the-case-analysis}

We collect the hypotheses of the three recurring arguments here so that each later reference has a specific algebraic meaning. Their proofs appear at the indicated locations. In this paragraph, \(H,T\) and the \(M_i\) are actual straddled gadget matrices for the same fixed \(f\).

\textbf{Jordan mixing (Lemma 3).} If \(H=\lambda I+N/2\), where \(\lambda\ne0\), \(N\ne0\), \(N^2=0\), and \(\operatorname{tr}(TN)\ne0\), then every fixed algebraic right unary is individually simulable. The nonzero trace moves the nilpotent column and row out of their respective invariant directions.

\textbf{Distinct-spectrum mixing (Proposition 7).} Suppose \(H\) has distinct nonzero eigenvalues with non-root-of-unity ratio, and

\[\det(HT-TH)\ne0,\qquad
(\operatorname{tr}T,\operatorname{tr}(HT))\ne(0,0).\]

Then every fixed right unary is individually simulable. If \(f\) is neither rank one nor generalized equality, Lemma 4 supplies a hard binary contraction and hence hardness. Neither mixing criterion requires \(T\) to be invertible.

\textbf{Projective orbits (Lemma 10).} Suppose \(M_1,\ldots,M_k\) are invertible, and the group generated by their projective classes is infinite and preserves neither a point nor an unordered pair of points. For a non-rank-one, non-equality \(f\), the problem is hard. The proof finds an infinite-order positive word, avoids the three column absorption zeros, and obtains more row directions than the finite tractable-binary bound allows. This argument needs enough candidate rows, rather than the simultaneous availability of unrelated unary types.

\section{4. The repeated-root surface and the general mixing lemma}\label{the-repeated-root-surface-and-the-general-mixing-lemma}

This section classifies \(\Delta_f=0\), where the first matrix \(A_f\) has a repeated eigenvalue. A nonzero repeated eigenvalue permits Jordan interpolation, but the resulting rank-one factor must still pass the absorption test; Sections 4.2--4.4 establish the required mixing and binary-contraction arguments. Section 4.5 applies them with \(H=A_f\) and \(T\in\{B,D\}\). When the repeated eigenvalue is zero, powers of \(A_f\) do not provide an interpolating recurrence, so Section 4.6 uses \(D\) instead. Section 4.7 closes the endpoint cases. The distinct-spectrum criterion in Section 4.8 is a separate reusable tool, not an additional assumption on the repeated-root theorem.

\subsection{4.1 The precise result}\label{the-precise-result}

Fix a symmetric Boolean ternary signature \(f=[f_0,f_1,f_2,f_3]\) whose entries are algebraic complex numbers. The input is a finite cubic bipartite multigraph, with \(f\) on the left and ternary equality \(e=[1,0,0,1]\) on the right. There are no freely available unary signatures. All reductions below are exact polynomial-time Turing reductions in a fixed number field.

The actual two-vertex straddled gadget has matrix

\[A_f=\begin{pmatrix}f_0&f_2\\f_1&f_3\end{pmatrix}.\]

Define its discriminant

\[\Delta_f=(f_0-f_3)^2+4f_1f_2.\]

\begin{holantstatement}{Theorem 1 (Repeated-root surface over the complex numbers)}

Suppose \(\Delta_f=0\). Then \(\operatorname{Holant}(f\mid e)\) is in FP at \(f=0\) and at nonzero scalar multiples of the following seven projective signatures:

\[[1,0,0,1],\qquad [1,a,-a^{-1},-1]\quad(a^6=1).\]

It is \#P-hard at every other point of the surface \(\Delta_f=0\).

\end{holantstatement}

Here ``seven projective signatures'' means that two nonzero signatures differing by a scalar count as one; the equation \(a^6=1\) has six distinct complex roots. This classifies the repeated-root surface, which is one branch of Theorem A.

The six points split into familiar classes. If \(a^3=-1\), then \(-a^{-1}=a^2\) and the signature is \((1,a)^{\otimes3}\). If \(a^3=1\), then

\[[1,a,-a^{-1},-1]=[1,a,-a^2,-1]\]

is a cube-root diagonal change of \([1,1,-1,-1]\). These are the rank-one and diagonally transformed affine cases already included in \(\mathcal T\).

The theorem includes weights that cannot be made real by the equality-preserving cube-root diagonal gauges. One explicit hard example is

\[f=\left[1,1+i,\frac{-1+i}{2},3\right],\qquad
A_f=\begin{pmatrix}1&(-1+i)/2\\1+i&3\end{pmatrix},\qquad
\det(tI-A_f)=(t-2)^2.\]

Multiplication of the second entry by any cube root of unity cannot make \(1+i\) real. After reversal and normalization the second entry is \((-1+i)/6\), which also cannot be made real by such a multiplication. Thus this example is not just a cube-root diagonal twist or reversal of a real signature.

\subsection{4.2 Binary hardness and the three complex absorption zeros}\label{binary-hardness-and-the-three-complex-absorption-zeros}

The reductions in this section retain the cubic bipartite constructions of Cai--Fan--Liu \cite{ref5}. The changes needed for complex weights concern the spectral parameters and the possible zeros of a column contraction.

We use the full complex binary classification of Kowalczyk--Cai. For a symmetric binary \(g=[g_0,g_1,g_2]\), the problem \(\operatorname{Holant}(g\mid e)\) is tractable exactly when one of the following holds:

\begin{enumerate}
\def\labelenumi{\arabic{enumi}.}
\tightlist
\item
  \(g_0g_2=g_1^2\);
\item
  \(g_1=0\);
\item
  \(g_0=g_2=0\);
\item
  \(g_1\ne0\), \(g_0g_2=-g_1^2\), and \((g_0/g_1)^{12}=1\).
\end{enumerate}

It is \#P-hard otherwise. This formulation follows from their theorem for \([x,1,y]\): the last condition is the twelve possibilities \(x^{12}=1,y=-x^{-1}\). Equivalently, these are the exceptional points obtained from the invariants \(X=xy\) and \(Y=x^3+y^3\) in Kowalczyk--Cai \cite[Theorem 4.19 of the preprint]{ref7}.

The main extra issue over the reals is absorption. A column \(u=(u_0,u_1)^T\) can be absorbed in triples at equality if

\[Q(u)=u_0^3+u_1^3\ne0.\]

Over \(\mathbb R\) there is only one forbidden direction, \((1,-1)^T\). Over \(\mathbb C\) there are \textbf{three} forbidden directions. Thus the real proof's assertion ``a column moved off that line is absorbable'' cannot simply be reused. The next lemma replaces it by a finite, exact construction.

\subsection{4.3 Four candidates suffice for complex absorption}\label{four-candidates-suffice-for-complex-absorption}

\begin{holantstatement}{Lemma 2 (Absorption along a nonconstant projective sequence)}

Let \(u,w\in\mathbb C^2\) be linearly independent, and let \(\gamma\ne0\). At least one of

\[w,\quad w+\gamma u,\quad w+2\gamma u,\quad w+3\gamma u\]

has nonzero \(Q\).

\end{holantstatement}

\begin{proof}[Proof]

The four vectors have four distinct projective directions: equality of the directions of \(w+j\gamma u\) and \(w+k\gamma u\) forces equality of the coefficients of \(w\), and then \(j=k\). The nonzero homogeneous cubic \(Q(X,Y)=X^3+Y^3\) vanishes on exactly three projective directions over \(\mathbb C\). Thus it cannot vanish at all four.
\end{proof}

We now explain precisely how this lemma interacts with the graph reductions.

\begin{holantstatement}{Lemma 3 (A nontrivial Jordan gadget supplies unaries)}

Let \(H\) and \(T\) be actual straddled gadget matrices for fixed \(f\). Suppose

\[H=\lambda I+N/2,\qquad \lambda\ne0,\quad N\ne0,\quad N^2=0,
\quad\operatorname{tr}(TN)\ne0.\]

Then every fixed complex algebraic right unary can be simulated one type at a time, by polynomial-time Turing reduction to \(\operatorname{Holant}(f\mid e)\).

\end{holantstatement}

\begin{proof}[Proof]

Factor the rank-one matrix \(N=uv\). Here \(u\) is a nonzero column and \(v\) is a nonzero row. Because \(N^2=0\), we have \(vu=0\). Put \(\gamma=vTu=\operatorname{tr}(TN)\ne0\). The columns \(u,Tu\) are independent, since a proportionality \(Tu=\alpha u\) would imply \(vTu=\alpha vu=0\).

First, \(N\) itself is available by interpolation: at \(m\) marked locations use the same actual chain

\[H^k=\lambda^k\left(I+\frac{k}{2\lambda}N\right),\quad k=1,\ldots,m+1.\]

After dividing the network value by \(\lambda^{km}\), the result is a polynomial of degree at most \(m\). Its leading coefficient, multiplied by \((2\lambda)^m\), is the all-\(N\) network value. This uses actual chains only; it does not assert that subtraction of gadgets is a legal graph operation.

For each fixed \(j\in\{0,1,2,3\}\), the same coefficient extraction with actual chains \(H^j T H^k\) simulates

\[H^jTN=(H^jTu)v.\]

Its column is

\[H^jTu=\lambda^j\left(Tu+\frac{j\gamma}{2\lambda}u\right).\]

Lemma 2 gives a fixed \(j\) for which \(Q(H^jTu)\ne0\). The choice depends only on \(f,H,T\), not on the input graph.

To simulate one right unary \(v\), replace each of its \(M\) occurrences by the rank-one matrix \((H^jTu)v\). Degree counting gives \(M\equiv0\pmod3\): a legal graph with \(L\) ternary left vertices, \(R\) ternary right vertices and \(M\) right unary vertices satisfies \(3L=3R+M\). Group the leftover \(M\) columns into triples at fresh right equality vertices. Divide the result by the nonzero factor \(Q(H^jTu)^{M/3}\). Every query is a legal cubic bipartite graph after the rank-one boxes have been removed by interpolation. Thus \(v\) is simulable.

Attach the actual matrix \(T\) to this row, giving \(s=vT\). Then \(sN=\gamma v\ne0\). The rows \(s\) and \(v\) are independent: \(su=\gamma\ne0\) but \(vu=0\). Therefore

\[sH^k=\lambda^k\left(s+\frac{k\gamma}{2\lambda}v\right)\]

ranges over infinitely many distinct row directions. For any fixed target row \(q=\alpha s+\beta v\), an input with \(M\) copies of \(q\) can be evaluated from \(M+1\) evaluations with every copy replaced by the same \(sH^k\). Expand the resulting degree-\(M\) polynomial in the coefficient of \(v\); the recovered coefficients give the target value at \((\alpha,\beta)\), including \(\alpha=0\). Each such evaluation contains only the one unary type \(v\) after the attached chains are removed, so its established simulation applies. The number of queries, graph size and exact arithmetic lengths are polynomial.
\end{proof}

The lemma does not require \(T\) to be invertible. The trace pairing is the exact condition used.

\subsection{4.4 Why arbitrary unaries give hardness here}\label{why-arbitrary-unaries-give-hardness-here}

For \(f=[1,a,b,c]\), attachment of the right unary row \((r,s)\) produces

\[F_f(r,s)=[r+as,\ ar+bs,\ br+cs].\]

The quadratic determinant test for obtaining a nondegenerate binary contraction from three distinct unary directions also appears in Li, Huang, and Zheng \cite[Lemma 9]{ref20}. Nondegeneracy alone does not imply hardness: a rank-two binary can still be diagonal, disequality, or affine. The next lemma counts all of these tractable possibilities.

\begin{holantstatement}{Lemma 4 (A hard binary exists)}

If \(f=[1,a,b,c]\) is neither degenerate nor a generalized equality, then at most six projective unary directions yield tractable binary signatures. If \(c-ab\ne0\), the bound improves to five. Thus simulation of every fixed unary proves \(\operatorname{Holant}(f\mid e)\) \#P-hard.

\end{holantstatement}

\begin{proof}[Proof]

The linear map \(F_f\) has columns \((1,a,b)^T\) and \((a,b,c)^T\). They are dependent exactly when \(b=a^2,c=a^3\), the excluded degenerate case. Hence \(F_f\) is injective and its projectivized image is a line in the projective plane of binary signatures.

The rank-one binary signatures form the nonsingular conic \(g_0g_2-g_1^2=0\). A projective line meets a nonsingular conic at at most two points, so this class contributes at most two unary directions. The diagonal equation \(ar+bs=0\) is nonzero because the generalized equality case \(a=b=0\) was excluded; it contributes at most one. Disequality is one projective point and contributes at most one. Finally, all twelve complex affine points lie on the other nonsingular conic \(g_0g_2+g_1^2=0\), contributing at most two directions. Thus the total is at most \(2+1+1+2=6\).

For an elementary justification of the conic intersection bound, a line contained in \(g_0g_2\pm g_1^2=0\) would give a rank-two linear map all of whose outputs have determinant zero as symmetric two-by-two matrices. If one of \(g_0,g_2\) vanishes identically, the equation also forces \(g_1\) to vanish identically, leaving rank at most one. Otherwise parameterize the line in coordinates where one endpoint is a free coordinate; substitution gives a nonzero quadratic. Equivalently, the quadratic forms \(g_0g_2\pm g_1^2\) have matrix rank three and therefore cannot factor into two linear forms; a line component would require such a factorization.

When \(c-ab\ne0\), the endpoint pair is \((r,s)A_f\). A nonzero row cannot give two zero endpoints, so the disequality point has no preimage and the count drops to five. The count of twelve affine rays separately would be needlessly weak; their common conic is essential to this sharper bound.

Choose a fixed row outside this finite set, simulate that row alone, and replace each vertex of the hard binary problem by one \(f\) vertex with the row attached to its third port. This is the reduction

\[\operatorname{Holant}(F_f(r,s)\mid e)\le_T\operatorname{Holant}(f\mid e).\]

The source binary theorem supplies the hardness.
\end{proof}

\subsection{4.5 Nonzero repeated eigenvalue: all non-scalar cases are hard}\label{nonzero-repeated-eigenvalue-all-non-scalar-cases-are-hard}

\begin{holantstatement}{Proposition 5}

Let \(f=[1,a,b,c]\) have \(\Delta_f=0\) and \(c-ab\ne0\). Then \(\operatorname{Holant}(f\mid e)\) is in FP if \(a=b=0,c=1\), and is \#P-hard otherwise.

\end{holantstatement}

\begin{proof}[Proof]

Write

\[A=\begin{pmatrix}1&b\\a&c\end{pmatrix},\quad
B=\begin{pmatrix}1+ab&a^2+bc\\a+b^2&ab+c^2\end{pmatrix},\]

\[D=\begin{pmatrix}1+2a^3+b^3&a^2+2ab^2+bc^2\\
a+2a^2b+b^2c&a^3+2b^3+c^3\end{pmatrix}.\]

These are actual gadgets: \(A\) is the double-edge construction, \(B\) is the four-vertex straddled construction, and \(D\) closes two ports of the ternary gadget formed by three copies of \(f\) attached to the same two equality vertices.

Set \(\lambda=(1+c)/2\) and \(N=2A-(1+c)I\). Then \(\lambda\ne0\), \(N^2=\Delta_f I=0\). If \(N=0\) we have \(f=[1,0,0,1]\), whose network is evaluated componentwise. Assume \(N\ne0\).

We claim at least one of \(T=B,D\) has \(\operatorname{tr}(TN)\ne0\). If \(c=1\), then \(ab=0\) and

\[\operatorname{tr}(BN)=2(a^3+b^3).\]

Exactly one of \(a,b\) is nonzero, so this expression is nonzero even over \(\mathbb C\).

If \(c\ne1\), put

\[t=(c-1)/2,\qquad a=tr,\quad b=-t/r,\quad R=r^3.\]

The repeated-root equation justifies these substitutions; \(t,r,R\) are nonzero. Define

\[P_0=t(R^2+2R-1)+2R,\] \[Q_0=t^2(R^2-8R+3)-t(R^2+8R-1)-4R.\]

Direct multiplication gives

\[\operatorname{tr}(BN)=2t^2P_0/R,
\qquad\operatorname{tr}(DN)=-2t^2Q_0/R.\]

If both traces vanished, \(P_0=0\) would give \(L=R^2+2R-1\ne0\) and \(t=-2R/L\). Substitution then gives

\[Q_0=-\frac{2R(R-1)^4}{L^2}.\]

Thus \(R=1\), whence \(t=-1\) and \(\lambda=1+t=0\), a contradiction. These are algebraic identities; no ordering or complex-conjugation argument is used.

With this actual \(T\), Lemma 3 supplies every fixed unary. Also \(a,b\) are not both zero: together with \(\Delta_f=0\) that would force the previously removed scalar case. Lemma 4 proves hardness.
\end{proof}

\subsection{4.6 Zero repeated eigenvalue: six tractable points}\label{zero-repeated-eigenvalue-six-tractable-points}

\begin{holantstatement}{Proposition 6}

For \(a\ne0\), let \(f=[1,a,-a^{-1},-1]\). Then \(\operatorname{Holant}(f\mid e)\) is in FP when \(a^6=1\), and is \#P-hard when \(a^6\ne1\).

\end{holantstatement}

\begin{proof}[Proof of hardness]

Assume \(a^6\ne1\). Set

\[\ell=a^3-a^{-3},\quad
u=\begin{pmatrix}1\\-a^{-2}\end{pmatrix},\quad v=(a,1),\quad
P=\frac{uv}{vu}.\]

Here \(\ell\ne0\), \(vu=a-a^{-2}\ne0\), and \(P^2=P\). The actual \(D\) gadget defined above satisfies

\[D=\ell(I+P).\]

Its eigenvalues are \(\ell,2\ell\), so the non-root-of-unity ratio 2 permits ordinary spectral interpolation of \(P\). Its column absorption factor is

\[Q(u)=1-a^{-6}\ne0.\]

Consequently the row \(v=(a,1)\) is simulable. Its binary contraction is

\[g=[2a,a^2-a^{-1},-2].\]

The two endpoints are nonzero; the middle entry is nonzero because \(a^3\ne1\). Its determinant is

\[g_0g_2-g_1^2=-\frac{(a^3+1)^2}{a^2}\ne0.\]

It remains to exclude all twelve complex affine exceptions, not only the two real ones. If \(g_0g_2=-g_1^2\), then \(g_1^2=4a\), and hence

\[\left(\frac{g_0}{g_1}\right)^{12}
=\left(\frac{4a^2}{4a}\right)^6=a^6\ne1.\]

Thus the last tractable binary condition also fails. The known binary theorem proves hardness, transferred by the one-unary contraction.
\end{proof}

\begin{proof}[Proof of tractability]

If \(a^3=-1\), the signature is \((1,a)^{\otimes3}\) and is evaluated as a product of unary contributions. If \(a^3=1\), its value at a Hamming-weight-\(k\) input is \(a^k h_k\), where \(h=[1,1,-1,-1]\). In any surviving assignment of a cubic bipartite graph, the number of 1-labelled edges is three times the number of right vertices assigned 1. Therefore the product of the extra factors is

\[a^{\sum_{v\in L}k_v}=a^{3m}=1.\]

So \(Z_f(G)=Z_h(G)\) exactly. The signature \(h(i,j,k)=(-1)^{ij+ik+jk}\) is affine; Gaussian elimination of its quadratic phase gives a polynomial-time evaluation.
\end{proof}

This last graph argument explains the equality-preserving cube-root diagonal gauge directly. More generally, replacing \(f_k\) by \(\omega^k f_k\) with \(\omega^3=1\) leaves every closed instance value unchanged.

\subsection{4.7 Completion at zero endpoints}\label{completion-at-zero-endpoints}

\begin{proof}[Proof of Theorem 1]

If \(f_0\ne0\), normalize \(f_0=1\). If \(\det A_f\ne0\), Proposition 5 applies. If \(\det A_f=0\) and \(\Delta_f=0\), then \(1+c=0\) and \(ab=-1\), giving exactly Proposition 6.

If \(f_0=0\) but \(f_3\ne0\), reverse the two domain labels first. This preserves equality, the discriminant equation and complexity. The seven listed projective points are closed under reversal: equality is fixed, and reversal sends the normalized parameter \(a\) to \(a^{-1}\) in the six-point family.

Finally, if \(f_0=f_3=0\), the discriminant equation gives \(f_1f_2=0\). Apart from \(f=0\), the signature is a nonzero multiple of Exact-One \([0,1,0,0]\) or its reversal. The rational base theorem already proves this problem \#P-hard on precisely the cubic bipartite model. Scaling and reversal transfer that hardness. These cases exhaust the surface.
\end{proof}

\subsection{4.8 Additional generic complex hardness criterion}\label{additional-generic-complex-hardness-criterion}

\begin{holantstatement}{Proposition 7}

Let \(f=[1,a,b,c]\) be complex algebraic, neither degenerate nor a generalized equality. Suppose an actual straddled gadget \(H\) has distinct nonzero eigenvalues whose ratio is not a root of unity. Let \(T\) be any actual straddled gadget such that

\[\det(HT-TH)\ne0,\qquad
(\operatorname{tr}T,\operatorname{tr}(HT))\ne(0,0).\]

Then \(\operatorname{Holant}(f\mid e)\) is \#P-hard. In particular, one can use \(H=A_f\) whenever these hypotheses hold for the first gadget.

\end{holantstatement}

\begin{proof}[Proof]

In an \(H\) eigenbasis write \(H=\operatorname{diag}(\lambda_1,\lambda_2)\) and \(T=\left(\begin{smallmatrix}p&q\\r&s\end{smallmatrix}\right)\). Write \(\delta=\lambda_1-\lambda_2\ne0\). Then

\[HT-TH=\begin{pmatrix}0&\delta q\\-\delta r&0\end{pmatrix},\qquad
\det(HT-TH)=\delta^2qr.\]

Hence \(q,r\ne0\). Also

\[\binom{\operatorname{tr}T}{\operatorname{tr}(HT)}
=\begin{pmatrix}1&1\\\lambda_1&\lambda_2\end{pmatrix}\binom ps,\]

so the two traces vanish together exactly when \(p=s=0\). At least one of \(p,s\) is therefore nonzero. For the first coordinate column, if \(p\ne0\), the column \(Te_1=(p,r)^T\) already has both entries nonzero. If \(p=0\), then \(s\ne0\) and \(T^2e_1=(qr,rs)^T\) has both nonzero. The same calculation for \(e_2\) and for coordinate rows proves the other three assertions. This explicit calculation permits singular \(T\) as well.

Interpolate either eigenprojector \(P=uv\). Choose \(j=1\) or 2 so \(T^ju\) has both eigencomponents nonzero. The four columns \(H^kT^ju\), \(k=0,1,2,3\), have distinct directions, because the eigenvalue ratio is not a root of unity. At least one has nonzero \(Q\). Thus one of the simulable rank-one matrices \(H^kT^jP\) supplies the row \(v\) by absorption.

Now choose \(j'=1\) or 2 so \(s_0=vT^{j'}\) has both row eigencomponents nonzero. The actual chains \(s_0H^k\) interpolate every fixed unary by a Vandermonde system. Lemma 4 finishes the reduction. The original first matrix \(A_f\) and the mixing matrix \(T\) may be singular. Only the recurrence matrix \(H\) is required to have the two nonzero distinct eigenvalues in the statement.
\end{proof}

The following sections treat the remaining matrix configurations explicitly.

\section{5. The singular first-matrix surface}\label{the-singular-first-matrix-surface}

Here the normalized parameters satisfy \(c=ab\), so the actual first matrix already factors as \(A=(1,a)^T(1,b)\). Matrix rank one does not force the ternary tensor \(f\) to be rank one. After removing the stated tractable signatures, Section 5.2 makes the row \((1,b)\) available, moving its unused column with \(B\) or \(D\) if needed. The first binary contraction in Section 5.3 is hard except on \(a=-b^2\), \(ab=-1\), or \(b^3=1\). Sections 5.4--5.6 settle those three families and their intersections. In particular, their ninth-root and purely imaginary residual points require new recurrences, rather than an inference from the first easy binary output.

\subsection{5.1 Statement and familiar tractable signatures}\label{statement-and-familiar-tractable-signatures}

We study the same exact problem \(\operatorname{Holant}(f\mid e)\) on cubic bipartite multigraphs, with fixed algebraic complex symmetric ternary \(f\) and \(e=[1,0,0,1]\). There are no free unary signatures. Normalizing \(f_0=1\), singularity of the first straddled gadget means

\[f=[1,a,b,ab],\qquad
A=\begin{pmatrix}1&b\\a&ab\end{pmatrix}
=\begin{pmatrix}1\\a\end{pmatrix}(1,b).\]

Write \(\Gamma_\omega[f_0,f_1,f_2,f_3]=[f_0,\omega f_1,\omega^2f_2,f_3]\) for \(\omega^3=1\). On every closed legal graph the additional factors multiply to 1, because the total number of 1-edges is a multiple of three. Thus \(\Gamma_\omega\) preserves the partition function exactly.

\begin{holantstatement}{Theorem 8 (Singular surface)}

Let \(a,b\) be fixed algebraic complex numbers. The problem with \(f=[1,a,b,ab]\) is in FP if either

\[b=a^2,\]

or \(f\) belongs to the cube-root diagonal orbits of the following six signatures:

\[[1,0,1,0],\quad[1,0,-1,0],\] \[[1,1,-1,-1],\quad[1,-1,-1,1],\] \[[1,i,1,i],\quad[1,-i,1,-i].\]

There are eighteen distinct projective points in these six orbits. The problem is \#P-hard at every other point of the singular surface.

\end{holantstatement}

The first class is rank one. The six displayed signatures are ordinary complex affine signatures; the diagonal gauges are already a familiar holographic operation. The assertion of Theorem 8 is that no further tractable signatures occur on the surface \(\det A_f=0\).

For clarity, the affine formulas are

\[[1,0,1,0](x,y,z)=\mathbf1[x+y+z\equiv0\pmod2],\] \[[1,0,-1,0](x,y,z)=\mathbf1[x+y+z\equiv0\pmod2](-1)^{xy+xz+yz},\] \[[1,\varepsilon,-1,-\varepsilon](x,y,z)=(-1)^{xy+xz+yz}\varepsilon^{x+y+z},\quad\varepsilon=\pm1,\] \[[1,\varepsilon i,1,\varepsilon i](x,y,z)
=(\varepsilon i)^{x+y+z}(-1)^{xy+xz+yz},\quad\varepsilon=\pm1.\]

Gaussian elimination evaluates these affine partition functions in polynomial time. Rank-one signatures are products of unary contributions, and the cube-root gauges preserve the value as explained above. This proves the tractable side of Theorem 8.

\begin{proof}[Proof of the hardness assertion of Theorem 8]

We retain \(f=[1,a,b,ab]\) and remove the rank-one family and the eighteen normalized gauge images of affine representatives covered by Proposition P.1. Sections 5.2--5.6 exhaust the remaining parameter values.

\subsection{5.2 Make the row of A available}\label{make-the-row-of-a-available}

\begin{figure}[H]
\centering
\begin{tikzpicture}[x=1.05cm,y=1.05cm]
\coordinate (l) at (0,0);
\coordinate (r) at (3,0);
\coordinate (i) at (-1.5,0);
\coordinate (j) at (4.5,0);
\draw (i) to[] node[midway,above,lab] {$i$} (l);
\draw (r) to[] node[midway,above,lab] {$j$} (j);
\draw (l) to[bend left=28] node[midway,above,lab] {$p$} (r);
\draw (l) to[bend right=28] node[midway,below,lab] {$q$} (r);
\node[L] at (l) {$h$};
\node[R] at (r) {$=_3$};
\end{tikzpicture}
\caption{The gadget $A_h$. Squares are left vertices carrying $h$; circles are right equality vertices. The two curved lines are distinct parallel edges. Only $i,j$ are external inputs; $p,q$ are summed over.}
\label{fig:gadget_a-9}
\end{figure}
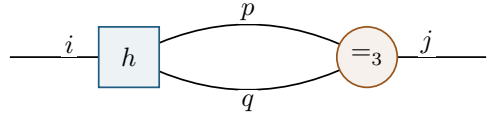

Put \(u=(1,a)^T\), \(v=(1,b)\), so \(A=uv\). The first absorption attempt has factor

\[Q(u)=1+a^3.\]

If it is nonzero, \(v\) is simulable by grouping unused columns in triples. Suppose \(a^3=-1\). Use a cube-root gauge to make \(a=-1\): choose \(\omega=-1/a\), which satisfies \(\omega^3=1\), and rename the transformed second parameter \(b\). The transformed signature is still \([1,-1,b,-b]\).

For this signature the actual \(B\) gadget obeys

\[Bu=(b-1)\begin{pmatrix}b\\1\end{pmatrix},
\qquad Q(Bu)=(b-1)^3(b^3+1).\]

Thus \(BA=(Bu)v\) supplies the same row \(v\) unless \(b=1\) or \(b^3=-1\). The point \(b=1\) is rank one and was already declared tractable. The point \(b=-1\) is the listed affine signature \([1,-1,-1,1]\).

For the two remaining roots \(b^3=-1,b\ne-1\), use the actual \(D\) gadget. Direct substitution gives

\[Du=2(b-1)(b+1)\begin{pmatrix}1\\-b\end{pmatrix},\]

and, using \(b^3=-1\),

\[Q(Du)=48b(b+1)\ne0.\]

Hence \(DA=(Du)v\) supplies \(v\) in these cases too. We have shown that, after removing already known tractable points and allowing the value-preserving gauges, the row \(v=(1,b)\) is always available. Only one row type is introduced in each reduction.

\subsection{5.3 The first binary leaves only three families}\label{the-first-binary-leaves-only-three-families}

Contract \(v\) with \(f\) and put \(p=1+ab\), \(t=a+b^2\). The result is

\[g=F_f(v)=[p,t,bp],\qquad
g_0g_2-g_1^2=(a^2-b)(b^3-1).\]

The complex binary theorem says this is hard unless it is rank one, diagonal, disequality, or one of the twelve affine points. The first three cases give respectively

\[b=a^2\ \text{or}\ b^3=1;\qquad a=-b^2;\qquad ab=-1.\]

To inspect the last case assume \(p,t\ne0\) and set \(\sigma=t/p\). The two affine conditions imply

\[\sigma^{12}=1,\qquad b=-\sigma^2,\]

and the equation \(a+b^2=\sigma(1+ab)\) becomes

\[a(1+\sigma^3)=\sigma(1-\sigma^3).\]

Because \(\sigma^3\in\{1,-1,i,-i\}\), this has four elementary cases. If \(\sigma^3=-1\), the equation is impossible. If \(\sigma^3=1\), it forces \(a=0,b^3=-1\), already a gauge of \([1,0,-1,0]\). If \(\sigma^3=\pm i\), it gives \(a=\mp i\sigma\), hence \(b=a^2\) (and in fact \(p=0\), contradicting the current assumption). Thus no other family arises.

After removing rank one and the listed affine points, it remains to prove hardness in

\[a=-b^2,\qquad ab=-1,\qquad b^3=1.\]

These families can intersect. Each proof below removes its own tractable intersection points before dividing by any factor.

\subsection{\texorpdfstring{5.4 The curve a = \(-b^{2}\), including the ninth roots}{5.4 The curve a = -b\^{}\{2\}, including the ninth roots}}\label{the-curve-a--b2-including-the-ninth-roots}

If \(b=0\), the signature is rank one. If \(b^3=1\), then \(a=-b^2\) and \(b=a^2\), also rank one. If \(b^3=-1\), apply \(\omega=b^{-2}\); then \(\omega a=-1\) and \(\omega^2b=-1\), so the transformed signature is \([1,-1,-1,1]\). These are the cases \(b^6=1\).

Assume from now on \(b\ne0\) and \(b^6\ne1\). We apply the left-unary absorption of Lemma P.3a to the actual factorization \(A=uv\), where \(u=(1,-b^2)^T\) and \(v=(1,b)\). Its required contraction factor is

\[\sum_{x,y,z\in\{0,1\}}f(x,y,z)v_xv_yv_z=1-b^6\ne0.\]

Two copies of this same left unary attached to a right equality vertex produce the right unary row \((1,b^4)\): if the remaining bit is \(j\), both attached bits must equal \(j\), so the contribution is \(u_j^2\). Thus any number of copies of this row reduces to copies of the one already simulated left unary. Attach this row to \(f\), and divide out \(1-b^3\ne0\), obtaining

\[h=[1+b^3,-b^2,b(1+b^3)].\]

The endpoints and the middle entry are nonzero. Its determinant is

\[h_0h_2-h_1^2=b(1+b^3+b^6).\]

Set \(x=b^3\). If the last affine binary class applied, its cross-relation would give

\[(1+x)^2=-x.\]

Its twelfth-power relation would then give

\[1=\left(\frac{1+x}{-b^2}\right)^{12}
=\frac{x^6}{x^8}=x^{-2}.\]

Thus \(x^2=1\), inconsistent with \((1+x)^2=-x\). Consequently the binary \(h\) is hard unless

\[b^6+b^3+1=0.\]

These six primitive ninth roots are a genuine extra exception to the real proof: over the reals the displayed polynomial never vanishes. We handle all six exactly.

Under \(b^6+b^3+1=0\), the actual \(D\) and \(B\) matrices satisfy

\[\det D=-6,\qquad (\operatorname{tr}D)^2=-27,\]

\[\det(DB-BD)=-9(b^3+1)\ne0,\qquad\operatorname{tr}B=-3b^3\ne0.\]

The eigenvalues of \(D\) are nonzero and distinct. If their ratio is \(\rho\), then

\[2+\rho+\rho^{-1}=
\frac{(\operatorname{tr}D)^2}{\det D}=\frac92,\]

so \(\rho\in\{2,1/2\}\). Apply Proposition 7 above with \textbf{\(H=D\) and \(T=B\)}. It simulates every fixed unary using actual gadgets of the original \(f\). The original \(f\) is not rank one because \(b-a^2=b(1-b^3)\ne0\), and is not a generalized equality because \(b\ne0\). Therefore the general form of Lemma 4 applies even though the original \(A\) is singular. This proves hardness at the primitive ninth roots and completes the curve.

\subsection{\texorpdfstring{5.5 The curve ab = \(-1\)}{5.5 The curve ab = -1}}\label{the-curve-ab--1}

Here \(f=[1,a,-a^{-1},-1]\) with \(a\ne0\). Proposition 6 above is already a complete complex classification: the problem is hard if \(a^6\ne1\); for \(a^3=-1\) it is rank one; and for \(a^3=1\) it is a cube-root gauge of \([1,1,-1,-1]\). Thus this entire curve is settled without invoking a real-weight theorem.

\subsection{\texorpdfstring{5.6 The three lines \(b^{3}\) = 1}{5.6 The three lines b\^{}\{3\} = 1}}\label{the-three-lines-b3-1}

Choose the equality-preserving gauge \(\omega=b\) to make the new second-even entry \(\omega^2b=1\). The transformed parameter is \(a'=ab\). Rename it \(a\). We now classify

\[f=[1,a,1,a].\]

The points \(a=\pm1\) are rank one; \(a=0\) is \([1,0,1,0]\); and \(a=\pm i\) are the two listed complex affine signatures. Assume these five values have been excluded.

Here the actual \(B\) gadget itself, without spectral interpolation, factors as

\[B=(1+a)\begin{pmatrix}1\\1\end{pmatrix}(1,a).\]

The column absorption factor is \(2(1+a)^3\ne0\), so the right unary row \((1,a)\) is available. Its contraction is

\[g=[1+a^2,2a,1+a^2].\]

The endpoints and the middle are nonzero, and its determinant is \((a^2-1)^2\ne0\). Therefore it is hard unless it lies in the binary affine class. That cross-relation is

\[a^4+6a^2+1=0.\]

All four solutions are purely imaginary: \(a=it\) with

\[t\in\{\sqrt2+1,\sqrt2-1,-\sqrt2+1,-\sqrt2-1\}.\]

In particular \(t\ne0\) and \(t^2\ne1\). The actual \(D\) gadget is

\[D=2(1+a)\begin{pmatrix}a^2-a+1&a\\a&a^2-a+1\end{pmatrix}.\]

Its eigenvectors are \((1,1)^T,(1,-1)^T\), with eigenvalues

\[\lambda_+=2(1+a)(a^2+1),\qquad
\lambda_-=2(1+a)(a-1)^2.\]

Both are nonzero. At the four residual points,

\[\left|\frac{\lambda_+}{\lambda_-}\right|
=\frac{|1-t^2|}{1+t^2}<1.\]

Thus their ratio is not a root of unity. The already available row \((1,a)\) has both row eigencomponents nonzero because \(a\ne\pm1\). The chains \((1,a)D^k\) interpolate every fixed right unary by the ordinary Vandermonde argument. Lemma 4 proves hardness, since \(a\ne\pm1\) makes \(f\) nondegenerate. This finishes the four residual points and hence all three lines \(b^3=1\). The three families left in Section 5.3 have all been classified, proving the hardness assertion.
\end{proof}

\subsection{5.7 Completion and unnormalized boundary}\label{completion-and-unnormalized-boundary}

\begin{proof}[Proof of Theorem 8]

The row-availability argument applies outside the listed tractable points, after any required equality-preserving gauge. Its binary contraction is hard unless the signature lies in one of the three families proved in Sections 5.4--5.6. Each of those family proofs explicitly recognizes its tractable intersections and proves hardness everywhere else. Their union with rank one and the affine points therefore exhausts all cases.
\end{proof}

The normalized theorem extends to the complete homogeneous surface

\[f_0f_3-f_1f_2=0.\]

If an endpoint is nonzero, normalize it after reversal if necessary, and use Theorem 8. If both endpoints vanish, then \(f_1f_2=0\): the zero signature is tractable, and every nonzero remaining signature is Exact-One or its reversal, which is \#P-hard by the rational 3--3 base theorem. The tractable classes in homogeneous form are rank one, the listed cube-root affine orbits, their reversals, and nonzero scalings.

\section{6. Every distinct-spectrum infinite-order first matrix}\label{every-distinct-spectrum-infinite-order-first-matrix}

The first matrix now has distinct nonzero eigenvalues with non-root-of-unity ratio. Its two eigenprojectors are therefore simulable. After removing generalized equality, Section 6.1 first tries the mixing pair \((A,B)\). If it does not apply, the standing condition is \(\det(AB-BA)=0\), which does not assert \(AB=BA\). Sections 6.2--6.4 then distinguish two, one, or no absorbable column eigenlines. Two available eigenrows are handled by comparing their binary contractions; one available eigenrow leaves explicit exceptional families; when neither column is absorbable, a holographic transformation and degree counting give perfect-matching hardness.

\begin{holantstatement}{Theorem 9}

Let \(f=[1,a,b,c]\) have algebraic complex entries. Suppose

\[A=\begin{pmatrix}1&b\\a&c\end{pmatrix}\]

has two distinct nonzero eigenvalues whose ratio is not a root of unity. If \(a=b=0\), then \(f\) is a generalized equality and \(\operatorname{Holant}(f\mid e)\) is in FP. In every other case the problem is \#P-hard on cubic bipartite multigraphs.

\end{holantstatement}

Together with Proposition 5 above, this handles every first gadget of infinite projective order. The matrices are actual \(f\mid e\) gadgets; no free unaries are assumed.

\begin{proof}[Proof of Theorem 9]

The hypotheses of Theorem 9 remain in force throughout the following four cases. We first reduce to a zero commutator determinant and then distinguish how many of the two column eigenlines have zero absorption factor.

\subsection{6.1 Noncommuting eigenlines}\label{noncommuting-eigenlines}

Use the same actual matrix

\[B=\begin{pmatrix}1+ab&a^2+bc\\a+b^2&ab+c^2\end{pmatrix}.\]

The identity

\[2\operatorname{tr}B=\Delta_A+(\operatorname{tr}A)^2\]

shows that \(\operatorname{tr}B\ne0\) under the theorem's hypotheses. Indeed, otherwise the nonzero distinct eigenvalues would be proportional to \(1+i,1-i\), with ratio \(i\) or \(-i\). If their sum were zero as well, both eigenvalues would be zero. Neither alternative is allowed.

Thus if \(\det(AB-BA)\ne0\), Proposition 7 above applies with \(H=A,T=B\) and proves hardness. From now on assume

\[\det(AB-BA)=0.\]

The two eigenprojectors of \(A\) can be simulated. A projector \(uv\) supplies its row when \(Q(u)=u_0^3+u_1^3\ne0\). We divide into zero, one, or two forbidden column eigenlines.

\subsection{6.2 Both column eigenlines are absorbable}\label{both-column-eigenlines-are-absorbable}

If \(a=0\), then \(b\ne0\). The row \((0,1)\) is an eigenrow of \(A\), and its paired column has nonzero \(Q\) by the current case assumption. Thus \((0,1)\) is available and gives the hard binary \([0,b,c]\); here \(c\ne0\) because \(A\) is invertible. So assume \(a\ne0\).

Write the two row eigenvectors as \((1,x),(1,y)\). Then

\[x\ne y,\qquad b=-axy,\qquad c=1+a(x+y),\] \[\lambda_x=1+ax\ne0,\qquad\lambda_y=1+ay\ne0.\]

The column paired with row \((1,x)\) is \((-y,1)^T\), and vice versa. The present absorption assumption is therefore \(x^3\ne1,y^3\ne1\). Both rows are available, and their binary contractions are

\[g_x=[\lambda_x,a(1-x^2y),x\lambda_x],\quad
g_y=[\lambda_y,a(1-xy^2),y\lambda_y].\]

The first entry is nonzero in both. Suppose, for contradiction, both binaries are tractable. Their possible types are rank one, diagonal, and the twelve affine points. We check all six unordered pairs.

\textbf{Two rank-one outputs.} Write their middle-to-first ratios as \(p,r\), so \(x=p^2,y=r^2\). Neither \(p\) nor \(r\) is zero, and \(p^2\ne r^2\). Their equations are

\[a(1-p^4r^2-p^3)=p,\qquad a(1-p^2r^4-r^3)=r.\]

Eliminating \(a\) gives \((p-r)(pr^2+1)(p^2r+1)=0\). The first factor is nonzero. The equation \(pr^2=-1\) forces \(a=p\) and \(\lambda_y=0\); the other factor forces \(\lambda_x=0\). Both are contradictions.

\textbf{Two diagonal outputs.} The equations \(x^2y=xy^2=1\) give \(x=y\), a contradiction.

\textbf{Diagonal and rank one.} Suppose \(g_x\) is diagonal. Then \(y=x^{-2}\), and the other middle-to-first ratio is \(\varepsilon/x\) for \(\varepsilon=\pm1\). The negative sign forces \(\lambda_x=0\). The positive sign gives

\[a=\frac{x^2}{x^3-2},\quad b=-\frac{x}{x^3-2},\quad
c=\frac{2x^3-1}{x^3-2}.\]

All displayed denominators are nonzero. Substitution yields

\[\det(AB-BA)=-\frac{4x^3(x^6-1)^2}{(x^3-2)^6}.\]

If this determinant is zero, then \(x^6=1\), whence \(y^3=x^{-6}=1\), contrary to absorption.

\textbf{Pairs involving an affine output.} Suppose \(g_x\) is affine. Set \(\sigma=a(1-x^2y)/\lambda_x\). Then \(\sigma^{12}=1\) and \(x=-\sigma^2\). Since \(x^3\ne1\), it follows that \(x^3=-1\) and \(\sigma^6=1\). An equality-preserving cube-root diagonal gauge therefore makes \(x=-1\) and \(\sigma=\varepsilon\in\{1,-1\}\). The other slope, now denoted \(t\), still satisfies \(t^3\ne1\) and \(t\ne-1\).

For \(\varepsilon=1\) the affine equation gives \(a=1/(2-t)\); for \(\varepsilon=-1\) it gives \(a=1/t\). Also \(b=at,c=1+a(t-1)\). Define

\[P_+(t)=t^3+t^2-t+1,\qquad P_-(t)=t^3-t^2+t+1.\]

The commutator determinants are respectively

\[-\frac{2(t-1)^2P_+(t)}{(t-2)^6},\qquad
-\frac{2(t-1)^2P_-(t)}{t^5}.\]

The prefactors are nonzero because \(t^3\ne1\) and the affine equations forbid the displayed zero denominators.

\begin{itemize}
\tightlist
\item
  If \(g_y\) is also affine, then \(t^3=-1\). But \(P_+(t)=t(t-1)\) and \(P_-(t)=t(1-t)\) at such a root, both nonzero. This contradicts the assumed zero commutator determinant.
\item
  If \(g_y\) is diagonal, its middle entry forces \(1+t^2=0\). Then \(P_+(t)=-2t\ne0\) and \(P_-(t)=2\ne0\), again a contradiction.
\item
  If \(g_y\) is rank one, its determinant forces respectively \(t^3+t^2+3t-1=0\) or \(t^3-3t^2-t-1=0\). In the first case subtraction from \(P_+=0\) gives \(t=1/2\), but \(P_+(1/2)=7/8\). In the second case subtraction gives \(t^2+t+1=0\), hence \(t^3=1\), forbidden by absorption.
\end{itemize}

Thus at least one available eigenrow contracts to a hard binary, completing this case. Only that one row is used by the hardness reduction.

\subsection{6.3 Exactly one column eigenline has zero Q}\label{exactly-one-column-eigenline-has-zero-q}

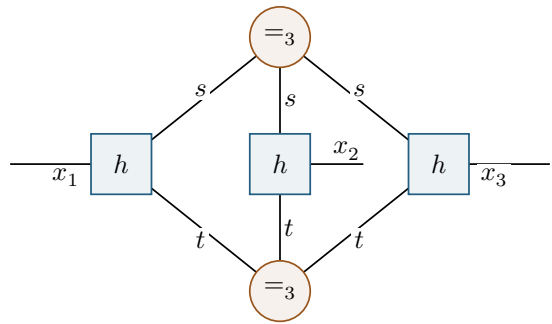
\begin{figure}[H]
\centering
\begin{tikzpicture}[x=1.05cm,y=1.05cm]
\coordinate (l0) at (-2,0);
\coordinate (l1) at (0,0);
\coordinate (l2) at (2,0);
\coordinate (r0) at (0,1.6);
\coordinate (r1) at (0,-1.6);
\coordinate (x0) at (-3.4,0);
\coordinate (x1) at (1.05,0);
\coordinate (x2) at (3.4,0);
\draw (x0) to[] node[midway,below,lab] {$x_1$} (l0);
\draw (l1) to[] node[pos=0.8,above,lab] {$x_2$} (x1);
\draw (l2) to[] node[midway,below,lab] {$x_3$} (x2);
\draw (l0) to[] node[midway,above,lab] {$s$} (r0);
\draw (l0) to[] node[midway,below,lab] {$t$} (r1);
\draw (l1) to[] node[midway,right,lab] {$s$} (r0);
\draw (l1) to[] node[midway,right,lab] {$t$} (r1);
\draw (l2) to[] node[midway,above,lab] {$s$} (r0);
\draw (l2) to[] node[midway,below,lab] {$t$} (r1);
\node[L] at (l0) {$h$};
\node[L] at (l1) {$h$};
\node[L] at (l2) {$h$};
\node[R] at (r0) {$=_3$};
\node[R] at (r1) {$=_3$};
\end{tikzpicture}
\caption{The ternary gadget $\Phi(h)$: three left $h$ vertices and two right equality vertices. All three edges at the upper circle carry $s$; all three at the lower circle carry $t$. The three dangling inputs are $x_1,x_2,x_3$. The central dangling stub ends inside the drawing and does not touch the right square.}
\label{fig:gadget_phi-10}
\end{figure}

A cube-root diagonal gauge sends the forbidden column to \((1,-1)^T\). The eigenrow paired with the other column is then \((1,1)\), and the signature parameters satisfy

\[c=1+a-b.\]

The two eigenvalues are \(1+a\) and \(1-b\), both nonzero and distinct. The row \((1,1)\) is available because its paired column is the other, absorbable eigenline. Its binary contraction is

\[g=[1+a,a+b,1+a].\]

The first entry is nonzero; \(a+b\ne0\) because the eigenvalues are distinct. The complex binary classification says \(g\) is hard except for the following cases:

\[b=1,\quad b=-1-2a,\quad
b=-a+i(1+a),\quad b=-a-i(1+a).\]

The case \(b=1\) makes \(A\) singular and is excluded. If \(b=-1-2a\), the Hadamard basis gives

\[fH^{\otimes3}=4[0,0,1+a,-1-3a],\qquad
(H^{-1})^{\otimes3}e=\tfrac14[1,0,1,0].\]

Counting 1-edges on each side forces every vertex to have exactly two incident 1-edges. Complementation identifies these assignments with perfect matchings of the original cubic bipartite graph. Therefore

\[Z_f(G)=(1+a)^{|L|}\#\operatorname{PM}(G),\]

and \(1+a\ne0\) transfers \#P-hardness. This degree-counting argument is valid over complex parameters too.

We now handle the two genuinely complex affine lines. Conjugating all fixed weights conjugates the answer in the fixed number field, so it suffices to use

\[b=-a+i(1+a),\qquad c=1+a-b.\]

Here \(a\ne-1\) and the eigenvalue ratio of \(A\) is \(1/(1-i)\) or its reciprocal, so it satisfies the required interpolation condition. Put

\[a_0=\frac{-1+2i}{5},\qquad
R(a)=5a^2+5(1-i)a+1-2i.\]

Direct factorization gives

\[\det(AB-BA)
=-\frac{(1+i)(2-i)^3}{5}(a+1)^2(a-a_0)^2R(a).\]

Our standing assumption that this determinant vanishes therefore leaves \(a=a_0\) or \(R(a)=0\).

At a root of \(R\), use the actual \(D\) gadget. Polynomial remainders modulo \(R\) are

\[\det(AD-DA)\equiv\frac{704-128i}{25}
\left(a+\frac25-\frac{i}{20}\right),\]

\[\operatorname{tr}D\equiv\frac{-6-8i}{5}
\left(a+\frac25-\frac{i}{5}\right).\]

Neither expression vanishes at a root of \(R\): evaluating \(R\) at the respective possible zeros gives \(3/80+i/20\) and \(3/5+i/5\), both nonzero. Proposition 7 therefore applies with \(H=A,T=D\) and proves hardness.

At \(a=a_0\), we have \(b=a,c=1\). The row \((1,1)\) is already available. The actual ternary gadget \(\Phi(f)\) is a nonzero scalar multiple of

\[h=\left[1,\frac{-3+4i}{25},\frac{-3+4i}{25},1\right].\]

Attach that same unary to \(h\). The resulting binary is proportional to

\[[11+2i,-3+4i,11+2i].\]

All entries are nonzero. The expressions for rank one and for the affine cross-relation are respectively \(124+68i\) and \(110+20i\), both nonzero. Thus this binary is hard. The ternary gadget and the one already simulated unary transfer hardness to the original \(f\). Complex conjugation settles the other affine line.

\subsection{6.4 Both column eigenlines have zero Q}\label{both-column-eigenlines-have-zero-q}

The dual row eigenvectors have slopes that are two distinct cube roots of unity. A cube-root gauge and exchange of the two eigenvalue labels put them at \(1,\omega\), where \(\omega^2+\omega+1=0\). Thus

\[b=-a\omega,\qquad c=1+a(1+\omega),\]

with \(a\ne0,-1,-\omega^2\) by distinctness and invertibility.

Let

\[S=\begin{pmatrix}-\omega&-1\\1&1\end{pmatrix}.\]

Then \(S^{-1}AS=\operatorname{diag}(1+a,1+a\omega)\). The two off-diagonal entries of \(S^{-1}BS\) vanish respectively at

\[a=\frac{1}{1-2\omega},\qquad
a=\frac{\omega^2}{1-2\omega^2},\]

apart from the excluded \(a=0\). Thus the zero-commutator-determinant assumption leaves exactly these two values. Exchanging eigenvalue labels and applying a cube-root gauge interchanges them while replacing \(\omega\) by \(\omega^2\). It suffices to handle the first.

For \(a=1/(1-2\omega)\), apply the holographic basis with left transformation \(S^{-1}\) and right transformation \(S^T\) in column-tensor convention. The transformed signatures are

\[f'=\frac{1}{(1-2\omega)(1-\omega)^3}
[6(1-\omega),-3(\omega+2),0,0],\]

\[e'=[0,1-\omega^2,1-\omega,0].\]

Every nonzero left term has at most one incident 1-edge; every nonzero right term has at least one. There are equally many vertices on the two sides. Consequently every vertex has exactly one incident 1-edge, and these edges form a perfect matching. If \(v\) is the weight-one entry of \(f'\) and \(p=1-\omega^2\) the weight-one entry of \(e'\), then

\[Z_f(G)=(vp)^{|L|}\#\operatorname{PM}(G).\]

The scalar \(vp\) is nonzero. This proves hardness at both exceptional parameters.

All three absorption cases are exhausted, completing Theorem 9.
\end{proof}

\section{7. Projective group orbits and large noncommuting orders}\label{projective-group-orbits-and-large-noncommuting-orders}

We now turn to cases in which the powers of \(A\) repeat projectively. Products of several actual matrices can nevertheless have infinite projective order. Section 7.2 proves the group criterion stated in Section 3.6, including the passage from group orbits to legal positive-word gadgets and one-type unary simulation. Section 7.3 applies it, or a finite-orbit variant for singular \(B\), to orders \(n\ge6\) with \(\det(AB-BA)\ne0\). Section 9 will complete those orders when this determinant vanishes.

\subsection{7.1 Two classical group facts}\label{two-classical-group-facts}

For an invertible matrix \(M\), write \([M]\) for its class in \(\operatorname{PGL}_2(\mathbb C)=\operatorname{GL}_2(\mathbb C)/\mathbb C^{\times}\). This class acts on column lines by \([u]\mapsto[Mu]\). A group has an invariant point if every element preserves one common line. It has an invariant pair if every element preserves a fixed two-element set of lines, possibly interchanging the two. These are different conditions: a diagonal group fixes two points separately, while a dihedral group can interchange them. All group arguments below concern these explicit actions.

Over \(\mathbb C\), the finite subgroups of \(\operatorname{PGL}_2\) are cyclic, dihedral, \(A_4\), \(S_4\), and \(A_5\). In particular, a finite subgroup containing an element of order at least six is cyclic or dihedral. The cyclic subgroup of rotations in a dihedral group fixes an unordered pair of projective points, and every element outside it interchanges those points. This is the characteristic-zero algebraically closed case of Beauville \cite{ref16}.

A finitely generated torsion subgroup of \(\operatorname{GL}_d(\mathbb C)\) is finite. One may deduce this from Selberg's lemma: it has a torsion-free subgroup of finite index, which for a torsion group is trivial. See Serre \cite[Section 1.2]{ref17} for Selberg's lemma and its characteristic-zero setting. It applies to \(\operatorname{PGL}_2(\mathbb C)\) because conjugation on the three-dimensional space of trace-zero two-by-two matrices gives a faithful linear representation: its kernel in \(\operatorname{GL}_2\) consists exactly of scalar matrices.

\subsection{7.2 A group lemma with explicit absorption}\label{a-group-lemma-with-explicit-absorption}

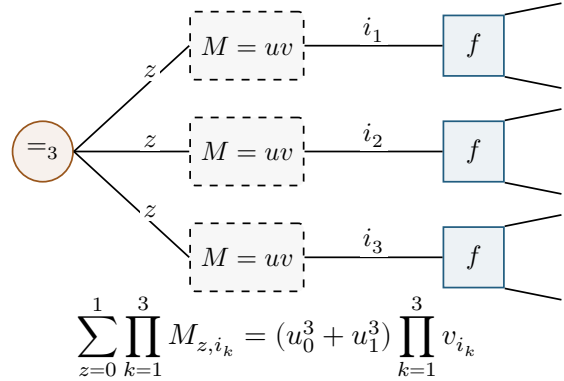
\begin{figure}[H]
\centering

\begin{tikzpicture}[x=1cm,y=1cm]
\node[R] (e) at (-2.7,0) {$=_3$};
\foreach \k/\yy in {1/1.4,2/0,3/-1.4}{
 \node[interpolated] (m\k) at (0,\yy) {$M=uv$};
 \node[L] (f\k) at (3,\yy) {$f$};
 \draw (e.east)--node[pos=0.66,above,lab] {$z$}(m\k.west);
 \draw (m\k.east)--node[above,lab] {$i_{\k}$}(f\k.west);
 \draw (f\k.north east)--++(0.75,0.15);
 \draw (f\k.south east)--++(0.75,-0.15);
}
\node[align=center] at (0.4,-2.45) {$\displaystyle\sum_{z=0}^1\prod_{k=1}^3 M_{z,i_k}
 =(u_0^3+u_1^3)\prod_{k=1}^3 v_{i_k}$};
\end{tikzpicture}

\caption{Absorbing three leftover columns. The three right-hand squares represent attachment sites in the rest of the instance. One new equality vertex contracts the L ports of three copies of $M$. This reproduces three RHS unaries, multiplied by $Q(u)$. All copies of the dashed matrix are eliminated together by the assumed reduction for networks containing $M$. For a spectral projector, this reduction is the chain interpolation already proved.}
\label{fig:absorption-11}
\end{figure}

\begin{holantstatement}{Lemma 10}

Fix an algebraic complex symmetric ternary \(f=[1,a,b,c]\) that is neither rank one nor generalized equality. Let \(M_1,\ldots,M_k\) be invertible actual straddled gadget matrices for \(f\mid e\). Suppose the group generated by their projective classes is infinite and preserves neither a single projective point nor an unordered pair of projective points. Then \(\operatorname{Holant}(f\mid e)\) is \#P-hard.

\end{holantstatement}

\begin{proof}[Proof]

First, some positive word in the \(M_i\) has infinite projective order. If a generator already has infinite order, use it. Otherwise every generator has finite projective order, and its inverse projective class is a positive power of itself. Hence the positive monoid of projective classes equals the generated group. If every positive word had finite order, the finitely generated torsion theorem would make the whole group finite, a contradiction. Choose one such actual positive word \(H\) once and for all.

An invertible two-by-two matrix of infinite projective order is either diagonalizable with a non-root-of-unity eigenvalue ratio, or has a nontrivial Jordan block. In the first case ordinary spectral interpolation recovers either rank-one eigenprojector. In the second case the leading coefficient of \(H^m\) recovers its rank-one nilpotent part, by Lemma P.2(2). Thus a fixed nonzero rank-one straddled signature

\[M=uv\]

is simulable by polynomial-time Turing reduction. Here simulation means recovery of a network containing any number of copies of that same matrix, not a claim that \(M\) is an actual graph.

Every projective column orbit of the group is infinite. An orbit of size one or two is excluded by hypothesis. If an orbit were finite of size at least three, the action on that finite set would be faithful: a Möbius transformation fixing three distinct points is the identity. The group would then embed into a finite symmetric group, contradicting infinitude. Row orbits are also infinite, because a row line corresponds to its one-dimensional column kernel, and the row action corresponds to inverse action on that kernel.

The positive-word orbit of every row and column is infinite as well. If a positive orbit were finite, each invertible generator would map that finite set injectively into itself and hence bijectively onto itself. Its inverse would therefore also preserve the set, making it a finite group orbit, which has just been excluded.

The cubic \(Q(u)=u_0^3+u_1^3\) has only three projective zeros. Choose an actual positive word \(W\) for which \(Q(Wu)\ne0\). The signature \(WM=(Wu)v\) is simulable by attaching the actual \(W\) gadget before every interpolated box. Absorption groups leftover columns in triples and supplies the single right unary row \(v\).

Now choose seven actual positive words \(V_1,\ldots,V_7\) such that the rows \(vV_j\) have distinct projective directions. This is possible because their positive orbit is infinite. They are individually simulable by attaching the relevant word to copies of the one available row \(v\). Lemma 4 above gives at most six tractable binary contractions for a nondegenerate, non-equality ternary signature. Choose one row outside this set and use only that row in the final binary reduction. If \(A_f\) is invertible, six words already suffice because the bound is five.

All choices of matrices and words depend only on the fixed signature. Interpolation uses polynomially many queries in a fixed algebraic number field. At no step are several unrelated unary types assumed to be jointly available.
\end{proof}

\subsection{7.3 Every order at least six in the noncommuting branch}\label{every-order-at-least-six-in-the-noncommuting-branch}

\begin{holantstatement}{Theorem 11}

Let \(f=[1,a,b,c]\) be fixed algebraic complex. Let

\[A=\begin{pmatrix}1&b\\a&c\end{pmatrix},\qquad
B=\begin{pmatrix}1+ab&a^2+bc\\a+b^2&ab+c^2\end{pmatrix}.\]

Suppose \(A\) is invertible and its two eigenvalues have ratio of finite multiplicative order \(n\ge6\). If

\[\det(AB-BA)\ne0,\]

then \(\operatorname{Holant}(f\mid e)\) is \#P-hard.

\end{holantstatement}

\begin{proof}[Proof]

Suppose first that \(B\) is singular. The commutator condition implies \(B\ne0\), so \(B=uv\) has rank one. In an \(A\) eigenbasis, write \(u=(u_1,u_2)^T\) and \(v=(v_1,v_2)\). The two off-diagonal entries of \(B\) are \(u_1v_2,u_2v_1\), and their product is nonzero. Thus all four factor coordinates are nonzero. Neither \(u\) nor \(v\) is an \(A\) eigenvector.

The columns \(A^j u\) and rows \(vA^j\) each have exactly \(n\) distinct projective directions for \(j=0,\ldots,n-1\). At least one column has nonzero \(Q\) because \(n>3\), so the actual rank-one gadget \(A^jB\) supplies \(v\) by absorption. At least one of the \(n\ge6\) rows has a hard binary contraction, because \(A_f=A\) is invertible and there are at most five tractable directions. Its actual \(A\) chain and the established single-unary simulation transfer that hardness.

\textbf{The case of invertible \(B\).} Let \(G=\langle[A],[B]\rangle\le\operatorname{PGL}_2(\mathbb C)\). The identity

\[2\operatorname{tr}B=\Delta_A+(\operatorname{tr}A)^2\]

implies \(\operatorname{tr}B\ne0\): zero trace would force \(A\)'s projective order to be four, contradicting \(n\ge6\).

The group \(G\) has no common invariant point because \(\det(AB-BA)\ne0\). To see this directly, diagonalize \(A\); a common invariant line would make one of the off-diagonal entries of \(B\) zero, and hence the commutator determinant zero.

It also preserves no unordered pair. Since \(n>2\), if \(A\) preserves a two-point set it fixes both points separately: a two-by-two matrix swapping two lines is off-diagonal in that basis and has projective order two. Thus the pair must be the two eigenlines of \(A\). If \(B\) preserves the pair, it is either diagonal or off-diagonal in the same basis. The diagonal alternative contradicts the nonzero commutator determinant; the off-diagonal alternative has trace zero, also excluded.

Finally \(G\) is infinite. If finite, the classical classification and \(n\ge6\) force it to be cyclic or dihedral. Such a group preserves a pair of projective points, which was just ruled out. Thus Lemma 10 applies. The hypotheses of Lemma 10 on \(f\) also hold: an invertible \(A_f\) excludes rank one, and generalized equality would make \(A\) and \(B\) diagonal. This proves hardness.
\end{proof}

\section{8. The reversal-symmetric and anti-reversal families}\label{the-reversal-symmetric-and-anti-reversal-families}

These special families arise when the actual matrices preserve common directions. The results in this section are used by the common-line analysis in Section 9 and by the low-order cases; they are not another branch of the spectral partition. The recurring construction uses the symmetric rows \((1,1)\) and \((1,-1)\). If the initial recurrence has root-of-unity ratio, a ternary replacement supplies a new recurrence or a previously classified signature.

\begin{holantstatement}{Theorem 12}

For every fixed algebraic complex \(a\), the cubic bipartite problem

\[\operatorname{Holant}([1,a,-a,-1]\mid=_3)\]

is in FP for \(a\in\{0,1,-1\}\) and is \#P-hard otherwise. The same classification holds after any equality-preserving cube-root diagonal gauge. Equivalently, this classifies the entire family \(f=[1,a,b,-1]\) with \(a^3+b^3=0\).

\end{holantstatement}

The equality in the last sentence includes \(a=b=0\). Otherwise choose \(\omega=-a/b\), for which \(\omega^3=1\), so the transformed entries obey \(b'=-a'\). The three tractable parameter values are generalized equality, the affine signature \([1,1,-1,-1]\), and rank one \([1,-1,1,-1]\).

\begin{proof}[Proof of Theorem 12]

Proposition P.1 supplies the algorithms for \(a\in\{0,1,-1\}\). The hardness proof occupies Sections 8.1--8.2; those three values are excluded whenever a denominator requires it.

\subsection{8.1 Two available rows when the B recurrence works}\label{two-available-rows-when-the-b-recurrence-works}

Put \(f_a=[1,a,-a,-1]\). Its first gadget is

\[A=\begin{pmatrix}1&-a\\a&-1\end{pmatrix},\qquad A^2=(1-a^2)I.\]

For \(a\ne\pm1\), its projective order is two. Thus the infinite-order-first-gadget theorem is not applicable. The actual second gadget is

\[B=\begin{pmatrix}1-a^2&a(1+a)\\a(1+a)&1-a^2\end{pmatrix}.\]

Its eigenvectors are \((1,1)^T,(1,-1)^T\), with eigenvalues

\[\lambda_+=1+a,\qquad\lambda_-=(1+a)(1-2a).\]

Assume \(a\ne-1\), \(a\ne1/2\), and \(z=1-2a\) is not a root of unity. Both eigenvalues are nonzero, and spectral interpolation supplies the plus projector

\[P_+=\tfrac12\begin{pmatrix}1&1\\1&1\end{pmatrix}.\]

Its column has nonzero cubic absorption factor, so the right unary row \((1,1)\) is simulable. Attaching the actual \(A\) gives

\[(1,1)A=(1+a)(1,-1),\]

so the row \((1,-1)\) is simulable too, one type at a time. Its binary contraction is

\[F_{f_a}(1,-1)=[1-a,2a,1-a].\]

The complex binary classification makes this hard except at

\[a\in\{0,1,-1,1/3,(1+2i)/5,(1-2i)/5\}.\]

Indeed, zero middle entry gives \(a=0\); two zero endpoints give \(a=1\); rank one requires \((1-a)^2=4a^2\), giving \(a=-1\) or \(a=1/3\). The last affine class requires \((1-a)^2=-4a^2\), giving the two complex values. Their endpoint-to-middle ratio is then \(\pm i\), which satisfies the twelfth-power condition.

The three already declared tractable values are set aside. At \(a=1/3\), the actual ternary gadget \(\Phi(f_a)\) is a nonzero scalar multiple of \(f_{1/7}\). Attach the \textbf{same already available} minus row; the resulting binary is proportional to \([3,1,3]\), which is hard.

At \(a=(1+2i)/5\), the same ternary construction is a nonzero multiple of \(f_{(7+4i)/13}\). Its minus contraction is proportional to

\[[3-2i,7+4i,3-2i].\]

The determinant and affine cross-expression are \(-28-68i\) and \(38+44i\), respectively, both nonzero; all three entries are nonzero. Thus it is hard. Conjugation handles \(a=(1-2i)/5\). No new unary simulation is needed for these substitutions.

The omitted singular-B value \(a=1/2\) is also hard. There \(B=(3/4)(1,1)^T(1,1)\) is an actual rank-one matrix, so absorption directly supplies the plus row; attaching \(A\) supplies the minus row. Its binary \([1/2,1,1/2]\) is hard.

We have therefore proved the theorem whenever \(z\) is zero or is not a root of unity, with the listed tractable values handled separately.

\subsection{8.2 Root-of-unity recurrences escape after one ternary construction}\label{root-of-unity-recurrences-escape-after-one-ternary-construction}

Now suppose \(z=1-2a\) is a root of unity. The values \(z=1,-1\) are respectively \(a=0,1\), already tractable. Assume \(z\ne\pm1\).

The actual ternary construction has the form

\[\Phi(f_a)=[w,x,-x,-w],\quad
w=1+a^3,\quad x=a-a^2-2a^3.\]

Here \(w\ne0\). If \(a^3=-1\), then \(a=-1\) or \(a=(1\pm i\sqrt3)/2\), giving \(z=3\) or \(z=\mp i\sqrt3\), none of modulus one. After dividing by \(w\), the new parameter is

\[a'=\frac{a(1-2a)}{1-a+a^2},\qquad
z'=1-2a'=\frac{5z^2-4z+3}{z^2+3}.\]

The denominator is nonzero on \(|z|=1\). On this circle complex conjugation sends \(z\) to \(z^{-1}\), so direct multiplication gives

\[|5z^2-4z+3|^2-|z^2+3|^2
=\frac{4(z-1)^2(3z^2-2z+3)}{z^2}.\]

Since \(z\ne1\), equality of the two moduli would force

\[z+z^{-1}=2/3.\]

But a root of unity and its inverse are algebraic integers. Their rational sum must therefore be an integer; it cannot equal \(2/3\). Consequently \(|z'|\ne1\). Thus \(z'\) is zero or is not a root of unity, precisely the case already proved in Section 8.1.

To apply that result, check the three tractable outputs cannot occur. The equation \(a'=0\) gives \(z(z-1)=0\); the equation \(a'=-1\) gives \((z-3)(z+1)=0\); and the equation \(a'=1\) gives \(3z^2-2z+3=0\). All are excluded by \(|z|=1\), \(z\ne\pm1\), and the algebraic-integer argument. Hence \(f_{a'}\) is hard.

The actual gadget \(\Phi(f_a)=w f_{a'}\), with \(w\ne0\), reduces its hard problem to the original one. This proves hardness in every remaining root-of-unity case and completes Theorem 12.
\end{proof}

The rational case belongs to the dichotomy of Cai--Fan--Liu \cite{ref5}. The rational map \(z\mapsto(5z^2-4z+3)/(z^2+3)\) proves the required escape simultaneously for roots of unity of every order; the binary reductions use the complex classification \cite{ref7}.

\subsection{8.3 The reversal-symmetric companion family}\label{the-reversal-symmetric-companion-family}

\begin{holantstatement}{Theorem 13}

For every fixed algebraic complex \(a\), the problem \(\operatorname{Holant}([1,a,a,1]\mid=_3)\) is in FP at \(a\in\{0,1,-1\}\) and is \#P-hard otherwise. This also holds for its equality-preserving cube-root gauges.

\end{holantstatement}

\begin{proof}[Proof]

The three listed signatures are generalized equality, rank one, and affine. Exclude them. Here

\[A=\begin{pmatrix}1&a\\a&1\end{pmatrix},\quad
B=\begin{pmatrix}1+a^2&a+a^2\\a+a^2&1+a^2\end{pmatrix}.\]

If the eigenvalue ratio \((1+a)/(1-a)\) is not a root of unity, Theorem 9 applies. Suppose instead that it is a root of unity. Its modulus is one, which gives \(|1+a|=|1-a|\) and hence \(\operatorname{Re}a=0\). Write \(a=it\) with \(t\in\mathbb R\smallsetminus\{0\}\).

The eigenvalues of \(B\), with plus and minus eigenvectors respectively, are

\[\mu_+=1+a+2a^2,\qquad\mu_-=1-a.\]

They are nonzero for such \(a\). Their squared moduli differ by

\[|\mu_+|^2-|\mu_-|^2=4t^2(t^2-1).\]

If \(t\ne\pm1\), the ratio is not a root of unity, so \(B\) interpolates the plus projector and its absorbable column gives the right unary row \((1,1)\). The binary contraction \([1+a,2a,1+a]\) is hard unless

\[a\in\{0,1,-1,-1/3,(-1+2i)/5,(-1-2i)/5\}.\]

None of these is nonzero purely imaginary. Thus this case is hard.

If \(a=i\), the ternary gadget \(\Phi([1,i,i,1])\) is a nonzero multiple of

\[\left[1,\frac{1-2i}{5},\frac{1-2i}{5},1\right].\]

Its first gadget has distinct nonzero eigenvalues with unequal moduli, because the new parameter has nonzero real part and is not \(\pm1\). Theorem 9 proves its hardness, which transfers through \(\Phi\). Conjugation handles \(a=-i\). This exhausts all cases.
\end{proof}

The two symmetry families are useful because their small matrices may generate only a dihedral or commuting projective group. The direct unary contractions and the explicit action of \(\Phi\) settle them without assuming that such a group must be infinite.

\section{9. Common invariant lines and completion of all orders at least six}\label{common-invariant-lines-and-completion-of-all-orders-at-least-six}

Section 7 settled orders at least six when the commutator determinant is nonzero. Here we treat its complement. A common invariant line of \(A,B,D\) gives parameter equations handled using Section 8; otherwise \(D\) moves the line preserved by \(A,B\). An invertible \(D\) can enlarge the generated group enough for Lemma 10, while a singular nonzero \(D\) supplies an actual rank-one matrix. Section 9.2 also handles the situation in which both eigenrows are individually available. These alternatives are assembled in Section 9.3 to cover every remaining order \(n\ge6\).

\subsection{9.1 Common invariant lines of the three actual gadgets}\label{common-invariant-lines-of-the-three-actual-gadgets}

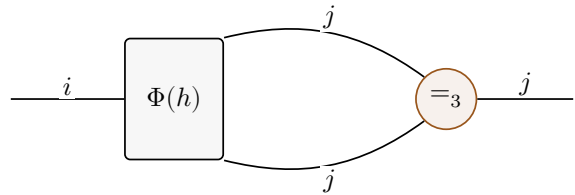
\begin{figure}[H]
\centering

\begin{tikzpicture}[x=1.2cm,y=1cm]
\node[boxg,minimum height=16mm] (p) at (0,0) {$\Phi(h)$};
\node[R] (e) at (3,0) {$=_3$};
\draw (-1.8,0) -- node[above,lab] {$i$} (p.west);
\draw (p.north east) to[bend left=25] node[above,lab] {$j$} (e.north west);
\draw (p.south east) to[bend right=25] node[below,lab] {$j$} (e.south west);
\draw (e.east) -- node[above,lab] {$j$} (4.4,0);
\end{tikzpicture}

\caption{$D_h=A_{\Phi(h)}$. The box abbreviates the ternary gadget with three left $h$ vertices and two shared right equality vertices; it is not a new freely available signature. Attach its second and third ports to one new equality vertex. Expanding the box gives three original left vertices and three right vertices.}
\label{fig:gadget_d-12}
\end{figure}

\begin{holantstatement}{Lemma 14}

Suppose \(A_f\) is invertible, \(f=[1,a,b,c]\) is not generalized equality, and the actual matrices \(A,B,D\) have a common invariant projective column line. Then, after an equality-preserving cube-root diagonal gauge, \(f\) belongs to one of the two families

\[[1,t,t,1],\qquad [1,t,-1-2t,2+3t].\]

Both are already \#P-hard under the invertibility and non-equality assumptions: the first by Theorem 13, the second by the direct weighted-perfect-matching identity of Theorem 9.

\end{holantstatement}

\begin{proof}[Proof]

The annihilator of a common column line is a common row eigendirection, including eigenvalue zero for a singular matrix. A common row \((1,0)\) forces \(b=0\) from \(A\) and then \(a=0\) from \(B\), which is generalized equality. A common row \((0,1)\) similarly forces \(a=b=0\). Thus write the common row as \((1,x)\) with \(x\ne0\).

The \(A\) eigenrow equation gives

\[c=1+ax-b/x,\qquad 1+ax\ne0.\]

The row-\(B\) equation is

\[R_B=a^2x^3+a^2-abx+ax^2-b^2x^2-b=0.\]

Let \(R_D=x((1,x)D)_1-x^2((1,x)D)_0\), with coordinates indexed \(0,1\). Polynomial division in \(b\) gives the following remainder of \(R_D\) modulo \(R_B\):

\[R_D\equiv\frac{x^3-1}{x^2}R_{\mathrm{line}},\qquad
R_{\mathrm{line}}=-(a^2x^2+1)b-a^3x+a^2x^3+a^2+ax^2.\]

If \(x^3=1\) we keep that case. Otherwise \(R_{\mathrm{line}}=0\). If \(a=0\), this equation forces \(b=0\), the removed generalized equality case. Thus assume \(a\ne0\). Put

\[L=a^2x^2+1,\qquad U=-a^3x+a^2x^3+a^2+ax^2.\]

If \(L=0\), then \(ax=\pm i\) and

\[U=(\pm i-1)(x^3+1)/x^2.\]

The equation \(R_{\mathrm{line}}=0\) now gives \(x^3=-1\). If \(L\ne0\), substitute \(b=U/L\) into \(R_B\); direct expansion and cancellation give

\[L^2R_B(U/L)=-a^2x^2(x-a^2)(x^3+1)(ax+1)^2.\]

Every prefactor except \((x-a^2)(x^3+1)\) is nonzero. Thus \(x^3=-1\) or \(x=a^2\). In the latter case \(R_{\mathrm{line}}=(a^6+1)(a^2-b)\). If \(a^6=-1\), again \(x^3=-1\). Otherwise \(b=a^2\), and the \(A\) row equation gives \(c=a^3\), making \(A\) singular, a contradiction.

We have proved \(x^3=1\) or \(x^3=-1\). A cube-root diagonal gauge makes \(x=1\) or \(x=-1\), respectively. At \(x=1\), the equations become

\[c=1+a-b,\qquad R_B=(a-b)(2a+b+1)=0.\]

They give exactly the two displayed families. At \(x=-1\) they become

\[c=1-a+b,\qquad R_B=(a-b)(b+1)=0.\]

The choice \(a=b\) gives the first family; \(b=-1\) gives \(c=-a\), making \(A\) singular. Thus no other case remains.
\end{proof}

We will also use a simpler commutation observation. Under invertible \(A\) and non-generalized-equality \(f\), the equation \(AB=BA\) forces \(c=1,b^3=a^3\), which is a cube-root gauge of the first family above. Indeed the off-diagonal commutator entries are

\[(AB-BA)_{01}=(1-c)(a^2-b),\qquad
(AB-BA)_{10}=(c-1)(b^2-ac).\]

If \(c\ne1\), they force \(b=a^2\) and either \(a=b=0\) or \(c=a^3\), both excluded. If \(c=1\), a diagonal commutator entry is \(b^3-a^3\).

\subsection{9.2 Two eigenrows suffice in the finite-order common-line branch}\label{two-eigenrows-suffice-in-the-finite-order-common-line-branch}

\begin{holantstatement}{Lemma 15}

Let \(f=[1,a,b,c]\), \(A=A_f\), and \(B=B_f\). Suppose \(A\) is invertible with distinct eigenvalues of finite projective order, \(\det(AB-BA)=0\), and \(f\) is not generalized equality. If the two row eigendirections of \(A\) are individually simulable, then the problem is \#P-hard.

\end{holantstatement}

\begin{proof}[Proof]

If \(a=0\), then \(b\ne0\) and the row \((0,1)\) gives the hard binary \([0,b,c]\). Assume \(a\ne0\). Parameterize the two rows as \((1,x),(1,y)\), with \(b=-axy,c=1+a(x+y)\) and eigenvalues \(\lambda_x=1+ax,\lambda_y=1+ay\). Their binary contractions are

\[g_x=[1+ax,a(1-x^2y),x(1+ax)],\qquad g_y=[1+ay,a(1-xy^2),y(1+ay)].\] This time no absorption restriction \(x^3,y^3\ne1\) is assumed: the two row simulations are a hypothesis.

Suppose both binaries are tractable. Two rank-one outputs and two diagonal outputs are excluded by exactly the algebraic equations in Theorem 9, without any absorption assumption. A diagonal/rank-one pair, after excluding a zero eigenvalue, gives

\[y=x^{-2},\quad a=x^2/(x^3-2),\quad
\lambda_x=2(x^3-1)/(x^3-2),\quad
\lambda_y=(x^3-1)/(x^3-2).\]

Their ratio is 2, contradicting finite projective order.

If an affine output has slope \(x\), then \(x^3=\pm1\). When \(x^3=1\), gauge it to \(x=1\). Its middle-to-first ratio is \(\pm i\), so

\[c=1+a-b,\quad b=-a\pm i(1+a).\]

The \(A\) eigenvalue ratio is \(1/(1\mp i)\) or its reciprocal, again not a root of unity. Thus an affine slope must satisfy \(x^3=-1\), and a gauge makes \(x=-1\), with middle-to-first ratio \(\varepsilon=\pm1\).

Write the other slope as \(t\). The affine equations give \(a=1/(2-t)\) for \(\varepsilon=1\) and \(a=1/t\) for \(\varepsilon=-1\). Here \(t\ne1\) because that would make \(a=1\) and \(\lambda_x=0\). For these two signs the vanishing commutator determinant gives respectively

\[P_+(t)=t^3+t^2-t+1=0,\qquad P_-(t)=t^3-t^2+t+1=0.\]

The omitted factors are respectively \(-2(t-1)^2/(t-2)^6\) and \(-2(t-1)^2/t^5\), both nonzero under the present assumptions.

If the other output is diagonal, then \(t^2=-1\) and the commutator polynomial is nonzero. If it is affine, \(t^3=1\) was already excluded by finite order, while \(t^3=-1\) also makes the commutator polynomial nonzero. If it is rank one, the positive-sign case gives the same impossible \(t=1/2\) substitution as in Theorem 9. The negative-sign case forces \(t^2+t+1=0\), but then

\[\lambda_x=1-t^{-1},\quad\lambda_y=2,\qquad
|\lambda_x/\lambda_y|=\sqrt3/2\ne1.\]

Every pairing is impossible. At least one of the two individually available rows therefore has a hard binary contraction, and using only that row proves hardness.
\end{proof}

\subsection{9.3 All finite projective orders at least six}\label{all-finite-projective-orders-at-least-six}

\begin{holantstatement}{Theorem 16}

Let \(f=[1,a,b,c]\) be fixed and algebraic complex. If the first actual gadget \(A_f\) has distinct nonzero eigenvalues whose ratio has finite order \(n\ge6\), then \(\operatorname{Holant}(f\mid e)\) is in FP for generalized equality and is \#P-hard otherwise.

\end{holantstatement}

\begin{proof}[Proof]

Generalized equality is tractable. Exclude it, and remove the already hard cases of Lemma 14 and of \(AB=BA\). Theorem 11 handles \(\det(AB-BA)\ne0\). Thus assume the commutator is nonzero but has determinant zero. Throughout, \(\operatorname{tr}B\ne0\) because zero trace would force order four.

\textbf{Case 1: B is invertible.} In an \(A\) eigenbasis, \(B\) is triangular with nonzero diagonal entries and exactly one nonzero off-diagonal entry. The group \(G_0=\langle[A],[B]\rangle\) is infinite: its group commutator is a nontrivial upper- or lower-unitriangular matrix. It has a unique common invariant column line \(L\). It preserves no unordered pair, because a pair preserved by \(A\) must be its eigenlines, while this \(B\) is neither diagonal nor off-diagonal.

If \(D\) preserves \(L\), Lemma 14 applies, one of the removed hard cases. Thus \(D(L)\not\subseteq L\).

If \(D\) is invertible, the group generated by \(A,B,D\) is infinite, has no common invariant point, and has no invariant pair because its subgroup \(G_0\) already has none. Lemma 10 proves hardness.

If \(D\) is singular, it is nonzero and factors as \(D=uv\). The condition \(D(L)\not\subseteq L\) says \(u\notin L\) and \(v\) does not annihilate \(L\). Both its column orbit and its row orbit under \(G_0\) are infinite: the only one-point column orbit is \(L\), the only one-point row orbit is its annihilator, there are no two-point orbits, and a larger finite orbit would make \(G_0\) finite. Positive-word orbits have the same finiteness property. Choose an actual word in \(A,B\) moving \(u\) off the three absorption-zero lines, absorb to obtain \(v\), and choose six positive-word row images. Since \(A_f\) is invertible, at most five give tractable binaries. One hard binary proves hardness. The starting rank-one matrix here is the actual \(D\), so no spectral extraction is needed.

\textbf{Case 2: B is singular.} In the same eigenbasis it is triangular, nonzero, has nonzero trace, and does not commute with \(A\). Its rank-one factorization \(B=uv\) therefore has exactly one eigenfactor: either \(u\) is a column eigenvector and \(v\) is not a row eigenvector, or \(v\) is a row eigenvector and \(u\) is not a column eigenvector. This follows directly from a triangular rank-one matrix with one off-diagonal entry nonzero and exactly one diagonal entry nonzero.

In the first alternative, \(vA^j\) has \(n\ge6\) directions. If \(Q(u)\ne0\), absorb \(B\) and finish by the five-direction bound. If \(Q(u)=0\), use \(D\) to move this eigenline. If \(D\) preserved it, \(A,B,D\) would have a common invariant line, already removed. Thus \(Du\) is nonzero and not on the original eigenline. If it is not an \(A\) eigenvector, its \(A\) orbit has \(n\ge6>3\) directions and one of \(A^jDB\) is absorbable. If it is the other eigenvector and that column has nonzero \(Q\), \(DB\) is already absorbable.

The only remaining possibility would be that both \(A\) column eigenlines have zero \(Q\). This cannot occur together with finite projective order and \(\det(AB-BA)=0\). As computed in Section 6.4, gauge the row slopes to \(1,\omega\), where \(\omega^2+\omega+1=0\). The two zero-commutator parameter values are

\[a_1=\frac1{1-2\omega},\qquad a_2=\frac{\omega^2}{1-2\omega^2},\qquad b=-a\omega,\quad c=1+a(1+\omega).\]

The row eigenvalues are \(1+a\) and \(1+a\omega\), and direct substitution gives

\[\frac{1+a_1}{1+a_1\omega}=2,\qquad \frac{1+a_2}{1+a_2\omega}=\frac12.\]

Neither ratio has finite order. Thus absorption succeeds, and the \(n\) row directions finish the proof.

In the second alternative, \(u\) is not an eigenvector. Its \(n\ge6\) column directions contain one with nonzero \(Q\), so an actual \(A^jB\) supplies the row eigenvector \(v\). If \(vD\) is zero or proportional to \(v\), then \(A,B,D\) share that row eigendirection (and hence its kernel column line), again a removed case. Otherwise \(vD\) is available and either is not an \(A\) eigenrow, in which case its \(n\) directions finish by the five-direction bound, or is the other eigenrow. In the latter case both row eigendirections are individually available, and Lemma 15 proves hardness.

These cases exhaust \(B\), and prove the theorem.
\end{proof}

\section{10. Complete treatment of projective order two}\label{complete-treatment-of-projective-order-two}

For an invertible nonscalar \(2\times2\) matrix, projective order two means trace zero. Thus the normalized first matrix has \(c=-1\), with its determinant still required to be nonzero. The proof separates singular \(B\), which can directly supply a unary after absorption, from invertible \(B\). In the latter case it distinguishes common directions from a finite group with no invariant point or pair. The additional actual gadget \(D\) resolves the finite-group obstruction. All exclusions made before division are stated in the individual lemmas.

\subsection{10.1 Statement and the actual matrices}\label{statement-and-the-actual-matrices}

Write

\[f=[1,a,b,-1],\qquad e=[1,0,0,1].\]

The three actual straddled gadgets used below have matrices

\[A=\begin{pmatrix}1&b\\a&-1\end{pmatrix},\qquad
B=\begin{pmatrix}1+ab&a^2-b\\a+b^2&1+ab\end{pmatrix},\]

\[D=\begin{pmatrix}
1+2a^3+b^3&a^2+2ab^2+b\\
a+2a^2b-b^2&a^3+2b^3-1
\end{pmatrix}.\]

Here \(A^2=(1+ab)I\). Thus, when \(ab\ne-1\), the two eigenvalues of \(A\) are nonzero and opposite: the projective order of \(A\) is two. The case \(ab=-1\) was classified by the singular-surface theorem.

\begin{holantstatement}{Theorem 17}

Let \(a,b\) be algebraic complex numbers with \(ab\ne-1\). Then

\[\operatorname{Holant}([1,a,b,-1]\mid e)\]

is polynomial-time computable when \(a=b=0\), and is \#P-hard otherwise.

\end{holantstatement}

Two previously proved mechanisms will be used, with their hypotheses stated explicitly:

\begin{itemize}
\tightlist
\item
  If invertible actual gadgets generate an infinite subgroup of \(\operatorname{PGL}_2(\mathbb C)\) with no invariant point and no invariant unordered pair of points, hardness follows from the orbit-and-absorption lemma. Its proof extracts a rank-one matrix from an infinite-order word, moves its column away from the three zeros of \(u_0^3+u_1^3\), and moves its row into a hard binary contraction. At most five row directions give tractable binaries because \(A\) is invertible.
\item
  If \(A,B,D\) have a common invariant column line, the common-line classification gives, after a cube-root diagonal gauge, either \([1,t,t,1]\) or \([1,t,-1-2t,2+3t]\). The reversal-symmetric theorem and the perfect-matching reduction classify these two families. Under invertible \(A\) and outside generalized equality, they are hard.
\end{itemize}

For the second mechanism, a diagonal gauge preserves the endpoint \(c=-1\), so the reversal-symmetric family cannot occur here. The matching family has \(2+3t=-1\), hence \(t=-1\). Its matching coefficient \(1+t\) vanishes, so this particular intersection has singular \(A\) and is also excluded. Consequently, in the present theorem \(A,B,D\) actually cannot share a line. We nevertheless retain the general formulation when explaining the group argument.

\subsection{10.2 A singular B gives a direct unary gadget}\label{a-singular-b-gives-a-direct-unary-gadget}

\begin{figure}[H]
\centering
\begin{tikzpicture}[x=1.05cm,y=1.05cm]
\coordinate (l0) at (0,0);
\coordinate (r0) at (4,0);
\coordinate (r1) at (1,-1.8);
\coordinate (l1) at (3,-1.8);
\coordinate (i) at (-1.3,0);
\coordinate (j) at (5.3,0);
\draw (i) to[] node[midway,above,lab] {$i$} (l0);
\draw (r0) to[] node[midway,above,lab] {$j$} (j);
\draw (l0) to[] node[midway,above,lab] {$j$} (r0);
\draw (l0) to[] node[midway,above,lab] {$t$} (r1);
\draw (l1) to[] node[midway,above,lab] {$j$} (r0);
\draw (r1) to[bend left=28] node[midway,above,lab] {$t$} (l1);
\draw (r1) to[bend right=28] node[midway,below,lab] {$t$} (l1);
\node[L] at (l0) {$h$};
\node[R] at (r0) {$=_3$};
\node[R] at (r1) {$=_3$};
\node[L] at (l1) {$h$};
\end{tikzpicture}
\caption{The gadget $B_h$. Vertex shape specifies the bipartition, even when a square is drawn to the right of a circle. Equality has already forced the top and right edges to carry $j$, and the three lower-left incident edges to carry the same bit $t$. Sum over $t$.}
\label{fig:gadget_b-13}
\end{figure}
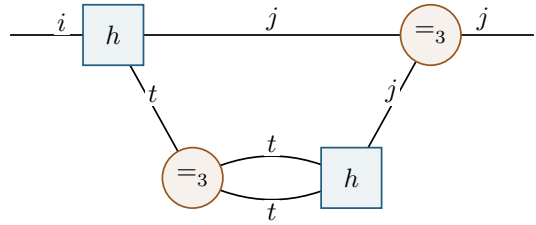

\begin{holantstatement}{Lemma 17.1}

Let \(f=[1,a,b,-1]\) with \(ab\ne-1\), and let \(B=\left(\begin{smallmatrix}1+ab&a^2-b\\a+b^2&1+ab\end{smallmatrix}\right)\). If \(\det B=0\), then \(\operatorname{Holant}(f\mid e)\) is \#P-hard.

\end{holantstatement}

\begin{proof}[Proof]

Put \(x=ab\). Direct expansion gives

\[\det B=1+3ab-a^3+b^3
=-\prod_{\omega^3=1}(a-\omega b-\omega^2).\]

For a cube root of unity \(\omega\), replacing \(f\) by

\[[1,\omega a,\omega^2b,-1]\]

does not change the value of any closed instance: all three edges incident to a right equality vertex have the same bit, and its gauge factor is \(\omega^{3j}=1\). A zero factor in the product therefore permits us to assume

\[b=a-1.\]

In these coordinates,

\[A=\begin{pmatrix}1&a-1\\a&-1\end{pmatrix},\quad
\det A=-(a^2-a+1),\quad
B=(a^2-a+1)\begin{pmatrix}1&1\\1&1\end{pmatrix}.\]

The scalar \(d=a^2-a+1\) is nonzero because \(A\) is invertible. The rank-one column \((1,1)^T\) has absorption value \(1^3+1^3=2\ne0\). Hence the row \((1,1)\) is individually simulable, with all its copies introduced through the same absorption construction.

Contracting this row against one input of \(f\) gives the symmetric binary

\[g=[1+a,\ 2a-1,\ a-2].\]

The complete complex binary classification tests the following four possibilities: rank one, diagonal, disequality, and the twelve affine rays. For this \(g\):

\[g_0g_2-g_1^2=-3(a^2-a+1)\ne0,\]

so it is not rank one. Its two endpoints cannot both vanish, so it is not disequality. It is diagonal only at \(a=1/2\).

For an affine binary, it is necessary that

\[g_0g_2+g_1^2=5a^2-5a-1=0,\]

and, since \(g_1\ne0\), that \((g_0/g_1)^{12}=1\). The displayed quadratic has two real roots. Thus \(g_0/g_1\) is real, so the twelfth-power condition forces \(g_0/g_1=1\) or \(-1\). These equations give \(a=2\) or \(a=0\), respectively, neither of which solves the quadratic. No affine case occurs.

Consequently \(g\) is hard unless \(a=1/2\). At that value \(f=[1,1/2,-1/2,-1]\) belongs to the already proved anti-reversal theorem: \([1,t,-t,-1]\) is hard for \(t\notin\{0,1,-1\}\). This completes the singular-\(B\) case.
\end{proof}

\subsection{10.3 A finite irreducible group is broken by D}\label{a-finite-irreducible-group-is-broken-by-d}

For an invertible matrix \(M\), let

\[\mathcal I(M)=\frac{(\operatorname{tr}M)^2}{\det M}.\]

This quantity is unchanged by conjugation or nonzero scaling. If the eigenvalue ratio is \(z\), then \(\mathcal I(M)=2+z+z^{-1}\). Every finite-order ratio therefore gives a real value in \([0,4]\).

\begin{holantstatement}{Lemma 17.2}

Let \(f=[1,a,b,-1]\) with \(ab\ne-1\), and let \(A,B,D\) be the three actual matrices displayed in Section 10.1. Assume \(A,B\) are invertible, do not commute, and generate a finite projective group preserving neither a point nor an unordered pair. Assume also \(a^3+b^3\ne0\). Then \(D\) is invertible and has infinite projective order.

\end{holantstatement}

\begin{proof}[Proof]

A finite subgroup of \(\operatorname{PGL}_2(\mathbb C)\) with these properties is tetrahedral, octahedral, or icosahedral. Its nonidentity element orders belong to \(\{2,3,4,5\}\).

Write

\[x=ab,\quad p=a^3,\quad q=b^3,\quad pq=x^3,\]

\[t=\mathcal I(B),\qquad u=\mathcal I(AB).\]

The trace identities are

\[\operatorname{tr}B=2(1+x)\ne0,\qquad
\operatorname{tr}(AB)=p+q\ne0.\]

Neither \(B\) nor \(AB\) is projectively the identity: either possibility would make \(A\) and \(B\) commute. Their nonzero traces also exclude order two. Therefore

\[t,u\in S_0:=\left\{1,2,\frac{3+\sqrt5}{2},\frac{3-\sqrt5}{2}\right\}.\]

Now

\[\det A=-(1+x),\quad
\det B=1+3x-p+q=\frac{4(1+x)^2}{t}.\]

Put \(M=p-q\) and \(v=p+q\). We obtain

\[M=1+3x-\frac{4(1+x)^2}{t},\qquad
v^2=-\frac{4u(1+x)^3}{t}.\]

The second identity follows from \(u=v^2/(\det A\det B)\). Direct expansion of the actual matrix \(D\) gives

\[\operatorname{tr}D=3v,\]

\[\det D=2v^2-2(x+1)M-(x+1)(3x^2+1).\]

Substitute the two preceding formulas. All dependence on the unknown point \(a,b\) cancels from the trace invariant:

\[\det D=\frac{(1+x)^3}{t}(8-8u-3t),\qquad
\boxed{\mathcal I(D)=\frac{36u}{8u+3t-8}}.\]

We check both invertibility and infinite order without solving for \(a,b\).

First, \(8u+3t-8\ne0\) for \(t,u\in S_0\). If both are rational, the possibilities are \(3,6,11,14\). If exactly one is irrational, its nonzero \(\sqrt5\) coefficient cannot cancel. If both are irrational, the denominator is

\[\frac{17+(8\varepsilon+3\eta)\sqrt5}{2},\qquad \varepsilon,\eta\in\{1,-1\},\]

which is nonzero because \(8\varepsilon+3\eta\in\{11,5,-5,-11\}\) and \(17^2\) equals neither \(5\cdot5^2\) nor \(5\cdot11^2\). Thus \(D\) is invertible.

Second, \(u>0\). If the denominator is negative, \(\mathcal I(D)<0\). If it is positive, then

\[\mathcal I(D)-4
=\frac{4(u-3t+8)}{8u+3t-8}>0,\]

because

\[u-3t+8\ge
\frac{3-\sqrt5}{2}-3\frac{3+\sqrt5}{2}+8
=5-2\sqrt5>0.\]

In every case \(\mathcal I(D)\notin[0,4]\). Its eigenvalue ratio cannot be a root of unity.
\end{proof}

\subsection{10.4 Proof of Theorem 17}\label{proof-of-theorem-17}

\begin{proof}[Proof]

Generalized equality, \(a=b=0\), is tractable. Assume \((a,b)\ne(0,0)\). Lemma 17.1 treats singular \(B\), so \(B\) is now invertible.

If \(a^3+b^3=0\), a cube-root gauge makes \(b=-a\). The anti-reversal theorem gives hardness, since \(a=0\) is generalized equality and \(a=\pm1\) would make \(A\) singular. We may therefore assume \(a^3+b^3\ne0\).

The matrices \(A\) and \(B\) do not commute. Indeed their off-diagonal commutator entries are \(2(a^2-b)\) and \(-2(b^2+a)\). If both vanish, then \(b=a^2\) and \(a(a^3+1)=0\). The first possibility is generalized equality; the second gives \(ab=-1\), contrary to invertibility.

Let \(G_0=\langle[A],[B]\rangle\).

\textbf{Suppose first that G0 has a common invariant point.} In an \(A\) eigenbasis, the noncommuting \(B\) is triangular with one nonzero off-diagonal entry and nonzero diagonal entries. Its group commutator with \(A\) is a nonidentity unipotent matrix. This proves that \(G_0\) is infinite; its unique common fixed point is the corresponding triangular coordinate line \(L\). The unipotent element also shows that \(G_0\) cannot preserve an unordered pair: a nonidentity unipotent transformation has only one finite orbit, its fixed point.

If \(D\) preserved \(L\), the common-line classification would apply. As explained after the theorem statement, no such point survives the present hypotheses. Thus \(D(L)\not\subseteq L\).

If \(D\) is invertible, \(\langle[A],[B],[D]\rangle\) is infinite and preserves neither a point nor a pair, so the orbit-and-absorption lemma proves hardness. If \(D\) is singular, it is nonzero because \(D(L)\not\subseteq L\), so factor it as \(D=uv\). The condition on \(L\) says \(u\notin L\) and that \(v\) does not annihilate \(L\). Both the column orbit of \(u\) and the row orbit of \(v\) under \(G_0\) are infinite; the unique fixed row line is the annihilator of the fixed column line \(L\). Choose a positive word moving \(u\) away from the three absorption zeros, absorb the resulting actual rank-one gadget, and then choose six distinct row images of \(v\). At most five yield easy binaries, so one gives hardness. This is the same rank-one absorption construction as in the lemma, without needing spectral interpolation.

\textbf{Suppose next that G0 has no common invariant point.} It has no invariant unordered pair either. To see this, use the two points of a hypothetical pair as a basis. A matrix preserving the pair is diagonal or off-diagonal. Since \(\operatorname{tr}B=2(1+ab)\ne0\), \(B\) must be diagonal. If \(A\) were diagonal as well, they would commute. Thus \(A\) would be off-diagonal, making \(AB\) off-diagonal and giving \(\operatorname{tr}(AB)=a^3+b^3=0\), already excluded.

If \(G_0\) is infinite, the orbit-and-absorption lemma applies directly. If it is finite, Lemma 17.2 gives an invertible infinite-order \(D\). The larger group generated by \(A,B,D\) is infinite; it inherits the absence of invariant points and pairs from \(G_0\). The same lemma proves hardness.

All possibilities have been exhausted.
\end{proof}

\section{11. Complete treatment of projective order four}\label{complete-treatment-of-projective-order-four}

The order-four relation is \((\operatorname{tr}A)^2=2\det A\), with \(A\) invertible. The singular-\(B\) case, the finite group with no invariant point or pair, and the invariant-pair case require different constructions. Sections 11.2, 11.4, and 11.5 address them separately; the trace identities in Section 11.3 are used to show when adding \(D\) escapes the initial group. The completion proof specifies which earlier common-line result applies to the remaining cases.

\subsection{11.1 Matrices and statement}\label{matrices-and-statement}

The actual straddled gadgets are

\[A=\begin{pmatrix}1&b\\a&c\end{pmatrix},\quad
B=\begin{pmatrix}1+ab&a^2+bc\\a+b^2&ab+c^2\end{pmatrix},\]

\[D=\begin{pmatrix}
1+2a^3+b^3&a^2+2ab^2+bc^2\\
a+2a^2b+b^2c&a^3+2b^3+c^3
\end{pmatrix}.\]

The matrix \(D\) is also the first gadget of the actual ternary output

\[h=\Phi(f)=[1+2a^3+b^3,\ a+2a^2b+b^2c,\ a^2+2ab^2+bc^2,\ a^3+2b^3+c^3].\]

Thus \(\operatorname{Holant}(h\mid e)\le_T\operatorname{Holant}(f\mid e)\) by replacing each left \(h\) vertex with its fixed ternary gadget.

For invertible \(M\) write \(\mathcal I(M)=(\operatorname{tr}M)^2/\det M\). Projective order four of \(A\) is equivalent to

\[\mathcal I(A)=2,\quad C:=1+c\ne0,\quad
ab=-\frac{c^2+1}{2},\quad\det A=\frac{C^2}{2}.\]

In particular \(\operatorname{tr}B=0\) throughout this surface.

\begin{holantstatement}{Theorem 18}

Let \(f=[1,a,b,c]\) be fixed and algebraic complex. If \(A_f\) has projective order four, then \(\operatorname{Holant}(f\mid e)\) is polynomial-time computable for generalized equality (\(a=b=0\)) and is \#P-hard otherwise.

\end{holantstatement}

We use the established binary contraction

\[F_f(r,s)=[r+as,\ ar+bs,\ br+cs].\]

Because \(A\) is invertible, no nonzero row gives the zero binary or disequality. Among all row directions, at most five yield easy binaries: two rank-one directions, one diagonal direction, and at most two intersections with the affine conic \(g_0g_2+g_1^2=0\).

\subsection{11.2 Singular B: a nilpotent kernel gives a hard binary}\label{singular-b-a-nilpotent-kernel-gives-a-hard-binary}

\begin{holantstatement}{Lemma 18.1}

Let \(f=[1,a,b,c]\) and suppose that \(A_f\) has projective order four. Let \(A,B,D\) be the actual matrices displayed in Section 11.1. If \(B\) is singular and \(A,B,D\) have no common invariant column line, then \(\operatorname{Holant}(f\mid e)\) is \#P-hard.

\end{holantstatement}

\begin{proof}[Proof]

Since \(\operatorname{tr}B=\det B=0\), Cayley--Hamilton gives \(B^2=0\). Also \(B\ne0\). Indeed \(B=0\) would give \(ab=-1\), \(c^2=1\), and hence \(c=1\) because \(C\ne0\). The remaining equations \(a+b^2=0\), \(a^2+b=0\) contradict \(ab=-1\).

Factor this nonzero rank-one matrix as \(B=uv\), with column \(u\ne0\) and row \(v\ne0\). Nilpotence gives \(vu=0\) and \(vB=0\).

First we show that \(v\) is individually simulable. Absorbing a rank-one gadget with column \(w\) is valid whenever

\[Q(w)=w_0^3+w_1^3\ne0.\]

If \(u\) is not an \(A\) eigenvector, its \(A\) orbit has exactly four projective directions. The polynomial \(Q\) has only three projective zeros. Therefore one of the four actual gadgets \(A^jB\), \(0\le j\le3\), is absorbable and supplies \(v\).

If \(u\) is an eigenvector and \(Q(u)\ne0\), absorb \(B\) directly. In the remaining case \(u\) is an eigenvector with \(Q(u)=0\). If \(D\) preserves its line, \(A,B,D\) have a common invariant line, one of the previously classified hard cases that we removed. Otherwise \(Du\) is nonzero and outside the original eigenline. If \(Du\) is not an eigenvector, its four-point orbit again escapes the three zeros of \(Q\). If it is the other eigenvector, that eigenline has nonzero \(Q\): the calculation for two isotropic eigenlines and \(\det(AB-BA)=0\) gives eigenvalue ratio \(2\) or \(1/2\), whereas the present ratio has order four. Thus some \(A^jDB\) supplies \(v\) as well.

Now let \(g=F_f(v)=[g_0,g_1,g_2]\). The two equations \(vB=0\) have the particularly useful form

\[g_0+b g_1=0,\qquad a g_1+c g_2=0.\]

Suppose first that \(c\ne0\). Then \(g_1\ne0\): otherwise both endpoints vanish, which is impossible because \(A\) is invertible and \(v\ne0\). Consequently

\[\frac{g_0g_2}{g_1^2}=\frac{ab}{c}=-\frac{c^2+1}{2c}.\]

For rank one this ratio would be \(1\), forcing \((c+1)^2=0\), contrary to \(C\ne0\). For an affine binary it would be \(-1\), forcing \((c-1)^2=0\). The output is not diagonal since \(g_1\ne0\). Therefore \(g\) is hard whenever \(c\notin\{0,1\}\).

At \(c=0\), the equations \(ab=-1/2\) and \(\det B=0\) give \(a^3=-1/2\) and \(b=a^2\). A kernel row is \(v=(-a,1)\), and

\[vA=(0,1/2),\qquad F_f(0,1)=[a,b,0].\]

This binary has a nonzero middle entry and exactly one nonzero endpoint, so it is hard.

At \(c=1\), we have \(ab=-1\) and \(a^3=\pm1\). If \(a^3=1\), the row factor of \(B\) is \((1,0)\); multiplying it by \(A\) and contracting gives

\[F_f(1,b)=[0,2a,2b].\]

If \(a^3=-1\), the row factor is \((0,1)\), and

\[F_f(a,1)=[2a,2a^2,0].\]

Again each binary has a nonzero middle entry and exactly one nonzero endpoint. Only the one absorbed unary type and its actual \(A\) chain are needed. This completes the proof.
\end{proof}

\subsection{11.3 Trace formulas for D and AD}\label{trace-formulas-for-d-and-ad}

Assume from now on that \(B\) is invertible. Set

\[d=\frac{\det B}{C^4},\qquad L=\frac{\operatorname{tr}(AB)}{C^3}.\]

Exact expansion gives

\[\frac{\operatorname{tr}D}{C^3}=3L+1,\]

\[\frac{\det D}{C^6}=2L^2+2L+d+\frac38,\]

\[\frac{\operatorname{tr}(AD)}{C^4}=L-2d+\frac12.\]

These are identities for the actual matrices, not assumptions that arbitrary linear combinations are realizable.

\subsection{11.4 Finite irreducible group generated by A and B}\label{finite-irreducible-group-generated-by-a-and-b}

\begin{holantstatement}{Lemma 18.2}

Let \(f=[1,a,b,c]\), suppose that \(A=A_f\) has projective order four, and let the actual matrix \(B=B_f\) be invertible. If \(G_0=\langle[A],[B]\rangle\) is finite and preserves neither a point nor an unordered pair, then the problem is \#P-hard.

\end{holantstatement}

\begin{proof}[Proof]

A finite subgroup of \(\operatorname{PGL}_2(\mathbb C)\) containing an element of order four and preserving no pair is the octahedral group \(S_4\). Since \(\operatorname{tr}B=0\), the element \([B]\) has order two. In the permutation representation of \(S_4\), \([A]\) is a 4-cycle.

The element \([B]\) cannot be a double transposition. The double transpositions form the normal Klein four group; adjoining one to a 4-cycle generates a subgroup of the order-eight dihedral group, which preserves a pair. Thus \([B]\) is a transposition. The product of a 4-cycle with a transposition has order two or three. Order two would again make the generated group dihedral. Hence \([AB]\) has order three, and

\[\mathcal I(AB)=1,\qquad d=2L^2.\]

Put

\[R_4(L)=32L^2+16L+3.\]

If \(D\) is invertible, its two relevant invariants are

\[\mathcal I(D)=\frac{8(3L+1)^2}{R_4(L)},\qquad
\mathcal I(AD)=\frac{4(-8L^2+2L+1)^2}{R_4(L)}.\]

Every element of \(S_4\) has invariant in \(\{0,1,2,4\}\). No \(L\) satisfies both displayed requirements. Here is an exact resultant certificate. Define

\[P_j=8(3L+1)^2-jR_4(L),\qquad
Q_k=4(-8L^2+2L+1)^2-kR_4(L).\]

For rows and columns indexed by \(j,k=0,1,2,4\), the matrix of resultants \(\operatorname{Res}_L(P_j,Q_k)/4096\) is

\[\begin{pmatrix}
10000&1&9604&87616\\
3844&-16079&-29952&-39548\\
5184&-167&-5276&-14768\\
132496&83089&45540&6016
\end{pmatrix}.\]

Every entry is nonzero, so no pair of tests has a common root. Thus \([D]\notin G_0\). A finite projective group containing \(S_4\) cannot be larger than \(S_4\): cyclic and dihedral groups preserve pairs, \(A_4\) is smaller, and \(A_5\) has no element of order four. Therefore \(\langle[A],[B],[D]\rangle\) is infinite and preserves no point or pair. The orbit-and-absorption hardness lemma applies.

If \(D\) is singular, it is nonzero: \(\operatorname{tr}D=0\) would give \(L=-1/3\), but then \(R_4(L)=11/9\ne0\). Factor \(D=uv\). Every projective column or row orbit under \(S_4\) has at least six points. Indeed a finite Möbius group fixing a point is cyclic, and the largest cyclic subgroup of \(S_4\) has order four; the orbit-stabilizer formula gives at least \(24/4=6\) points. The column orbit therefore escapes the three zeros of \(Q\), and the row orbit contains a direction outside the five easy binary directions. All these group elements are positive words in the finite-order actual matrices \(A,B\), up to harmless nonzero scalars. Absorb one such actual rank-one gadget and use one hard row image.
\end{proof}

\subsection{11.5 The invariant-pair branch}\label{the-invariant-pair-branch}

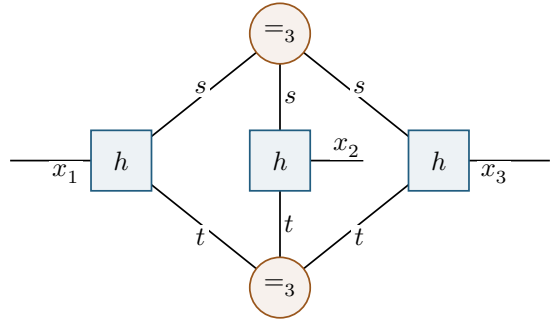
\begin{figure}[H]
\centering
\begin{tikzpicture}[x=1.05cm,y=1.05cm]
\coordinate (l0) at (-2,0);
\coordinate (l1) at (0,0);
\coordinate (l2) at (2,0);
\coordinate (r0) at (0,1.6);
\coordinate (r1) at (0,-1.6);
\coordinate (x0) at (-3.4,0);
\coordinate (x1) at (1.05,0);
\coordinate (x2) at (3.4,0);
\draw (x0) to[] node[midway,below,lab] {$x_1$} (l0);
\draw (l1) to[] node[pos=0.8,above,lab] {$x_2$} (x1);
\draw (l2) to[] node[midway,below,lab] {$x_3$} (x2);
\draw (l0) to[] node[midway,above,lab] {$s$} (r0);
\draw (l0) to[] node[midway,below,lab] {$t$} (r1);
\draw (l1) to[] node[midway,right,lab] {$s$} (r0);
\draw (l1) to[] node[midway,right,lab] {$t$} (r1);
\draw (l2) to[] node[midway,above,lab] {$s$} (r0);
\draw (l2) to[] node[midway,below,lab] {$t$} (r1);
\node[L] at (l0) {$h$};
\node[L] at (l1) {$h$};
\node[L] at (l2) {$h$};
\node[R] at (r0) {$=_3$};
\node[R] at (r1) {$=_3$};
\end{tikzpicture}
\caption{The ternary gadget $\Phi(h)$: three left $h$ vertices and two right equality vertices. All three edges at the upper circle carry $s$; all three at the lower circle carry $t$. The three dangling inputs are $x_1,x_2,x_3$. The central dangling stub ends inside the drawing and does not touch the right square.}
\label{fig:gadget_phi-14}
\end{figure}

\begin{holantstatement}{Lemma 18.3}

Let \(f=[1,a,b,c]\), and let \(A=A_f\), \(B=B_f\), and \(D=A_{\Phi(f)}\) be the actual matrices displayed in Section 11.1. Suppose \(A\) has projective order four and \(B\) is invertible. If \(AB\ne BA\) and \(A,B\) preserve an unordered pair of projective column lines, then \(\operatorname{Holant}(f\mid e)\) is \#P-hard.

\end{holantstatement}

\begin{proof}[Proof]

Since \(A\) has order four, the preserved pair must be its two eigenlines. In their basis \(B\) must be off-diagonal: a diagonal \(B\) would commute with \(A\). Thus

\[\operatorname{tr}(AB)=0,\qquad L=0,\]

and \(G_0\) is a dihedral group of order eight. The trace identities become

\[\operatorname{tr}D=C^3,\quad
\det D=C^6(d+3/8),\quad
\operatorname{tr}(AD)=C^4(1/2-2d).\]

We first show that an invertible \(D\) does not preserve this pair. Its nonzero trace excludes an off-diagonal form, so pair preservation would require \(AD=DA\). Let \(x=ab\), \(p=a^3\), \(q=b^3\). The condition \(L=0\) gives \(p+q=c(1+c)\). The diagonal commutator equation is

\[-p-xc^2+x+cq=0,\]

hence

\[q=c+x(c-1),\qquad p=c^2-x(c-1).\]

If \(x=0\), then \(c=\pm i\), and these formulas give \(pq=c^3\ne0=x^3\), a contradiction. If \(x\ne0\), dividing the upper off-diagonal commutator entry by \(b\) gives

\[\frac{(AD-DA)_{01}}b=\frac{c(c-1)(c+1)^2}{c^2+1}.\]

Its vanishing forces \(c=0\) or \(1\). But the identity \(pq=x^3\) becomes

\[0=-\frac{(c+1)^2}{8}(c^4-2c^3-2c^2-2c+1),\]

which fails at both values. Thus \(D\) breaks the pair.

If \(D\) is invertible and the enlarged group were finite, it would have to be \(S_4\): it contains an element of order four and no longer preserves a point or pair. Its invariants would satisfy

\[\mathcal I(D)=\frac8{8d+3}\in\{1,2,4\},\qquad
\mathcal I(AD)=\frac{4(1-4d)^2}{8d+3}\in\{0,1,2,4\}.\]

The three possibilities in the first condition are \(d=5/8,1/8,-1/8\). Their second invariants are \(9/8,1/4,9/2\), respectively, none allowed. The enlarged group is therefore infinite, and the orbit-and-absorption lemma proves hardness.

It remains to handle singular \(D\), namely \(d=-3/8\). We show directly that its ternary source \(h=\Phi(f)\) is a hard point of the already proved singular-\(A\) classification.

The value \(c=1\) is impossible here: \(L=0\) would give \(p+q=2\) and \(d=-1/8\). Put

\[z=\frac{c-1}{c+1}.\]

Solving \(L=0\) and \(d=-3/8\) for \(p,q\), and using \(pq=x^3\), gives

\[Z(z):=z^4+10z^2-4=0.\]

In particular \(z\ne0\). The endpoints of \(h\) satisfy

\[H_0:=\frac{h_0}{C^3}=-\frac{z^2-4z+2}{8z},\qquad
H_3:=\frac{h_3}{C^3}=\frac{z^2+4z+2}{8z}.\]

Neither endpoint is zero. Indeed their product, reduced using \(Z(z)=0\), is \((11z^2-4)/(32z^2)\), and substituting \(z^2=4/11\) into \(Z\) gives \(-28/121\ne0\).

Also \(H_0^4\ne H_3^4\). Their difference factors as

\[H_0^4-H_3^4=-\frac{(z^2+2)(z^4+20z^2+4)}{128z^3}.\]

Modulo \(Z\), the second factor is \(2(5z^2+4)\); neither \(z^2=-2\) nor \(z^2=-4/5\) solves \(Z\). Consequently \(h_3/h_0\notin\{1,-1,i,-i\}\).

Finally \(h\) is not rank one. The numerator of

\[\frac{h_1^3-h_0^2h_3}{C^9}
=\frac{z^8+5z^7+18z^6+53z^5+82z^4+10z^3+44z^2+20z-8}{512z^3}\]

has remainder \(16(z+1)^2\) modulo \(Z\). It is nonzero on \(Z=0\) because \(Z(-1)=7\).

The singular-\(A\) classification says that its only easy points are rank-one tensors, generalized equality, and cube-root gauges of six explicit affine signatures. Generalized equality is impossible here: it would give \(h_1=h_2=0\) and hence \(\det D=h_0h_3\ne0\), contrary to the current singular case. Every remaining affine signature in that list, when its first endpoint is nonzero, has endpoint ratio in \(\{0,1,-1,i,-i\}\). We have excluded every one of these possibilities. Thus \(\operatorname{Holant}(h\mid e)\) is hard, and the actual ternary replacement proves hardness for \(f\).
\end{proof}

\subsection{11.6 Completion of the order-four theorem}\label{completion-of-the-order-four-theorem}

\begin{proof}[Proof of Theorem 18]

Generalized equality is easy. Remove it and the already classified common-line families of \(A,B,D\); they are hard under the present hypotheses. Also remove \(AB=BA\): for invertible \(A\) outside generalized equality, commutation forces \(c=1,b^3=a^3\), a cube-root gauge of a reversal-symmetric signature \([1,t,t,1]\), already classified. On the order-four surface \(t=\pm i\), both hard.

Lemma 18.1 treats singular \(B\). Suppose \(B\) is invertible.

If \(G_0=\langle[A],[B]\rangle\) has a common invariant point, express \(B\) in an \(A\) eigenbasis. It is triangular and noncommuting, with both diagonal entries nonzero and exactly one off-diagonal entry nonzero. Its group commutator with \(A\) is a nonidentity unipotent. Thus \(G_0\) is infinite, has a unique common fixed line \(L_0\), and has no invariant pair. Since \(A,B,D\) have no common line, \(D(L_0)\not\subseteq L_0\).

For invertible \(D\), the larger group has no fixed point or pair and is infinite, so the orbit-and-absorption lemma applies. For singular \(D\), the condition \(D(L_0)\not\subseteq L_0\) first implies \(D\ne0\), so write \(D=uv\). The same condition implies \(u\notin L_0\) and that \(v\) does not annihilate \(L_0\). Their \(G_0\) orbits are infinite: the unique fixed row line is the annihilator of \(L_0\), the dual of the fixed column line. Move the column away from the three absorption zeros and take six distinct row images; one yields a hard binary. This uses an actual rank-one gadget, so no extraction step is required.

If \(G_0\) has no common point but preserves a pair, Lemma 18.3 applies. In the remaining case it preserves neither. If infinite, apply the orbit-and-absorption lemma; if finite, apply Lemma 18.2.

These alternatives exhaust the possibilities.
\end{proof}

\section{12. Finite irreducible groups at orders three and five}\label{finite-irreducible-groups-at-orders-three-and-five}

Sections 12 and 13 jointly finish orders three and five. This section assumes \(B\) is invertible and the group generated by \([A],[B]\) is finite with no invariant point or pair. Infinite groups with neither are already covered by Lemma 10. The finite subgroup classification reduces the current branch to finitely many exact trace-invariant possibilities. We test whether \(D\) leaves the finite group; the residual order-five parameters are generalized equality, while the residual order-three parameters yield an actual rank-one matrix from \(E=A_{\Psi(f)}\). Appendix A records the exact calculations used in these reductions.

\begin{holantstatement}{Proposition 19 (finite groups at orders three and five)}

Fix a complex algebraic symmetric ternary signature \(f=[1,a,b,c]\). Suppose the actual first matrix \(A\) is invertible and has projective order three or five, the actual second matrix \(B\) is invertible, and \(G=\langle[A],[B]\rangle\) is finite and preserves neither a point nor an unordered pair of points of \(\mathbb P^1(\mathbb C)\). Then \(\operatorname{Holant}(f\mid=_3)\) is \#P-hard.

\end{holantstatement}

All matrices below are actual straddled graph gadgets, so their positive products are actual gadgets. The proof never assumes free unary signatures. Matrix subtraction is used only to verify polynomial identities.

\begin{proof}[Proof of Proposition 19]

Throughout Sections 12.1--12.4, \(A\) and \(B\) are invertible and their projective group \(G\) is finite with no invariant point or pair. We first constrain the possible trace invariants and then treat the exceptional rows of the resulting finite table.

\subsection{12.1 Five trace identities}\label{five-trace-identities}

Use the established actual matrices

\[A=\begin{pmatrix}1&b\\a&c\end{pmatrix},\quad
B=\begin{pmatrix}1+ab&a^2+bc\\a+b^2&ab+c^2\end{pmatrix},\] \[D=\begin{pmatrix}1+2a^3+b^3&a^2+2ab^2+bc^2\\a+2a^2b+b^2c&a^3+2b^3+c^3\end{pmatrix}.\]

The actual gadget \(E=A_{\Psi(f)}\) from Section 2.4 is available.

\begin{figure}[htbp]
\centering
\definecolor{svgcolor0}{HTML}{344054}
\definecolor{svgcolor1}{HTML}{e7f1fb}
\definecolor{svgcolor2}{HTML}{20578a}
\definecolor{svgcolor3}{HTML}{fff1dc}
\definecolor{svgcolor4}{HTML}{996022}
\begin{tikzpicture}[x=0.463636pt,y=-0.463636pt]
\path[use as bounding box] (0,0) rectangle (600,550);
\path[fill=white,draw=none,line width=0.4636pt] (0,0) rectangle (600,550);
\path[fill=black,draw=svgcolor0,line width=1.1591pt] (300,245) -- (300,125);
\path[fill=black,draw=svgcolor0,line width=1.1591pt] (300,245) -- (405,320);
\path[fill=black,draw=svgcolor0,line width=1.1591pt] (300,245) -- (195,320);
\path[fill=black,draw=svgcolor0,line width=1.1591pt] (465,160) -- (300,125);
\path[fill=black,draw=svgcolor0,line width=1.1591pt] (465,160) -- (405,320);
\path[fill=black,draw=svgcolor0,line width=1.1591pt] (300,440) -- (405,320);
\path[fill=black,draw=svgcolor0,line width=1.1591pt] (300,440) -- (195,320);
\path[fill=black,draw=svgcolor0,line width=1.1591pt] (135,160) -- (195,320);
\path[fill=black,draw=svgcolor0,line width=1.1591pt] (135,160) -- (300,125);
\path[fill=black,draw=svgcolor0,line width=1.1591pt] (465,160) -- (530,115);
\node[anchor=base,inner sep=0pt,text=black,font=\rmfamily\fontsize{10.664}{12.796}\selectfont] at (530,106) {x};
\path[fill=black,draw=svgcolor0,line width=1.1591pt] (300,440) -- (300,510);
\node[anchor=base,inner sep=0pt,text=black,font=\rmfamily\fontsize{10.664}{12.796}\selectfont] at (300,537) {y};
\path[fill=black,draw=svgcolor0,line width=1.1591pt] (135,160) -- (70,115);
\node[anchor=base,inner sep=0pt,text=black,font=\rmfamily\fontsize{10.664}{12.796}\selectfont] at (70,106) {z};
\path[fill=svgcolor1,draw=svgcolor2,line width=0.9273pt] (280,225) rectangle (320,265);
\node[anchor=base,inner sep=0pt,text=black,font=\rmfamily\fontsize{10.664}{12.796}\selectfont] at (300,252) {f};
\path[fill=svgcolor3,draw=svgcolor4,line width=0.9273pt] (300,125) circle[radius=9.2727pt];
\node[anchor=base,inner sep=0pt,text=black,font=\rmfamily\fontsize{10.664}{12.796}\selectfont] at (300,132) {=};
\path[fill=svgcolor3,draw=svgcolor4,line width=0.9273pt] (405,320) circle[radius=9.2727pt];
\node[anchor=base,inner sep=0pt,text=black,font=\rmfamily\fontsize{10.664}{12.796}\selectfont] at (405,327) {=};
\path[fill=svgcolor3,draw=svgcolor4,line width=0.9273pt] (195,320) circle[radius=9.2727pt];
\node[anchor=base,inner sep=0pt,text=black,font=\rmfamily\fontsize{10.664}{12.796}\selectfont] at (195,327) {=};
\path[fill=svgcolor1,draw=svgcolor2,line width=0.9273pt] (445,140) rectangle (485,180);
\node[anchor=base,inner sep=0pt,text=black,font=\rmfamily\fontsize{10.664}{12.796}\selectfont] at (465,167) {f};
\path[fill=svgcolor1,draw=svgcolor2,line width=0.9273pt] (280,420) rectangle (320,460);
\node[anchor=base,inner sep=0pt,text=black,font=\rmfamily\fontsize{10.664}{12.796}\selectfont] at (300,447) {f};
\path[fill=svgcolor1,draw=svgcolor2,line width=0.9273pt] (115,140) rectangle (155,180);
\node[anchor=base,inner sep=0pt,text=black,font=\rmfamily\fontsize{10.664}{12.796}\selectfont] at (135,167) {f};
\node[anchor=base,inner sep=0pt,text=black,font=\rmfamily\fontsize{9.736}{11.684}\selectfont] at (300,38) {A planar ternary gadget with three exterior ports};
\end{tikzpicture}
\caption{The Psi construction used again in the order-three residual.}
\end{figure}
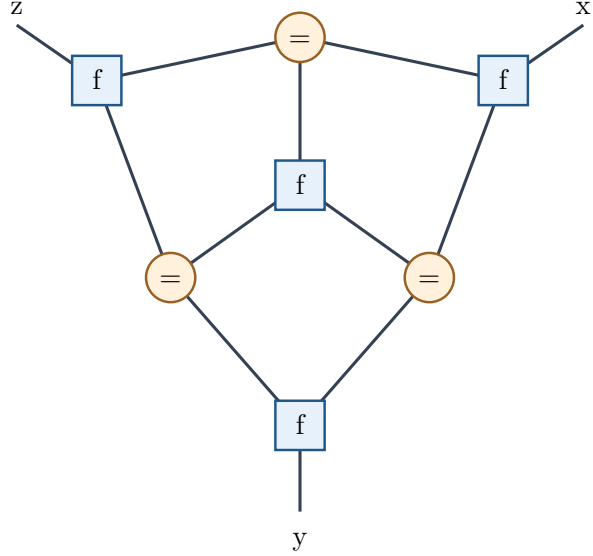

Its four entries are

\[E=\begin{pmatrix}w&y\\x'&z'\end{pmatrix},\] \[\begin{aligned}
w&=1+3a^3+3a^2b^2+b^3c,\\
x'&=a+a^4+2a^2b+a^2bc+2ab^3+b^2c^2,\\
y&=a^2+ab^2+2a^3b+b^4+2ab^2c+bc^3,\\
z'&=a^3+3a^2b^2+3b^3c+c^4.
\end{aligned}\]

Let \(C=1+c\), \(s=C^2/\det A\), \(d=\det B/C^4\), and \(L=\operatorname{tr}(AB)/C^3\). Here \(C\ne0\), because the projective order of \(A\) is neither two nor one. Direct polynomial expansion gives

\[\begin{aligned}
\operatorname{tr}D/C^3&=3L-2+6/s,\\
\det D/C^6&=2L^2+(-2+8/s)L+(2-2/s)d-2/s+4/s^2+3/s^3,\\
\operatorname{tr}(AD)/C^4&=L-2d-1/s+4/s^2,\\
\operatorname{tr}E/C^4&=1-4/s+6/s^2-4d,\\
\det E/C^8&=2L^2/s+2Ld-L/s+4L/s^2+4d^2-2d\\
&\quad+9d/s-11d/s^2-1/s+5/s^2-9/s^3+8/s^4.
\end{aligned}\]

Each identity follows by substituting the displayed entries and expanding after multiplication by its denominator. In particular, these are identities in \(\mathbb Q[a,b,c]\) after clearing powers of \(C\) and \(\det A\). They remain valid at \(c=1\). Appendix A gives an exact verification procedure; no division by \(c-1\) is used in these five identities.

\subsection{12.2 A finite exact calculation}\label{a-finite-exact-calculation}

Write \(I(M)=(\operatorname{tr}M)^2/\det M\), and put

\[\alpha=(3+\sqrt5)/2,\qquad\beta=(3-\sqrt5)/2,\qquad
\Omega=\{0,1,2,\alpha,\beta,4\}.\]

A finite subgroup of \(\operatorname{PGL}_2(\mathbb C)\) preserving neither a point nor a pair is \(A_4,S_4\), or \(A_5\). Every element in these groups has its invariant in \(\Omega\). If \(A\) has order five, the group is \(A_5\), which has no element of order four. A scalar \(B\) or \(AB\) would make the generated group cyclic, so \(I(B),I(AB)\ne4\).

Set \(t=I(B)\) and \(u=I(AB)\). Since

\[\operatorname{tr}B/C^2=1-2/s,\]

we have \(t\ne0\) for the present orders, and

\[d=(1-2/s)^2/t,\qquad L^2=ud/s.\]

The allowable sets, deliberately allowing some triples not realized by a finite group, are

{\def\LTcaptype{table} 
\begin{longtable}[]{@{}llll@{}}
\toprule\noalign{}
First order & \(s\) & \(t\) & \(u\) \\
\midrule\noalign{}
\endhead
\bottomrule\noalign{}
\endlastfoot
3 & \(1\) & \(1,2,\alpha,\beta\) & \(0,1,2,\alpha,\beta\) \\
5 & \(\alpha\) or \(\beta\) & \(1,\alpha,\beta\) & \(0,1,\alpha,\beta\) \\
\end{longtable}
}

For each of these 44 triples, substitute \(d\) into the expressions for \(D\). The polynomial \(L^2-ud/s\) has no common zero with \(\det D/C^6\). Its common zeros with

\[(3L-2+6/s)^2-v(\det D/C^6),\qquad v\in\Omega,\]

are exactly the following six rows:

{\def\LTcaptype{table} 
\begin{longtable}[]{@{}llll@{}}
\toprule\noalign{}
\(s\) & \(t=u\) & \(L\) & \(I(D)\) \\
\midrule\noalign{}
\endhead
\bottomrule\noalign{}
\endlastfoot
\(1\) & \(1\) & \(-1\) & \(1\) \\
\(1\) & \(2\) & \(-1\) & \(1\) \\
\(1\) & \(\alpha\) & \(-1\) & \(1\) \\
\(1\) & \(\beta\) & \(-1\) & \(1\) \\
\(\alpha\) & \(\beta\) & \(-\beta^2\) & \(\beta\) \\
\(\beta\) & \(\alpha\) & \(-\alpha^2\) & \(\alpha\) \\
\end{longtable}
}

This table is an exact finite polynomial calculation over \(\mathbb Q(\sqrt5)\), not a numerical root test. For each of the 44 triples, apply the Euclidean algorithm to the two quadratic polynomials specified above, once for singularity and once for each \(v\in\Omega\). The resulting 44 singularity gcds are constant. Of the 264 invariant gcds, exactly six are nonconstant; each is the linear polynomial with the root recorded in the table. Appendix A specifies the field representation and includes the complete finite calculation as supplementary data.

Outside the table, \(D\) is invertible and its invariant is outside \(\Omega\). Thus the group generated by \(A,B,D\) is infinite, and it still preserves neither a point nor a pair because its subgroup \(G\) does not. The existing infinite-group lemma proves hardness.

For the third and fourth rows, the displayed formula for \(AD\) gives

\[I(AD)=6-2\sqrt5\quad\text{or}\quad6+2\sqrt5,\]

both outside \(\Omega\). The same argument applies with the actual word \(AD\).

\subsection{12.3 The order-five residual is generalized equality}\label{the-order-five-residual-is-generalized-equality}

In the last two rows, \(s^2-3s+1=0\), \(d=s^{-2}\), and \(L=-s^{-2}\). Put

\[z=\frac{c-1}{c+1},\qquad
X=\frac{ab}{C^2}=\frac{1-z^2}{4}-\frac1s,\qquad
P=\frac{a^3}{C^3},\quad Q=\frac{b^3}{C^3}.\]

Then \(PQ=X^3\). The trace and determinant equations give

\[r:=P+Q=L-\frac{3+z^2}{4}+\frac2s,\] \[z(P-Q)=2d-\frac{X(1+3z^2)}2-\frac{(1-z^2)^2}{8}+r.\]

Under the residual conditions, reduction by \(s^2-3s+1=0\) yields \(r=X\) and

\[PQ-X^3=-\frac{X^2}{s^2z^2}\quad(z\ne0).\]

Therefore \(X=0\), and \(P+Q=PQ=0\) gives \(P=Q=0\). Thus \(a=b=0\), which is generalized equality; its diagonal matrices cannot generate the assumed irreducible group.

The value \(z=0\) cannot satisfy the residual equations either. Write \(k=z(P-Q)\), with \(k\) equal to the right side of the displayed determinant relation. Using \(r=P+Q\), the polynomial identity before dividing by \(z^2\) is

\[z^2\left(\frac{r^2}{4}-X^3\right)-\frac{k^2}{4}=-\frac{X^2}{s^2}.\]

On any actual solution, \(PQ=X^3\) and \(k=z(P-Q)\) make the left side zero. It follows that \(X=0\) even when \(z=0\). But there \(X=1/4-1/s\ne0\) for either root of \(s^2-3s+1\). This is a contradiction and treats the value excluded by the earlier division.

\subsection{12.4 The order-three residual supplies an actual rank-one gadget}\label{the-order-three-residual-supplies-an-actual-rank-one-gadget}

For \(s=1\), \(t=u\in\{1,2\}\), \(L=-1\), the formulas above give

\[\det E=0,\qquad \operatorname{tr}E=C^4(3-4/t)\ne0.\]

Thus \(E\) is an actual nonzero rank-one matrix. Write \(E=uv\).

Every projective column orbit of \(G\) has at least four points. Indeed \(G\) is one of \(A_4,S_4,A_5\), and a point stabilizer in a finite Möbius group is cyclic. Orbit-stabilizer gives minima four, six, and twelve respectively. The cubic \(Q(u)=u_0^3+u_1^3\) has only three projective zeros. Some positive word \(W\) in \(A,B\) therefore gives \(Q(Wu)\ne0\). Positive words realize every group element projectively because \(G\) is finite. The actual matrix \(WE\) supplies the single row \(v\) by the established absorption lemma.

If \(t=2\), then \(B\) has order four, so \(G=S_4\). The row orbit has at least six directions. At most five directions give tractable binary contractions because \(A\) is invertible. A fixed positive-word image of \(v\) therefore gives a hard binary, proving hardness with just that one row type.

It remains to handle \(t=1\). The invariant equations reduce to

\[ab=-c^2-c-1,\qquad a^3+b^3=c(c+1),\] \[a^3=\frac{(c^2+2c+2)(2c^2+c+1)}{c-1},\qquad
b^3=-\frac{(c^2+c+2)(2c^2+2c+1)}{c-1},\] \[R(c):=3c^4+5c^3+9c^2+5c+3=0.\]

Before dividing by \(c-1\), exclude \(c=1\) directly: the first two equations then give \(ab=-3\) and \(a^3+b^3=2\), so direct expansion gives \(\det B=-4\). But \(d=1\) requires \(\det B=C^4=16\), a contradiction. Also \(c=-1\) was already excluded by \(C\ne0\). The formulas follow by solving the linear equations for \(a^3,b^3\); their product equation gives \(R(c)=0\) after removing the nonzero factor \((c+1)^4/(c-1)^2\). Finally \(R\) has no common root with \(c^2+c+1\), so \(ab\ne0\) as required in the following computations.

Substitution shows \(E(a,b)^T=0\). As \(E\) has rank one, its row direction is precisely \((b,-a)\), which is therefore supplied by the absorption above. Attach one actual \(A\) to obtain

\[v_A=(b,-a)A=(b-a^2,b^2-ac).\]

Its binary contraction is

\[g=[v_{A,0}+av_{A,1},\;av_{A,0}+bv_{A,1},\;bv_{A,0}+cv_{A,1}].\]

On the displayed locus, each of \(g_1\), \(g_0g_2\), \(g_0g_2-g_1^2\), and \(g_0g_2+g_1^2\) is nonzero. To verify the four nonvanishing assertions, express each gauge-invariant quantity as a polynomial in \(ab,a^3,b^3,c\) and substitute the three displayed rational functions. For each resulting rational function, its numerator and denominator are relatively prime to \(R(c)\). Appendix A gives the resulting remainder table. This covers all choices of \(a,b\) with the prescribed cubes. The equation \(E(a,b)^T=0\) follows by the same substitution after multiplying its first and second coordinates by \(a^2\) and \(b^2\), respectively; these factors are nonzero.

Thus \(g\) is neither diagonal nor disequality, has rank two, and fails the necessary affine relation \(g_0g_2+g_1^2=0\). The complex binary dichotomy makes it \#P-hard. Only the one supplied row \(v_A\) is used in the final reduction. This completes the proposition.
\end{proof}

\section{13. Singular and common-direction boundaries at orders three and five}\label{singular-and-common-direction-boundaries-at-orders-three-and-five}

This section treats the orders-three-and-five cases not covered by Section 12 or the infinite-group lemma: either \(B\) is singular, or \(B\) is invertible and its group with \(A\) preserves a point or pair. The first task is column absorption; the second is to obtain a hard binary from the available row directions. At order five, a separate five-orbit argument is needed because a five-point orbit is not larger than the general bound of five easy directions. At order three, two explicit contractions suffice, with the remaining incompatible algebraic conditions certified in Appendix A. Sections 13.5--13.6 assemble these results without assuming unrelated unary types are jointly available.

\subsection{13.1 Statement and the exact previously established tools}\label{statement-and-the-exact-previously-established-tools}

Fix \(f=[1,a,b,c]\) with algebraic complex entries. Every input is a cubic bipartite multigraph, with \(f\) on the left and \(e=[1,0,0,1]\) on the right. There are no free unary signatures. Set

\[A=\begin{pmatrix}1&b\\a&c\end{pmatrix},\qquad
B=\begin{pmatrix}1+ab&a^2+bc\\a+b^2&ab+c^2\end{pmatrix},\]

\[D=\begin{pmatrix}1+2a^3+b^3&a^2+2ab^2+bc^2\\a+2a^2b+b^2c&a^3+2b^3+c^3\end{pmatrix}.\]

These matrices arise from the explicit bipartite graphs in Section 2. The present theorem concerns arbitrary cubic bipartite multigraphs, so there is no requirement that the dangling ports lie on the outer face of a planar drawing.

\begin{holantstatement}{Theorem 20}

Let \(f=[1,a,b,c]\), and let \(A,B,D\) be the actual matrices displayed above. Suppose \(A\) is invertible and has distinct eigenvalues whose ratio has multiplicative order three or five. Assume either:

\begin{enumerate}
\def\labelenumi{\arabic{enumi}.}
\tightlist
\item
  \(B\) is singular; or
\item
  \(B\) is invertible and \(\langle[A],[B]\rangle\le\operatorname{PGL}_2(\mathbb C)\) preserves a projective point or an unordered pair of projective points.
\end{enumerate}

Then \(\operatorname{Holant}(f\mid e)\) is in FP for generalized equality, \(a=b=0\), and is \#P-hard otherwise. The generalized-equality exception cannot occur in case 1.

\end{holantstatement}

We use the following established facts, giving the precise interfaces needed here.

\begin{itemize}
\item
  A rank-one straddled matrix \(uv\), actual or interpolated, supplies one fixed right unary row \(v\) when \(Q(u)=u_0^3+u_1^3\ne0\). For an instance with \(M\) copies of that row, degree counting gives \(3\mid M\); group the leftover columns in triples at equality and divide by \(Q(u)^{M/3}\). Multiple unrelated unary types are not granted. This is the absorption argument in Lemma P.3 in Section 3.2.
\item
  Contracting a row \(v=(r,s)\) against one edge of \(f\) gives

  \[F(v)=[r+as,ar+bs,br+cs].\]

  Since \(A\) is invertible, \(F(v)\) never has both endpoints zero for \(v\ne0\). The complex binary dichotomy says its tractable possibilities are rank one, diagonal, or one of twelve affine rays. These lie respectively on

  \[g_0g_2-g_1^2=0,\qquad g_1=0,\qquad g_0g_2+g_1^2=0.\]

  For \(a,b\) not both zero, there are at most five projective rows giving tractable binaries: at most two zeros of either nonzero homogeneous quadratic and at most one zero of the nonzero linear form. The rank-one quadratic has \(rs\) coefficient \(c-ab\ne0\). The affine quadratic has coefficients \(b+a^2,c+3ab,ac+b^2\); if all vanished, they would force \(a=b=c=0\), contradicting invertibility of \(A\). Thus neither quadratic vanishes identically. This is Lemma 4 in Section 4.4; its external input is \href{https://arxiv.org/pdf/1001.0464}{Kowalczyk--Cai's complex binary classification}.
\item
  An infinite projective group generated by invertible actual gadgets, with neither an invariant point nor an invariant pair, proves hardness by extracting a rank-one matrix and moving its column and row along infinite orbits. This is Lemma 10 in Section 7.2, including its proof that positive-word orbits suffice.
\item
  If \(A,B,D\) share a column line, invertible \(A\) and non-equality \(f\) imply hardness. The same holds when \(AB=BA\). The classification behind this statement is Lemma 14 in Section 9.1: up to an equality-preserving cube-root gauge, the common-line cases are \([1,t,t,1]\) and \([1,t,-1-2t,2+3t]\). Their hardness is proved there using the reversal-symmetric theorem and a direct perfect-matching reduction. Invertibility removes their degenerate endpoints.
\item
  If \(\det(AB-BA)=0\) and both eigenrows of a finite-order \(A\) are individually available, one gives a hard binary. This is Lemma 15 in Section 9.2. It needs no lower bound on that finite order.
\end{itemize}

Write \(C=1+c\) and \(h=\mathcal I(A)=C^2/(c-ab)\). In the order-three case \(h=1\); in the order-five case \(h^2-3h+1=0\). In both cases \(C\ne0\) and

\[\operatorname{tr}B=C^2\left(1-\frac2h\right)\ne0.\]

Thus a singular \(B\) is a nonzero rank-one matrix. The exclusions \(h\ne0,2\) are automatic in these two orders.

\subsection{13.2 Two elementary column facts}\label{two-elementary-column-facts}

\begin{holantstatement}{Lemma 20.1 (three-column absorption)}

Let \(f=[1,a,b,c]\), \(A=A_f\), and \(Q(u)=u_0^3+u_1^3\). Suppose \(A\) has projective order three, \(f\) is not generalized equality, and \(u\) is not a column eigenvector of \(A\). At least one of \(u,Au,A^2u\) has nonzero \(Q\).

\end{holantstatement}

\begin{proof}[Proof]

The three columns have distinct projective directions. If all had zero \(Q\), they would be exactly the three lines with slope \(u_1/u_0\in\{-1,-\omega,-\omega^2\}\), where \(\omega^3=1\) and \(\omega\ne1\). The action of \(A\) cyclically permutes those three points. A projective transformation is determined by its action on three points; the two possible three-cycles here are multiplication of the slope by \(\omega\) or \(\omega^2\). Their matrices are diagonal. Hence \(A\) would be diagonal, giving \(a=b=0\), a contradiction.
\end{proof}

For projective order five, an even simpler assertion holds: a non-eigen column has five directions, and the cubic \(Q\) has only three projective zeros. At least one is absorbable.

\begin{holantstatement}{Lemma 20.2 (two forbidden eigenlines cannot occur in the common-line branch)}

Let \(f=[1,a,b,c]\), \(A=A_f\), \(B=B_f\), and \(Q(u)=u_0^3+u_1^3\). Suppose \(A\) is invertible, has distinct eigenvalues of finite projective order, \(f\) is not generalized equality, and \(\det(AB-BA)=0\). Then its two column eigenlines cannot both have zero \(Q\).

\end{holantstatement}

\begin{proof}[Proof]

If both did, the dual row eigendirections would have two distinct cube-root-of-unity slopes. A cube-root gauge and choice of ordering put them at \(1,\omega\), with \(\omega^2+\omega+1=0\). Consequently

\[b=-a\omega,\qquad c=1+a(1+\omega),\qquad a\ne0.\]

Direct substitution gives

\[\det(AB-BA)=a^4(-7a^2+a\omega+\omega+1).\]

Its two possible nonzero parameter values are

\[a=\frac1{1-2\omega},\qquad a=\frac{\omega^2}{1-2\omega^2}.\]

The two eigenvalues of \(A\) are \(1+a\) and \(1+a\omega\). Their ratios at these values are respectively \(2\) and \(1/2\), neither a root of unity. This contradicts finite projective order.
\end{proof}

\begin{holantstatement}{Corollary 20.3 (absorption from a singular second matrix)}

Let \(f=[1,a,b,c]\) and let \(A=A_f\), \(B=B_f\), \(D=A_{\Phi(f)}\). Suppose that \(A\) has projective order three or five, that \(B=uv\) is a nonzero rank-one matrix, that \(AB\ne BA\), and that \(A,B,D\) have no common invariant column line. Then an actual positive word \(W\) in \(A,D\) gives \(Q(Wu)\ne0\), and the row \(v\) is available.

\end{holantstatement}

\begin{proof}[Proof]

If \(u\) is not an \(A\) eigenvector, use Lemma 20.1 or the five-direction count. If \(u\) is an eigenvector and \(Q(u)\ne0\), use \(W=I\). Otherwise let \(L=\mathbb C u\). Since \(B\) maps into \(L\), both \(A\) and \(B\) preserve it. In a basis beginning with \(L\), both matrices are triangular, so their commutator has zero determinant. The excluded common-line case means \(D(L)\not\subseteq L\), so \(Du\ne0\) lies on another line. If it is not an eigenline, an \(A\) power makes it absorbable. If it is the other eigenline, Lemma 20.2 says its \(Q\) is nonzero. Thus one of \(B,A^jB,DB,A^jDB\) supplies the original row \(v\), always with a nonzero fixed column contraction.
\end{proof}

\subsection{13.3 An order-five orbit always reaches a hard binary}\label{an-order-five-orbit-always-reaches-a-hard-binary}

\begin{holantstatement}{Lemma 20.4 (five-orbit lemma)}

Let \(f=[1,a,b,c]\), \(A=A_f\), and \(F=F_f\). Suppose \(A\) has projective order five and \(f\) is not generalized equality. For every non-eigen row \(v\), at least one of

\[F(v),F(vA),F(vA^2),F(vA^3),F(vA^4)\]

is a hard complex binary signature.

\end{holantstatement}

This assertion does not assume singularity of \(B\).

\begin{proof}[Proof]

We first reduce the possible binary types to two patterns. The five row directions are distinct. Assume all five outputs are tractable. Their endpoints are not simultaneously zero. The two rank-one directions, one diagonal direction, and two affine-conic directions must therefore account for exactly five different points; there can be no overlap. In particular there is exactly one diagonal direction. Reindex the orbit so it occurs at \(v_0\). Then \(v_0\) is proportional to \((b,-a)\).

In an \(A\) eigenbasis parameterize the row line by \(v(z)=w_1+zw_2\) so that the orbit occurs at \(z=1,\zeta,\zeta^2,\zeta^3,\zeta^4\), where \(\zeta\) is a primitive fifth root and the diagonal direction is \(z=1\). Put

\[R(z)=F(v(z))_0F(v(z))_2-F(v(z))_1^2,\] \[S(z)=F(v(z))_0F(v(z))_2+F(v(z))_1^2.\]

Both are degree-two polynomials with two roots among the four nonidentity fifth roots. The middle coordinate is a nonzero multiple of \(z-1\), so \(S-R\) is a nonzero multiple of \((z-1)^2\). Thus \(S(1)=R(1)\ne0\) and \(S'(1)=R'(1)\). Their logarithmic derivatives at one agree.

If \(u,v\) are the two roots of \(R\), this equality and

\[\sum_{j=1}^4\frac1{1-\zeta^j}=2\]

give \(1/(1-u)+1/(1-v)=1\). Multiplication by \((1-u)(1-v)\) yields \(uv=1\). Hence the rank-one pair is either the directions indexed \(\pm1\) or those indexed \(\pm2\). The affine-conic pair is the other pair. These are the only two patterns.

\textbf{Elimination of the two patterns.} Write

\[x=ab,\qquad p=a^3,\qquad q=b^3,\qquad pq=x^3,\]

\[h^2-3h+1=0,\qquad x=c-\frac{(1+c)^2}{h},\qquad
T=c^2+(2-h)c+1.\]

Thus \(x=-T/h\). For any row \(w\), abbreviate the binary tests by

\[\mathcal R(w)=F(w)_0F(w)_2-F(w)_1^2,\qquad
\mathcal C(w)=F(w)_0F(w)_2+F(w)_1^2.\]

Set \(v_0=(b,-a)\) and \(A^\#=\operatorname{adj}(A)\). Powers of \(A^\#\) have the same projective action as negative powers of \(A\). In the first pattern the four zero equations are

\[\mathcal R(v_0A)=\mathcal R(v_0A^\#)=0,\qquad
\mathcal C(v_0A^2)=\mathcal C(v_0(A^\#)^2)=0.\]

In the other pattern interchange \(\mathcal R\) and \(\mathcal C\). Denote these four polynomials, after multiplication respectively by \(h^3,h^3,h^4,h^4\), by \(E_1,E_2,E_3,E_4\). Rewrite their monomials using \(ab=x\), \(a^3=p\), \(b^3=q\), substitute the displayed value of \(x\), and use \(pq=x^3\). These are exact polynomial operations; multiplying by the powers of \(h\ne0\) clears the denominators.

For \textbf{both} patterns, the equations

\[(E_1-E_2)/(c+1)^2=0,\qquad
(E_3+h(c+1)^2E_1)/(c+1)^4=0\]

are the same two linear equations in \(p,q\):

\[\begin{pmatrix}
(5h-2)c-(3h-1)&-(3h-1)c+(5h-2)\\
(16h-6)c-(13h-5)&-(13h-5)c+(16h-6)
\end{pmatrix}\binom pq=\binom UV,\]

where

\[U=2\bigl(h(c^4+1)+(2-2h)(c^3+c)+(2h+1)c^2\bigr),\] \[V=(7h-3)(c^4+1)+(3-12h)(c^3+c)+(20h-10)c^2.\]

The determinant is \(-(c-1)(c+1)(34h-13)\). Since \(h\) is irrational and \(c+1\ne0\), it is nonzero when \(c\ne1\). Cramer's rule then gives

\[p=T((h-2)c+2h-5),\qquad
q=T((2h-5)c+h-2).\]

Substitution into \(pq=x^3\) gives

\[0=-(c+1)^2(11h-29)T^2.\]

The nonzero factors force \(T=0\), and then \(p=q=0\), or \(a=b=0\). This is the excluded generalized-equality case.

It remains \(c=1\). The first linear equation gives \(p+q=20h-50\). The equation \(E_2=0\) gives, respectively in the two patterns,

\[p^2+q^2=296h-774\quad\text{or}\quad p^2+q^2=1224h-3206.\]

But \(pq=x^3=(4h-11)^3\) and the known sum force

\[p^2+q^2=(20h-50)^2-2(4h-11)^3=4090-1560h.\]

The two possible differences are \(64(29h-76)\) and \(96(29h-76)\), both nonzero because \(h\) is irrational. Thus \(c=1\) is impossible as well. The assumption of five tractable outputs is contradicted in every case.
\end{proof}

The elimination uses only the displayed two-by-two linear system and polynomial substitution modulo \(h^2-3h+1\); it does not require a Gröbner basis. Appendix A records the exact identity verification.

\subsection{13.4 At order three, a singular B needs only two contractions}\label{at-order-three-a-singular-b-needs-only-two-contractions}

\begin{figure}[H]
\centering

\begin{tikzpicture}[x=1cm,y=1cm]
\node[L] (f) at (0,0) {$f$};\node[R,dashed] (v) at (3,0) {$v$};
\draw (-1.7,0.9)--node[above,lab] {$x$}(f.north west);
\draw (-1.7,-0.9)--node[below,lab] {$y$}(f.south west);
\draw (f.east)--node[above,lab] {$z$}(v.west);
\node[align=center] at (1,-1.7) {$F(v)(x,y)=\displaystyle\sum_{z=0}^1 f(x,y,z)v_z$};
\end{tikzpicture}

\caption{Attach one simulated RHS unary to a left ternary vertex. Its two other ports remain free, giving a binary signature. The dashed unary must later be eliminated by its established Turing reduction; it is not assumed available in the original input.}
\label{fig:contraction-16}
\end{figure}
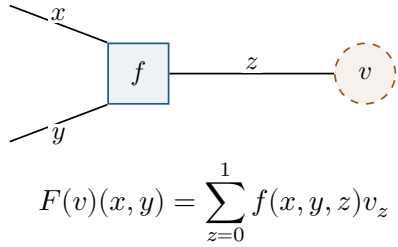

\begin{holantstatement}{Lemma 20.5}

Let \(f=[1,a,b,c]\), \(A=A_f\), \(B=B_f\), and \(F=F_f\). Suppose \(A\) has projective order three and \(B\) is singular. For any nonzero row \(v\) in the row space of \(B\), at least one of \(F(v)\) and \(F(vA)\) is a hard complex binary signature.

\end{holantstatement}

\begin{proof}[Proof]

First suppose \(a\ne0\) and the first coordinate of \(v\) is nonzero. Scale \(v=(1,t)\) and set

\[p=a^3\ne0,\qquad Y=at,\qquad x=ab=-c^2-c-1.\]

The two rows of \(B\) are proportional to \(v\) exactly when

\[p+xc-Y(1+x)=0,\qquad Y(p+x^2)-p(x+c^2)=0.\]

Here \(c+1\ne0\). Introduce \(z\) with \(zp(c+1)-1=0\) to encode these nonzero assumptions.

Write \(vA^k=(r_k,s_k/a)\), where

\[(r_0,s_0)=(1,Y),\qquad
(r_{k+1},s_{k+1})=(r_k+s_k,xr_k+cs_k).\]

The corresponding binary is

\[F(vA^k)=\left[r_k+s_k,\frac{pr_k+xs_k}{a^2},\frac{xr_k+cs_k}{a}\right].\]

Let \(J_k=pr_k+xs_k\) and \(G_k=p(r_k+s_k)(xr_k+cs_k)\). Multiplying by the nonzero powers of \(a\) shows that its three possible tractable tests require

\[R_k=G_k-J_k^2=0,\qquad D_k=J_k=0,\qquad C_k=G_k+J_k^2=0.\]

The third equation is only a necessary condition for the twelve affine rays. Using this larger set strengthens the exclusion.

For each of the nine pairs \((T_0,T_1)\in\{R,D,C\}^2\), form the five generators

\[p+xc-Y(1+x),\quad Y(p+x^2)-p(x+c^2),\quad zp(c+1)-1,\quad T_0,\quad T_1,\]

with \(x=-c^2-c-1\). For each pair there are rational polynomials \(H_1,\ldots,H_5\) such that \(\sum_{j=1}^5H_jG_j=1\), where \(G_1,\ldots,G_5\) are these five generators in their displayed order. The nine coefficient arrays are the supplementary certificates described in Appendix A; they contain 3,220 nonzero multiplier terms in total. The appendix specifies the monomial encoding and a verifier using rational addition and multiplication only. Evaluating such an identity at a hypothetical common zero gives \(0=1\). Hence no pair of tests can hold, and the two binaries cannot both be tractable.

\textbf{Case \(a=0\).} Order three implies \(c^2+c+1=0\), and singularity of \(B\) gives \(b^3=c\). In particular \(b\ne0\). Here

\[B=\binom1{b^2}(1,bc),\qquad Q\!\left(\binom1{b^2}\right)=1+b^6=-c\ne0.\]

The row \((1,bc)\) gives

\[g=[1,b^2c,-bc].\]

Its middle entry and both endpoints are nonzero. Its determinant is \(bc^2\ne0\), and its affine-conic expression is \(b(1-c)\ne0\). Thus \(g\) itself is hard.

\textbf{The row at infinity.} If \(v=(0,1)\), the first column of \(B\) vanishes. Hence \(ab=-1\), \(a+b^2=0\), and the order-three equation gives \(c=0\). Thus \(b^3=1\), and a cube-root gauge reduces the signature to \([1,-1,1,0]\). Its actual matrices satisfy

\[B=\begin{pmatrix}0&1\\0&-1\end{pmatrix},\qquad
AB=\begin{pmatrix}0&0\\0&-1\end{pmatrix}.\]

The latter has an absorbable column and supplies \((0,1)\). Its contraction \([-1,1,0]\) is hard. These charts cover every nonzero row of \(B\), proving the lemma.
\end{proof}

\subsection{13.5 Closing all singular-B cases at the two orders}\label{closing-all-singular-b-cases-at-the-two-orders}

\begin{proof}[Proof of Theorem 20 when B is singular]

Generalized equality has invertible \(B=\operatorname{diag}(1,c^2)\), so it cannot occur here. Remove the already hard common-line and commuting cases listed in Section 13.1. Factor \(B=uv\).

At order three, Corollary 20.3 supplies the row \(v\). Lemma 20.5 gives a hard contraction from \(v\) or \(vA\). Attach that fixed actual chain and use only the chosen unary type in the binary reduction.

Now suppose the order is five. Again Corollary 20.3 supplies \(v\). If \(v\) is not an \(A\) eigenrow, Lemma 20.4 selects a hard contraction among its five powers and finishes the proof.

If \(v\) is an eigenrow, examine \(vD\). Since \(B=uv\), the row \(v\) is also a \(B\) eigenrow, with eigenvalue \(vu\). Consequently \(A\) and \(B\) preserve the column line \(\ker v\) and \(\det(AB-BA)=0\). If \(vD\) is zero or proportional to \(v\), then \(A,B,D\) share that row eigenline and hence its annihilator column line, an already removed hard case. Otherwise \(vD\) is available by attaching the actual \(D\) gadget to the supplied row. If it is not an eigenrow, Lemma 20.4 applies to its five \(A\) images. If it is the other eigenrow, both eigenrows are individually available, and the finite-order two-eigenrow lemma stated in Section 13.1 applies with the just-verified commutator hypothesis and supplies a hard binary.

Every branch uses the one original rank-one matrix and its actual attached words. The selected hard row is fixed before processing an input graph. Thus no simultaneous availability of unrelated unaries is assumed.
\end{proof}

\subsection{13.6 Closing invertible B with an invariant point or pair}\label{closing-invertible-b-with-an-invariant-point-or-pair}

\begin{proof}[Proof of Theorem 20 when B is invertible]

The generalized-equality case is tractable; exclude it. If \(AB=BA\), the existing commuting-case classification proves hardness. Thus assume \(AB\ne BA\) and let \(G_0=\langle[A],[B]\rangle\).

An invariant unordered pair for \(G_0\) would have to be the two eigenlines of \(A\). Indeed, a matrix interchanging two lines is projectively an involution, whereas the order of \(A\) is three or five. In this eigenbasis, a matrix preserving the pair is diagonal or off-diagonal. Since \(\operatorname{tr}B\ne0\), \(B\) cannot be off-diagonal; it would be diagonal and commute with \(A\), a contradiction. Thus the noncommuting case has no invariant pair.

By the theorem's hypothesis it must have a common invariant column line \(L\). In an \(A\) eigenbasis, \(B\) is triangular, with nonzero diagonal entries and exactly one nonzero off-diagonal entry. The group commutator of \(A\) and \(B\) is a nonidentity unipotent transformation. Hence \(G_0\) is infinite, its unique common fixed point is \(L\), and it has no finite orbit other than \(L\).

If \(D(L)\subseteq L\), the common-line classification of \(A,B,D\) proves hardness. Otherwise \(D(L)\not\subseteq L\).

If \(D\) is invertible, \(\langle[A],[B],[D]\rangle\) remains infinite, has no common invariant point, and has no invariant pair because its subgroup \(G_0\) has none. The established orbit-and-absorption lemma gives hardness.

If \(D\) is singular, it is nonzero and factors as \(D=uw\). The condition \(D(L)\not\subseteq L\) says that \(u\notin L\) and \(w\) does not annihilate \(L\). Thus the column orbit of \(u\) and the row orbit of \(w\) under \(G_0\) are infinite. Their positive-word orbits are also infinite: if one were finite, every invertible generator would permute it, and so its inverse would preserve it as well, making the full group orbit finite.

Choose a positive word \(W\) in \(A,B\) with \(Q(Wu)\ne0\). The actual matrix \(WD=(Wu)w\) supplies \(w\). Choose six positive-word row images \(wV_j\) with distinct directions. At most five give tractable binaries, so one gives a hard binary. Only this one row is used in the final reduction. This proves hardness and completes Theorem 20.
\end{proof}

\section{14. Zero endpoints and complete coverage}\label{zero-endpoints-and-complete-coverage}

All normalized branches have now been classified. We first handle \(f=[0,x,y,0]\), the signatures whose endpoints both vanish. When \(xy\ne0\), the actual replacement \(\Phi(f)\) has a nonzero endpoint and can therefore be tested against the classifications already proved in Sections 4--13. This is the direction of the dependency: the endpoint argument invokes those completed branches, not Theorem A itself. The proof treats the possible complex zeros of \(2t^6+t^3+2\) explicitly. Section 14.1 then combines the endpoint result with all normalized branches, and Section 14.2 records the exhaustive coverage.

\begin{holantstatement}{Proposition 21 (both endpoints zero)}

Let \(x,y\) be fixed complex algebraic numbers. Then

\[\operatorname{Holant}([0,x,y,0]\mid e)\]

is in FP when \(x=y=0\) and is \#P-hard otherwise.

\end{holantstatement}

\begin{proof}[Proof]

The zero case is immediate. If exactly one of \(x,y\) is nonzero, scaling and reversal give Exact-One \([0,1,0,0]\). Its hardness in this cubic bipartite equality model is Fan--Cai \cite[Lemma 6.1]{ref4}.

Suppose \(xy\ne0\). Scale to \(f=[0,1,t,0]\), where \(t\ne0\). The actual ternary construction gives

\[g=\Phi(f)=[2+t^3,\ 2t,\ 2t^2,\ 1+2t^3].\]

Put \(u=t^3\). Its endpoints cannot both vanish, since this would require \(u=-2\) and \(u=-1/2\) simultaneously. Its middle entries are nonzero, so \(g\) is not generalized equality. Moreover

\[\det A_g=2u^2+u+2.\]

If this determinant is nonzero, reverse and normalize \(g\) if needed. Every invertible-first-matrix branch has been proved above: repeated eigenvalue, infinite projective order, and all finite orders. Their only easy case is generalized equality, already excluded. Thus \(g\) is hard.

If the determinant vanishes, \(u\notin\{-2,-1/2\}\), so both endpoints are nonzero. The signature is not rank one, because

\[g_0g_2-g_1^2=2t^5\ne0.\]

Let \(r=g_3/g_0=(1+2u)/(2+u)\). Solving for \(u\) and substituting in its quadratic gives

\[8r^2-11r+8=0.\]

This excludes \(r=0,1,-1,i,-i\) by direct substitution. The singular-\(A\) theorem's non-rank-one easy list consists precisely of cube-root gauges of the six normalized affine representatives, whose endpoint ratios belong to this excluded set. Thus \(g\) is hard in the singular case as well.

Replacing every left \(g\) vertex by the actual gadget \(\Phi(f)\) gives \(\operatorname{Holant}(g\mid e)\le_T\operatorname{Holant}(f\mid e)\). Hence \(f\) is hard in all nonzero cases.
\end{proof}

\subsection{14.1 Proof of Theorem A}\label{proof-of-theorem-a}

\begin{proof}[Proof]

Proposition P.1 gives independent algorithms for every signature in \(\mathcal T\). Suppose now \(f\notin\mathcal T\).

If both endpoints vanish, Proposition 21 applies. Otherwise reverse if necessary and divide by a nonzero endpoint to obtain \(f=[1,a,b,c]\). These operations preserve membership in \(\mathcal T\) and preserve complexity up to a known nonzero scalar factor on each instance.

If \(A_f\) is singular, Theorem 8 gives precisely the rank-one and affine easy exceptions already contained in \(\mathcal T\), so the present signature is hard. Suppose \(A_f\) is invertible.

If it has a repeated eigenvalue, Theorem 1 gives hardness unless it is generalized equality. If it has two distinct eigenvalues and their ratio is not a root of unity, Theorem 9 gives the same conclusion. These easy exceptions were excluded by \(f\notin\mathcal T\).

The remaining ratio is a root of unity of finite order \(n\ge2\). Order two is Theorem 17, order four is Theorem 18, and every order at least six is Theorem 16. Consider orders three and five. A singular \(B\) is covered by Theorem 20. For invertible \(B\), a group preserving a point or a pair is also covered by that theorem. If the group preserves neither and is infinite, Lemma 10 applies. If it is finite, Proposition 19 applies. Thus every possible case is hard.

The reductions use actual cubic bipartite gadgets or the explicitly proved one-type interpolation and absorption procedures. The arithmetic discussion in Section 3.5 makes all of them polynomial-time for fixed algebraic weights. This proves the dichotomy.
\end{proof}

\subsection{14.2 Coverage table}\label{coverage-table}

{\def\LTcaptype{table} 
\begin{longtable}[]{@{}
  >{\raggedright\arraybackslash}p{(\linewidth - 2\tabcolsep) * \real{0.5000}}
  >{\raggedright\arraybackslash}p{(\linewidth - 2\tabcolsep) * \real{0.5000}}@{}}
\toprule\noalign{}
\begin{minipage}[b]{\linewidth}\raggedright
Input branch after permitted normalization
\end{minipage} & \begin{minipage}[b]{\linewidth}\raggedright
Result used
\end{minipage} \\
\midrule\noalign{}
\endhead
\bottomrule\noalign{}
\endlastfoot
\(f_0=f_3=0\) & Proposition 21, proved without real-positivity assumptions \\
Singular \(A_f\) & Theorem 8 \\
Repeated eigenvalues & Theorem 1 \\
Distinct eigenvalues, non-root-of-unity ratio & Theorem 9 \\
Finite projective order 2 & Theorem 17 \\
Finite projective order 4 & Theorem 18 \\
Finite projective order at least 6 & Theorem 16 \\
Order 3 or 5, singular \(B\) & Theorem 20 \\
Order 3 or 5, invertible \(B\), invariant point or pair & Theorem 20 \\
Order 3 or 5, infinite group with neither & Lemma 10 \\
Order 3 or 5, finite group with neither & Proposition 19 \\
\end{longtable}
}

\section{15. The role of the strict degree restriction}\label{the-role-of-the-strict-degree-restriction}

The dichotomy is proved in Section 14. This final section explains its relation to broader classifications. We first record the port-count identity obeyed by every actual cubic bipartite gadget. We then give two examples in which allowing unrestricted signature placements or unrestricted edge orderings changes tractability. These examples show why the corresponding published hardness results cannot simply be restricted to the present input graphs. The reusable reductions and their required auxiliary signatures were specified in Section 1.4.

Theorem A completes the classification with the graph restrictions retained in every final oracle call. The interpolation constants, gadget words, and algebraic extensions depend on the fixed signature, while the number and size of oracle calls are polynomial in the input graph. The tractable side follows from component factorization and affine quadratic sums. The hardness side uses the complex binary dichotomy \cite{ref7}, fixed-degree perfect-matching hardness \cite{ref18}, and the Exact-One base \cite{ref4}. The group-theoretic inputs are the finite subgroup classification \cite{ref16} and the finitely generated torsion theorem, for which \cite{ref17} gives a characteristic-zero consequence of Selberg's lemma.

The distinction between an actual graph and an interpolated signature cannot be removed from the proof. If a gadget has \(L\) left ternary vertices, \(R\) right ternary vertices, \(m\) internal edges, and \(d_L,d_R\) dangling edges on the respective sides, then

\[3L=m+d_L,\qquad 3R=m+d_R,\qquad d_L-d_R=3(L-R).\]

In particular, no actual gadget over the original signatures has two dangling edges on one side and one on the other. This explains why a reduction using such a projector in another Holant model does not directly give a gadget in this model. Interpolation and triple absorption are compatible with this degree identity: the former replaces whole straddled boxes by chains, and the latter adds one equality vertex per three unused columns.

Huang--Lu's degree-divisible theorem \cite{ref8} and the later general treatment \cite{ref9} place the tractable diagonal transformations in a broader classification. The present theorem establishes hardness on the smaller exactly-three-occurrence domain. The following example also separates that domain from ordinary Holant, even for the same two signatures. It is useful when comparing Theorem A with ordinary classifications \cite{ref12,ref13,ref19}.

\begin{holantstatement}{Example 22 (why ordinary Holant hardness is not enough)}

Let \(\omega\) be a primitive cube root of unity, and set

\[q=[1,1,-1,-1],\qquad h=\Gamma_\omega q=[1,\omega,-\omega^2,-1].\]

Strict \(\operatorname{Holant}(h\mid e)\) is in FP. Ordinary \(\operatorname{Holant}(\{h,e\})\) is \#P-hard, with the usual native wires and unrestricted placements of the two signatures.

\end{holantstatement}

\begin{proof}[Proof]

Every satisfying assignment in a strict instance has a multiple of three one-valued incidences at the left vertices, because each right equality contributes either zero or three. Therefore the gauge factor is one. The strict value equals the value for \(q\), which has the affine phase

\[q(x,y,z)=(-1)^{xy+xz+yz}.\]

For the ordinary problem, ternary equality trees implement a variable of any degree at least three. A native wire implements degree two, and a self-loop on ternary equality gives the degree-one signature \([1,1]\). Consequently \(\#\operatorname{CSP}(h)\) reduces to ordinary \(\operatorname{Holant}(\{h,e\})\). The complex Boolean \#CSP dichotomy \cite{ref10} makes this source problem hard: \(h\) has full support, so product type would force a product of three unaries, whereas

\[h_0h_2=-\omega^2\ne\omega^2=h_1^2.\]

It is not affine either, since the ratio \(h(1,0,0)/h(0,0,0)=\omega\) is not a fourth root of unity. This proves the asserted hardness.
\end{proof}

The distinction between a fixed bipartite orientation and an arbitrary edge ordering can be seen without any exceptional gadget calculations.

\begin{holantstatement}{Example 23 (a restriction of an asymmetric spin problem)}

Let \(\omega\) be a primitive cube root of unity, and let

\[W=\begin{pmatrix}\omega&\omega\\1&-1\end{pmatrix}.\]

For a cubic bipartite multigraph \(G=(L,R,E)\) with every edge ordered from \(L\) to \(R\), define

\[Z_W^{L\to R}(G)=\sum_{\sigma:L\cup R\to\{0,1\}}
\prod_{uv\in E,\ u\in L,\ v\in R}W_{\sigma(u),\sigma(v)}.\]

This restricted problem is in FP, on both general and planar inputs. The spin problem with the same \(W\) on cubic graphs with arbitrary edge orderings is \#P-hard, even on planar inputs.

\end{holantstatement}

\begin{proof}[Proof]

Fix the spins at the right vertices and sum independently over the spin at each left vertex. If the three adjacent right spins contain \(j\) ones, the resulting left weight is

\[q_j=\omega^{3-j}\omega^j+1^{3-j}(-1)^j
=1+(-1)^j.\]

Thus \(q=[2,0,2,0]\), and \(Z_W^{L\to R}(G)\) is exactly the strict value for \(\operatorname{Holant}(q\mid e)\). Each left vertex contributes two when the adjacent bits have even parity and zero otherwise. The value is \(2^{|L|}\) times the number of solutions of a homogeneous linear system over \(\mathbb F_2\), hence is computable by Gaussian elimination. Repeated neighbors, if any, are counted with their edge multiplicities in the parity equations.

For the unrestricted spin problem, apply the cubic classification \cite[Theorem 2.4]{ref21}, which recalls \cite{ref11}. Write \((w,x,y,z)=(\omega,\omega,1,-1)\) for the four entries of \(W\). All entries are nonzero and \(wz-xy=-2\omega\ne0\), so none of the product-type cases applies. An affine case would require a common sign \(\epsilon\in\{1,-1\}\) with \(w^6=\epsilon z^6\) and \(x^2=\epsilon y^2\). The first equation forces \(\epsilon=1\), while the second would require \(\omega^2=1\). The additional planar case requires \(x=\epsilon y\), which would give \(\omega=\pm1\). Both are impossible. This proves the asserted hardness and shows why a hardness result allowing arbitrary edge orderings cannot simply be restricted to the fixed bipartite ordering.
\end{proof}

\section{Appendix A. Exact algebraic certificates}\label{appendix-a.-exact-algebraic-certificates}

The finite algebraic calculations in Sections 11--13 are part of the proof. This appendix specifies their inputs and verification procedures. All computations use rational coefficients or the fixed quadratic field \(\mathbb Q(\sqrt5)\); none of the assertions is inferred from a numerical approximation. Supplementary coefficient arrays encode polynomial identities rather than the output of a black-box satisfiability test.

\subsection{A.1 Trace identities and the finite-group table}\label{a.1-trace-identities-and-the-finite-group-table}

The five trace and determinant identities in Section 12.1 have denominators that are powers of \(C=1+c\) and \(\det A\). Multiply each identity by its displayed denominator, insert

\[A=\begin{pmatrix}1&b\\a&c\end{pmatrix},\qquad
B=\begin{pmatrix}1+ab&a^2+bc\\a+b^2&ab+c^2\end{pmatrix},\]

and the displayed polynomial matrices \(D,E\). Expansion in the ring \(\mathbb Q[a,b,c]\) gives zero coefficient for every monomial. This proves the identities under precisely the nonzero-denominator assumptions used in the text.

Here is a complete algorithm for the table of Section 12.2. Represent an element of \(K=\mathbb Q(\sqrt5)\) by the ordered pair \((r,s)\) for \(r+s\sqrt5\), with \(r,s\in\mathbb Q\). Multiplication uses \((\sqrt5)^2=5\), and inversion of a nonzero pair uses

\[\frac1{r+s\sqrt5}=\frac{r-s\sqrt5}{r^2-5s^2}.\]

For each triple \((s,t,u)\) in the two finite rows of the table, form the following polynomials in an indeterminate \(X\):

\[d=(1-2/s)^2/t,\qquad P(X)=X^2-ud/s,\]

\[D_0(X)=2X^2+(-2+8/s)X+(2-2/s)d-2/s+4/s^2+3/s^3,\]

\[D_v(X)=(3X-2+6/s)^2-vD_0(X),\qquad v\in\Omega.\]

Compute the monic greatest common divisors \(\gcd(P,D_0)\) and \(\gcd(P,D_v)\) by Euclidean division in \(K[X]\). There are \(20+12+12=44\) triples. All 44 gcds of the first kind are one. Of the 264 gcds of the second kind, exactly the six listed in Section 12.2 have degree one; their roots give the displayed values of \(L\). These equations, the finite input sets, and the six outputs specify the calculation without any choices of branch for square roots of \(ud/s\). The supplementary verifier implements these operations and records every triple.

The order-four resultant table in Section 11.4 is a separate rational calculation. It is obtained from the sixteen pairs of displayed polynomials \(P_j,Q_k\) by the determinant of the Sylvester matrix. Since every resulting determinant is nonzero, no common zero exists. This implication holds over all of \(\mathbb C\).

\subsection{A.2 The order-three rank-one residual}\label{a.2-the-order-three-rank-one-residual}

In the last case of Section 12.4, put

\[R(c)=3c^4+5c^3+9c^2+5c+3,\qquad X=-c^2-c-1,\]

\[P=\frac{(c^2+2c+2)(2c^2+c+1)}{c-1},\qquad
Q=-\frac{(c^2+c+2)(2c^2+2c+1)}{c-1}.\]

Every gauge-invariant monomial \(a^ib^jc^k\) has \(i+2j\equiv0\pmod3\). Remove \((ab)^{\min(i,j)}\); the remaining exponent of \(a\) or \(b\) is divisible by three. Replacing \(ab,a^3,b^3\) by \(X,P,Q\) therefore expresses the monomial as a rational function of \(c\). Apply this procedure to \(a^2(E(a,b)^T)_0\) and \(b^2(E(a,b)^T)_1\). Their numerators have zero remainder modulo \(R\). Since \(ab\ne0\), this proves that \((a,b)^T\) belongs to the kernel of \(E\).

For the row \(v_A=(b-a^2,b^2-ac)\), put

\[g=F_f(v_A),\qquad
U_1=g_1,\quad U_2=g_0g_2,\quad
U_3=g_0g_2-g_1^2,\quad U_4=g_0g_2+g_1^2.\]

The same monomial reduction gives the following explicit rational functions, with \(C=c+1\):

\[U_1=-\frac{C^2(3c^2+4c+3)}{c-1},\qquad U_2=C^4(c^2+c+1),\]

\[U_3=-\frac{C^4(2c^4+15c^3+16c^2+15c+2)}{(c-1)^2},\]

\[U_4=\frac{C^4(4c^4+13c^3+16c^2+13c+4)}{(c-1)^2}.\]

Every possible zero or pole is excluded by the following rational resultant calculation:

{\def\LTcaptype{table} 
\begin{longtable}[]{@{}ll@{}}
\toprule\noalign{}
Polynomial \(S(c)\) & \(\operatorname{Res}_c(R,S)\) \\
\midrule\noalign{}
\endhead
\bottomrule\noalign{}
\endlastfoot
\(c+1\) & \(5\) \\
\(c-1\) & \(25\) \\
\(3c^2+4c+3\) & \(225\) \\
\(c^2+c+1\) & \(1\) \\
\(2c^4+15c^3+16c^2+15c+2\) & \(140625\) \\
\(4c^4+13c^3+16c^2+13c+4\) & \(15625\) \\
\end{longtable}
}

All six resultants are nonzero. Therefore every \(U_j\) is defined and nonzero at every root of \(R\), exactly as required in the binary hardness test. This calculation uses the invariant quantities \(U_j\), so it also covers every cube-root choice of \(a\) and \(b\).

\subsection{A.3 Unit certificates for the singular order-three branch}\label{a.3-unit-certificates-for-the-singular-order-three-branch}

We spell out why the certificate in Lemma 20.5 proves the asserted exclusion. Work in \(\mathbb Q[z,Y,p,c]\) and set

\[x=-c^2-c-1,\]

\[G_1=p+xc-Y(1+x),\qquad
G_2=Y(p+x^2)-p(x+c^2),\qquad
G_3=zp(c+1)-1.\]

For \(k=0,1\), define \((r_0,s_0)=(1,Y)\) and

\[(r_{k+1},s_{k+1})=(r_k+s_k,xr_k+cs_k),\]

\[J_k=pr_k+xs_k,\qquad W_k=p(r_k+s_k)(xr_k+cs_k),\]

\[R_k=W_k-J_k^2,\qquad D_k=J_k,\qquad C_k=W_k+J_k^2.\]

For each ordered pair \(\tau=(\tau_0,\tau_1)\in\{R,D,C\}^2\), let \(G_4=\tau_0{}_0\) and \(G_5=\tau_1{}_1\). The supplementary certificate provides five polynomials \(H_{\tau,1},\ldots,H_{\tau,5}\) with rational coefficients satisfying

\[\boxed{\sum_{j=1}^5H_{\tau,j}G_j=1.}\]

Each polynomial is encoded as a finite list of monomials. An entry consists of four nonnegative integer exponents \((e_z,e_Y,e_p,e_c)\) and a nonzero rational coefficient \(q\), and denotes \(qz^{e_z}Y^{e_Y}p^{e_p}c^{e_c}\). Repeated exponent tuples are forbidden. The coefficient field and ordered variable list are included in the data. The nine arrays contain 3,220 nonzero multiplier terms in total.

Verification consists of three operations: reconstruct \(G_1,\ldots,G_5\) from the formulas above; compare them coefficientwise with the generators recorded in the certificate; and multiply and sum the five products to check that the result is the constant polynomial one. The supplementary verifier uses sparse maps from exponent tuples to rational numbers. Its arithmetic requires only addition, multiplication, and exact reduction of rational fractions; it does not run a Gröbner-basis algorithm.

\begin{proof}[Proof of the certificate implication]

Suppose the two contractions in Lemma 20.5 were tractable in the finite row chart with \(p(c+1)\ne0\). Their binary types would select one of the nine pairs \(\tau\). Their parameters would satisfy \(G_1=G_2=G_4=G_5=0\), and choosing \(z=1/(p(c+1))\) would also give \(G_3=0\). Evaluating the boxed identity would then give \(0=1\), a contradiction. The row at infinity and the case \(a=0\) are proved separately in the text. This completes the logical connection between the coefficient certificates and the entire parameter domain of that lemma.
\end{proof}

\subsection{A.4 Supplementary material}\label{a.4-supplementary-material}

The supplement contains the nine coefficient arrays and their independent rational-arithmetic verifier, the finite-group invariant calculation, the order-five two-pattern elimination, and the remaining symbolic matrix identities. The certificate data are mathematical inputs to the verification in Section A.3; the other programs provide redundant checks of identities displayed in the paper. The input signatures, their matrices, all denominators divided out, and the exceptional charts are specified in the text. Consequently the role of each calculation is a polynomial identity or a finite algebraic exclusion, while the graph substitutions and interpolation reductions have their separate proofs in the preceding sections.

\section{Appendix B. Real weights: simplified tools and instructive examples}\label{appendix-b.-real-weights-simplified-tools-and-instructive-examples}

Corollary A.1 is obtained by restricting the complex dichotomy and identifying its real tractable intersection. There are also useful simplifications in the proof tools when every matrix and unary under consideration is real. We record them with their own assumptions. The examples then explain three distinct issues: arbitrarily large finite eigenvalue orders, a ternary gadget that maps a hard signature to a tractable one, and a limitation of equality-preserving basis changes.

\subsection{B.1 Absorption and binary contractions over the reals}\label{b.1-absorption-and-binary-contractions-over-the-reals}

\begin{holantstatement}{Proposition B.1 (the smaller real exceptional sets)}

Fix a real algebraic symmetric Boolean ternary signature \(f\), and let \(e=[1,0,0,1]\). All gadgets and simulations in the first assertion use \(f\) on the left and \(e\) on the right. For a nonzero real column \(u\), the absorption factor \(Q(u)=u_0^3+u_1^3\) vanishes exactly on the line \(u_0+u_1=0\). Consequently, if an actual straddled gadget over this pair has a matrix \(H\) with distinct nonzero real eigenvalues and nonzero trace, at least one of its two row eigenvectors is simulable.

In addition, let \(f=[1,a,b,c]\) be real algebraic with \(c-ab\ne0\) and \((a,b)\ne(0,0)\). Among nonzero real rows up to scalar, at most four directions give a tractable binary contraction

\[F_f(r,s)=[r+as,ar+bs,br+cs].\]

Hence five individually simulable real row directions suffice to prove hardness.

\end{holantstatement}

\begin{proof}[Proof]

The factorization

\[Q(u)=(u_0+u_1)(u_0^2-u_0u_1+u_1^2),\qquad
u_0^2-u_0u_1+u_1^2=(u_0-u_1/2)^2+3u_1^2/4\]

shows that its second factor is positive for a nonzero real column. The two column eigenspaces of \(H\) are distinct, so at least one avoids the zero line. Its spectral projector \(uv\) is simulable: the eigenvalue ratio is real and is neither 1 nor \(-1\), since the eigenvalues are distinct, nonzero, and have nonzero sum. Lemmas P.2 and P.3 therefore supply its row factor \(v\).

For the second assertion, restrict the binary table in Section 3.4 to real entries. The extra affine points reduce to \([1,1,-1]\) and \([1,-1,-1]\), since the only real solutions to \(z^{12}=1\) are \(z=\pm1\). Write the output as \(g=[g_0,g_1,g_2]\). Its endpoints obey

\[ (g_0,g_2)=(r,s)\begin{pmatrix}1&b\\a&c\end{pmatrix}.\]

Invertibility excludes zero and disequality outputs for nonzero rows and makes \(F_f\) injective on directions. Rank one requires the nonzero homogeneous quadratic

\[(b-a^2)r^2+(c-ab)rs+(ac-b^2)s^2=0.\]

It has at most two projective roots: in the chart \(s\ne0\) it is a polynomial of degree at most two in \(r/s\), and the possible direction \(s=0\) contributes at most the missing degree. Diagonality requires \(ar+bs=0\), which gives one direction because \((a,b)\ne(0,0)\). Both extra affine points require

\[g_0+g_2=(1+b)r+(a+c)s=0.\]

Its coefficients cannot both vanish, since \(b=-1,c=-a\) would imply \(c-ab=0\). Thus the two affine types together contribute at most one further direction. The total is at most four. Of five individually available directions select one outside this set and apply Lemma P.5. Only that one unary is used in the final reduction.
\end{proof}

The first statement has no direct complex analogue with a single forbidden line: over \(\mathbb C\) the same cubic has three projective zeros, and both spectral columns may be forbidden. Similarly, the real affine binaries lie on the one line \(g_0+g_2=0\), whereas the twelve complex affine points require the conic argument of Lemma 4. These are specific reasons why a real proof does not transfer merely by changing its coefficient field.

\subsection{B.2 An elementary real-matrix recurrence}\label{b.2-an-elementary-real-matrix-recurrence}

\begin{holantstatement}{Lemma B.2 (a finite product with real spectrum)}

Let \(A\) be a real two-by-two matrix with nonreal eigenvalues, and let \(T\) be real and invertible with \(AT\ne TA\). Assume either

\[\det T>0,\]

or

\[\det T<0\quad\text{and}\quad
\bigl(\operatorname{tr}T\ne0\ \text{or}\ \operatorname{tr}(AT)\ne0\bigr).\]

Then a finite product \(H\) of copies of \(A,T\), with no inverse factors, satisfies

\[\det H\ne0,\qquad (\operatorname{tr}H)^2>4\det H,\qquad
\operatorname{tr}H\ne0.\]

If \(A,T\) are actual straddled matrices for fixed real algebraic \(f\), this product is an actual fixed gadget and its eigenvalue ratio is not a root of unity.

\end{holantstatement}

\begin{proof}[Proof]

In a real basis write

\[A=\begin{pmatrix}p&-q\\q&p\end{pmatrix},\qquad q\ne0,\qquad
T=\begin{pmatrix}r&s\\t&u\end{pmatrix}.\]

This form of \(A\) follows by taking real and imaginary parts of a complex eigenvector; the basis need not be orthogonal. Direct multiplication gives

\[\det(AT-TA)=-q^2\bigl((s+t)^2+(r-u)^2\bigr)<0.\]

Equality would make \(T\) commute with \(A\), contrary to the hypothesis.

Suppose first that \(T\) also has nonreal eigenvalues. Both determinants are positive. The two-by-two trace identity gives

\[\operatorname{tr}(ATA^{-1}T^{-1})
=2-\frac{\det(AT-TA)}{\det A\det T}>2.\]

To verify the identity, put \(X=AT,Y=TA\). They have the same trace \(\tau\) and determinant \(d=\det A\det T\). The formulas

\[\det(X-Y)=2d-\tau^2+\operatorname{tr}(XY),\qquad
\operatorname{tr}(XY^{-1})=\frac{\tau^2-\operatorname{tr}(XY)}d\]

give the result because \(XY^{-1}=ATA^{-1}T^{-1}\).

Normalize the matrices for this algebraic argument to

\[\widehat A=A/\sqrt{\det A},\qquad
\widehat T=T/\sqrt{\det T}.\]

Each is similar over the reals to a rotation. Nonnegative powers of a rotation approach its inverse: for an irrational rotation angle, the pigeonhole principle supplies unbounded positive exponents whose powers approach the identity, and decreasing those exponents by one gives the inverse in the limit. For a rational angle, the inverse is already a nonnegative power. Similarity by a fixed matrix preserves this convergence. This argument applies to each matrix separately and does not require them to share a rotation basis.

The normalized commutator has trace strictly greater than two. By continuity there are finite nonnegative integers \(m,n\) such that

\[\operatorname{tr}(\widehat A\widehat T\widehat A^m\widehat T^n)>2.\]

This finite matrix has determinant one and two distinct positive real eigenvalues. The actual product \(H=ATA^mT^n\) is a positive scalar multiple, so it satisfies the claimed inequalities. The inverse matrices were used only to prove existence of a finite word; they are not factors of the gadget.

Next suppose \(T\) has real eigenvalues and positive determinant. Distinct eigenvalues have the same sign and a nonzero sum, so take \(H=T\). If its eigenvalues coincide, noncommutation implies that \(T\) is not scalar. Write

\[T=\mu(I+N),\qquad \mu\ne0,\quad N\ne0,\quad N^2=0.\]

In a real basis with \(N=\left(\begin{smallmatrix}0&1\\0&0\end{smallmatrix}\right)\), the scalar \(\operatorname{tr}(AN)=A_{21}\) is nonzero. Otherwise \(A\) would be upper triangular over the reals and could not have nonreal eigenvalues. Therefore

\[\operatorname{tr}(A(I+kN))=\operatorname{tr}A+k\operatorname{tr}(AN)\]

is unbounded in absolute value, whereas \(\det(A(I+kN))=\det A>0\) is fixed. For a sufficiently large fixed positive integer \(k\), take \(H=AT^k\).

Finally, if \(\det T<0\) and \(\operatorname{tr}T\ne0\), take \(H=T\). If its trace is zero, the assumption gives \(\operatorname{tr}(AT)\ne0\); take \(H=AT\), whose determinant is negative. A negative determinant gives two nonzero real eigenvalues of opposite signs, and their nonzero sum excludes ratio \(-1\).

Every selected product uses actual factors when \(A,T\) are actual. Its length depends only on the fixed signature. If desired, enumerate words and test the displayed algebraic inequalities exactly until a qualifying word is found; its existence has just been proved. Subsequent oracle graphs use that fixed finite word, not a matrix limit or approximate signature.
\end{proof}

This lemma explains how a real proof can bypass arbitrarily large finite rotation orders without classifying them separately. It does not resolve every real branch by itself: a singular \(T\), or the negative-determinant case with both traces zero, needs additional actual matrices or direct contractions. The complex proof treats those boundaries as part of its full finite-order analysis.

\subsection{B.3 Finite orders are unbounded even for real algebraic weights}\label{b.3-finite-orders-are-unbounded-even-for-real-algebraic-weights}

\begin{holantstatement}{Example B.3 (arbitrary root-of-unity orders)}

For every integer \(m\ge3\), the real algebraic signature

\[f=[1,-t^2,1,1],\qquad t=\tan(\pi/m)\]

has a first straddled matrix with eigenvalue ratio of order exactly \(m\).

\end{holantstatement}

\begin{proof}[Proof]

Its first matrix is

\[A_f=\begin{pmatrix}1&1\\-t^2&1\end{pmatrix}.\]

The characteristic polynomial is \((\lambda-1)^2+t^2\), so the eigenvalues are \(1\pm it\). With \(\theta=\pi/m\),

\[\frac{1+i\tan\theta}{1-i\tan\theta}
=\frac{\cos\theta+i\sin\theta}{\cos\theta-i\sin\theta}
=e^{2\pi i/m}.\]

This is a primitive \(m\)th root of unity. The parameter is algebraic: for \(\zeta=e^{2\pi i/m}\), one has \(t=-i(\zeta-1)/(\zeta+1)\), and \(\zeta+1\ne0\) for \(m\ge3\). It is real by its defining trigonometric expression.
\end{proof}

For rational entries of a two-by-two matrix, the eigenvalues and their ratio lie in an extension of degree at most two over \(\mathbb Q\). A root of unity in such a field has order \(1,2,3,4\), or \(6\). The example shows why that finite list in the rational argument \cite{ref5} does not cover real algebraic weights. For any fixed finite order \(m\), the powers of this particular \(A_f\) supply only finitely many directions; they do not provide arbitrarily many interpolation nodes for graphs of unbounded size.

\subsection{B.4 A tractable ternary output from a hard input}\label{b.4-a-tractable-ternary-output-from-a-hard-input}

\begin{holantstatement}{Example B.4 (the first ternary replacement can lose hardness)}

Let \(r=2^{-1/3}\) and \(f=[1,-r^2,r,-1]\). The actual ternary gadget \(\Phi(f)\) of Section 2.2 is rank one, but \(\operatorname{Holant}(f\mid e)\) is \#P-hard. A direct proof uses the actual serial product \(H=A_fB_f\) and one right unary obtained by interpolation and absorption.

\end{holantstatement}

\begin{proof}[Proof]

Using \(r^3=1/2\) in the formula for the three-copy ternary gadget gives

\[\Phi(f)=[1,-r^2,r^4,-r^6]=(1,-r^2)^{\otimes3}.\]

Thus this particular replacement produces a tractable signature. For the original signature the two actual matrices and their product are

\[A=\begin{pmatrix}1&r\\-r^2&-1\end{pmatrix},\qquad
B=\begin{pmatrix}1/2&-r/2\\0&1/2\end{pmatrix},\qquad
H=AB=\begin{pmatrix}1/2&0\\-r^2/2&-1/4\end{pmatrix}.\]

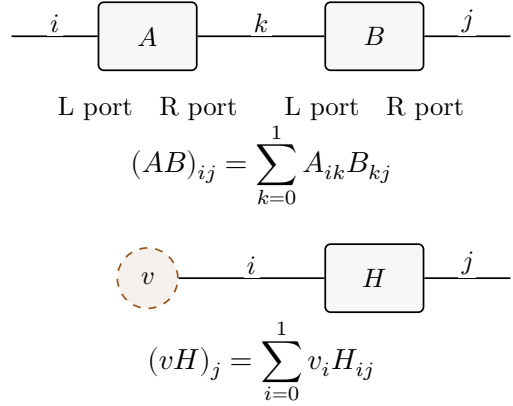
\begin{figure}[H]
\centering

\begin{tikzpicture}[x=1cm,y=1cm]
\node[boxg] (a) at (0,0) {$A$};\node[boxg] (b) at (3,0) {$B$};
\draw (-1.8,0)--node[above,lab] {$i$}(a.west);
\draw (a.east)--node[above,lab] {$k$}(b.west);
\draw (b.east)--node[above,lab] {$j$}(4.8,0);
\node[below=2mm of a] {\small L port\quad R port};
\node[below=2mm of b] {\small L port\quad R port};
\node at (1.5,-1.7) {$(AB)_{ij}=\displaystyle\sum_{k=0}^1 A_{ik}B_{kj}$};
\node[R,dashed] (v) at (0,-3.2) {$v$};\node[boxg] (h) at (3,-3.2) {$H$};
\draw (v.east)--node[above,lab] {$i$}(h.west);
\draw (h.east)--node[above,lab] {$j$}(4.8,-3.2);
\node at (1.5,-4.3) {$(vH)_j=\displaystyle\sum_{i=0}^1 v_iH_{ij}$};
\end{tikzpicture}

\caption{Port order determines matrix order. Each box has its L port on the left and R port on the right. Joining an R port to an L port is legal. The lower picture assumes that the RHS unary $v$ has already been simulated; its circular shape indicates its side, not that it is an equality vertex.}
\label{fig:composition-17}
\end{figure}

The eigenvalues of \(H\) are \(1/2,-1/4\), with ratio \(-2\). Its projector for \(1/2\) is

\[P=\frac{H+I/4}{3/4}
=\begin{pmatrix}1&0\\-2r^2/3&0\end{pmatrix}=uv,\qquad
u=(1,-2r^2/3)^T,\quad v=(1,0).\]

Lemma P.2 simulates this projector. Its column contraction is

\[Q(u)=1-8r^6/27=25/27\ne0,\]

so Lemma P.3 supplies the single right unary \(v\). Attaching it to one port of \(f\) gives

\[F_f(v)=[1,-r^2,r],\qquad
F_f(v)_0F_f(v)_2-F_f(v)_1^2=r-r^4=r/2\ne0.\]

\begin{figure}[H]
\centering

\begin{tikzpicture}[x=1cm,y=1cm]
\node[L] (f) at (0,0) {$f$};\node[R,dashed] (v) at (3,0) {$v$};
\draw (-1.7,0.9)--node[above,lab] {$x$}(f.north west);
\draw (-1.7,-0.9)--node[below,lab] {$y$}(f.south west);
\draw (f.east)--node[above,lab] {$z$}(v.west);
\node[align=center] at (1,-1.7) {$F(v)(x,y)=\displaystyle\sum_{z=0}^1 f(x,y,z)v_z$};
\end{tikzpicture}

\caption{Attach one simulated RHS unary to a left ternary vertex. Its two other ports remain free, giving a binary signature. The dashed unary must later be eliminated by its established Turing reduction; it is not assumed available in the original input.}
\label{fig:contraction-18}
\end{figure}
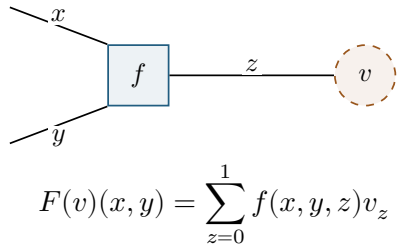

Both endpoints and the middle entry are nonzero, excluding disequality and diagonal cases. The endpoint sum \(1+r>0\) excludes the two real affine points, and the displayed determinant excludes rank one. The real specialization of the binary classification therefore makes this binary hard. Lemma P.5 completes the reduction to the original signature.
\end{proof}

This example concerns a reduction, not a contradiction to monotonicity of hardness: the actual ternary gadget gives \(\operatorname{Holant}(\Phi(f)\mid e)\le_T\operatorname{Holant}(f\mid e)\), which permits a tractable problem to reduce to a hard one. The successful binary contraction supplies the reduction in the direction needed for a hardness proof.

\subsection{B.5 An obstruction to a single equality-preserving basis change}\label{b.5-an-obstruction-to-a-single-equality-preserving-basis-change}

One might try to reduce the real problem to the known nonnegative classification by making \(f\) nonnegative under a holographic transformation while keeping equality on the other side. A negative entry in the original basis does not itself rule out that approach. The following invariant does rule it out for a particular signature.

\begin{holantstatement}{Proposition B.5 (a similarity invariant)}

Let \(f\) be a symmetric ternary signature with \(A_f\) invertible. Suppose an invertible, possibly complex matrix \(T\) gives

\[g=T^{\otimes3}f,\qquad
(T^{-T})^{\otimes3}e=\kappa e,\qquad\kappa\ne0,\]

using the column-tensor convention of Section 2.5. Then \(A_g\) is invertible and

\[\mathcal I(A_g)=\mathcal I(A_f),\qquad
\mathcal I(M)=\frac{(\operatorname{tr}M)^2}{\det M}.\]

For \(f=[1,-\sqrt2,1,1]\), no such transformation can make \(g\) entrywise nonnegative.

\end{holantstatement}

\begin{proof}[Proof]

Transform both vertices of the two-vertex straddled gadget. Cancellation on its two internal edges gives \(TA_fT^{-1}\). On the other hand, its transformed right vertex is \(\kappa e\), so it has the same signature as the gadget defining \(A_g\), multiplied by \(\kappa\) once. Consequently

\[\kappa A_g=TA_fT^{-1}.\]

The right side preserves determinant and trace under similarity; multiplying a two-by-two matrix by \(\kappa\) multiplies trace by \(\kappa\) and determinant by \(\kappa^2\). Thus \(A_g\) is invertible and the two scalar factors cancel in \(\mathcal I\).

For a nonnegative \(g=[w,x,y,z]\),

\[A_g=\begin{pmatrix}w&y\\x&z\end{pmatrix},\qquad
(\operatorname{tr}A_g)^2-4\det A_g=(w-z)^2+4xy\ge0.\]

If its determinant is positive, \(\mathcal I(A_g)\ge4\); if its determinant is negative, \(\mathcal I(A_g)\le0\). A zero determinant was excluded above. The proposed source has

\[A_f=\begin{pmatrix}1&1\\-\sqrt2&1\end{pmatrix},\qquad
\mathcal I(A_f)=\frac4{1+\sqrt2}=4(\sqrt2-1)\in(0,4),\]

contradicting the invariant required of the nonnegative target.
\end{proof}

The hypothesis that the transformed right signature remains a scalar equality is essential. The proposition excludes that single basis-change shortcut, including complex basis changes; it makes no claim against a reduction using additional gadgets. The matrix convention here is the same as throughout the paper, so the straddled similarity is \(TA_fT^{-1}\).


\clearpage
\begin{thebibliography}{99}
\bibitem{ref1}
Leslie G. Valiant. \emph{Holographic Algorithms}. SIAM Journal on Computing 37(5), 1565--1594 (2008). \href{https://doi.org/10.1137/070682575}{Publisher}.

\bibitem{ref2}
Jin-Yi Cai, Pinyan Lu, Mingji Xia. \emph{Holant Problems and Counting CSP}. Proceedings of STOC 2009, 715--724 (2009). \href{https://doi.org/10.1145/1536414.1536511}{Publisher}.

\bibitem{ref3}
Jin-Yi Cai, Sangxia Huang, Pinyan Lu. \emph{From Holant to \#CSP and Back: Dichotomy for Holant\(^c\) Problems}. Algorithmica 64, 511--533 (2012). \href{https://doi.org/10.1007/s00453-012-9626-6}{Publisher}, \href{https://pages.cs.wisc.edu/~jyc/papers/holant-c.pdf}{author manuscript}.

\bibitem{ref4}
Austen Z. Fan, Jin-Yi Cai. \emph{Dichotomy Result on 3-Regular Bipartite Non-negative Functions}. Theoretical Computer Science 949, 113745 (2023). \href{https://doi.org/10.1016/j.tcs.2023.113745}{Publisher}, \href{https://arxiv.org/abs/2011.09110}{preprint}.

\bibitem{ref5}
Jin-Yi Cai, Austen Z. Fan, Yin Liu. \emph{Bipartite 3-Regular Counting Problems with Mixed Signs}. Journal of Computer and System Sciences 135, 15--31 (2023). \href{https://doi.org/10.1016/j.jcss.2023.01.006}{Publisher}, \href{https://arxiv.org/abs/2110.01173}{preprint}.

\bibitem{ref6}
Jin-Yi Cai, Austen Z. Fan. \emph{Planar 3-way Edge Perfect Matching Leads to A Holant Dichotomy}. arXiv:\allowbreak{}2303.16705v1 (2023); also posted as SSRN 5102132 (2025). \href{https://arxiv.org/abs/2303.16705v1}{arXiv version}, \href{https://papers.ssrn.com/sol3/papers.cfm?abstract_id=5102132}{SSRN version}.

\bibitem{ref7}
Michael Kowalczyk, Jin-Yi Cai. \emph{Holant Problems for 3-Regular Graphs with Complex Edge Functions}. Theory of Computing Systems 59(1), 133--158 (2016). \href{https://doi.org/10.1007/s00224-016-9671-7}{Publisher}. The preprint, \emph{Holant Problems for Regular Graphs with Complex Edge Functions}, is \href{https://arxiv.org/abs/1001.0464}{arXiv:\allowbreak{}1001.0464}.

\bibitem{ref8}
Sangxia Huang, Pinyan Lu. \emph{A Dichotomy for Real Weighted Holant Problems}. Computational Complexity 25, 255--304 (2016). \href{https://doi.org/10.1007/s00037-015-0118-3}{Publisher}, \href{https://sangxia.github.io/pubs/pdf/real-holant.pdf}{author manuscript}. Conference version: CCC 2012, 96--106.

\bibitem{ref9}
Jiabao Lin. \emph{On the Complexity of \#CSP\(^d\)}. Proceedings of ITCS 2021, LIPIcs 185, 40:1--40:10 (2021). \href{https://doi.org/10.4230/LIPIcs.ITCS.2021.40}{Proceedings}.

\bibitem{ref10}
Jin-Yi Cai, Pinyan Lu, Mingji Xia. \emph{The Complexity of Complex Weighted Boolean \#CSP}. Journal of Computer and System Sciences 80(1), 217--236 (2014). \href{https://doi.org/10.1016/j.jcss.2013.07.003}{Publisher}, \href{https://pages.cs.wisc.edu/~jyc/papers/cw-csp.pdf}{author manuscript}.

\bibitem{ref11}
Jin-Yi Cai, Michael Kowalczyk, Tyson Williams. \emph{Gadgets and Anti-Gadgets Leading to a Complexity Dichotomy}. ACM Transactions on Computation Theory 11(2), Article 7 (2019). \href{https://arxiv.org/abs/1108.3383}{Preprint}.

\bibitem{ref12}
Boning Meng, Juqiu Wang, Mingji Xia, Jiayi Zheng. \emph{From an Odd Arity Signature to a Holant Dichotomy}. Proceedings of CCC 2025, LIPIcs 339, 23:1--23:20 (2025). \href{https://doi.org/10.4230/LIPIcs.CCC.2025.23}{Proceedings}.

\bibitem{ref13}
Chenghua Liu, Boning Meng, Juqiu Wang. \emph{A Full Complexity Dichotomy for Complex-Valued Boolean Holant Problems}. Undated preprint, available from the authors. \href{https://liuchliuch.github.io/files/complex_holant_lch_mbn_wjq.pdf}{Manuscript}, \href{https://liuchliuch.github.io/}{author publication page}.

\bibitem{ref14}
Yuan Huang, Zhiguo Fu. \emph{The Computational Complexity of Holant Problems on 4-regular Graphs from the Stable Subgroup Sequence of \(SL(2,\mathbb C)\)}. arXiv:\allowbreak{}2609.11175v1 (2026). \href{https://arxiv.org/abs/2609.11175v1}{Preprint}.

\bibitem{ref15}
Mingji Xia. \emph{The Framework to Unify All Complexity Dichotomy Theorems for Boolean Tensor Networks and Klein Group Upper Case}. arXiv:\allowbreak{}2603.09417v3 (2026). \href{https://arxiv.org/abs/2603.09417v3}{Preprint}.

\bibitem{ref16}
Arnaud Beauville. \emph{Finite Subgroups of \(\operatorname{PGL}_2(K)\)}. In Vector Bundles and Complex Geometry, Contemporary Mathematics 522, 23--29 (2010). \href{https://doi.org/10.1090/conm/522/10289}{Publisher}, \href{https://math.univ-cotedazur.fr/~beauvill/pubs/PGL\%282\%29.pdf}{author manuscript}.

\bibitem{ref17}
Jean-Pierre Serre. \emph{Bounds for the Orders of the Finite Subgroups of \(G(k)\)}. In Group Representation Theory, EPFL Press, 405--450 (2007). \href{https://www.college-de-france.fr/media/jean-pierre-serre/UPL3821667391778701726_6___Bounds_for_the_orders.pdf}{Author manuscript}, \href{https://arxiv.org/abs/1011.0346}{arXiv version}.

\bibitem{ref18}
Paul Dagum, Michael Luby. \emph{Approximating the Permanent of Graphs with Large Factors}. Theoretical Computer Science 102(2), 283--305 (1992). \href{https://doi.org/10.1016/0304-3975(92)90234-7}{Publisher}.

\bibitem{ref19}
Peng Yang, Yuan Huang, Zhiguo Fu. \emph{The Computational Complexity of Holant Problems on 3-Regular Graphs}. Theoretical Computer Science 982, 114256 (2024). \href{https://doi.org/10.1016/j.tcs.2023.114256}{Publisher}.

\bibitem{ref20}
Junda Li, Yuan Huang, Yan-Lin Zheng. \emph{Dichotomy for Non-negative Valued Holant Problems on 3-Regular Bipartite Graphs}. Theory of Computing Systems 69, Article 4 (2025). \href{https://doi.org/10.1007/s00224-024-10206-7}{Publisher}.

\bibitem{ref21}
Peng Yang, Yuan Huang, Zhiguo Fu. \emph{A complexity trichotomy for k-regular asymmetric spin systems with complex edge functions}. Theoretical Computer Science 1020, 114835 (2024). \href{https://doi.org/10.1016/j.tcs.2024.114835}{Publisher}.

\end{thebibliography}
\end{document}